\documentclass[pdftex,twoside]{book}
\usepackage{amsmath}
\usepackage{amsthm}
\usepackage{amssymb}
\usepackage{graphicx}
\usepackage{braket}
\usepackage{verbatim}
\usepackage[margin=1in]{geometry}
\usepackage{leftidx}
\usepackage{wrapfig}
\usepackage{multicol}
\usepackage{array}
\usepackage{subfig}
\usepackage[T1]{fontenc}
\usepackage{fancyhdr}

\makeatletter

\newenvironment{figurehere}
{\def\@captype{figure}}
{}
\makeatother

\numberwithin{equation}{chapter}

\title{Tunneling with Tamm-Dancoff method\protect\footnote{
Master thesis defended at Jagiellonian University in Krakow, Department of Physics, Astronomy and Applied Computer Science. Written under supervision of Prof. Jacek Wosiek.}}
\author{Zbigniew Ambrozi\'nski\footnote{zbigniew.ambrozinski@uj.edu.pl}}

\begin{document}
\maketitle
\section*{Acknowledgments}
I first express my gratitude towards my supervisor, Prof. Jacek Wosiek who gave me inspiration and guidance while I was working on the dissertation.
Also, I would like to thank Prof. Meurice Yannick, Prof. Michael Teper and Dr. Mithat \"Unsal for very helpful discussions and useful advices. I thank Foundation for Polish Science whose scholarship made it possible for me to continue dealing with physics. Finally, I am grateful to my wife who is always beside me.

This work was supported by Foundation for Polish Science MPD Programme co-financed by the European Regional Development Fund, agreement no. MPD/2009/6.

\tableofcontents
\chapter{Introduction}
The first sign of quantum tunneling effect, decay of metastable state, was observed by H. Becquerel in 1886 in radioactive uranium. Explanation of this phenomenon could be based on Louis de Broglie proposition that matter has properties of waves which was given in 1923. In this formalism particles penetrate potential barriers in the same way in which light is transmitted through absorbing medium. A quantitative result could not be given without Schr\"odinger equation and Born's probabilistic interpretation of quantum mechanical wavefunctions introduced in 1920s. In 1927 F. Hund studied an electron in potential with two or more minima separated by a barrier which is classically impenetrable. This was the first theoretical approach to the tunneling effect. Since then, the double well potential became a standard example of a system with tunneling and was considered as a toy model for more complex theories.

In a potential with multiple minima a few lowest energies are degenerate in the domain of perturbative calculus. Tunneling is responsible for splitting the energies. In general, this effect cannot be studied analytically. One can give a first approximation to the splitting using semiclassical--approximation (or WKB approximation) which was developed by G. Wentzel, H. Kramers and L. Brillouin in 1926. In this approach, one finds that the ground energy is shifted by a quantity which a nonperturbative function of action of the classical solution in Euclidean space. This classical solution is called an instanton.

Validity of the instanton calculus is limited to systems with widely separated minima divided by a large barrier of potential. Nevertheless, it has vast applications to modern field theories. In Yang--Mill theory with $SU(2)$ symmetry, classical fields which are constant at infinity may be considered as mappings of $SU(2)$ group onto itself (see \cite{ITEP}). These fields may have a nontrivial topological structure and thus can be divided into topological sectors. It turns out that all sectors can be labeled by a Pontryagin index which takes integer values and is a topological invariant. In each sector there can be found a topological vacuum $\ket n$. One vacuum cannot be continuously deformed into another due to different topological properties. Therefore, a perturbation analysis about one of vacua does not take into account different sectors. The Pontryagin index plays an analogous role to the impenetrable potential barrier in quantum mechanics. An instanton in Yang--Mills theory is a classical field satisfying equations of motion in Euclidean space which connects two such vacua. It was discovered by A. Belavin, A. Polyakov, A. S. Schwartz and Yu. S. Tyupkin in 1975 \cite{BPST} and named a BPST instanton. Presence of the BPST instanton in the theory has serious consequences for the structure of vacuum in QCD. Semiclassically, trajectories which satisfy equations of motion in the Euclidean space form instanton liquids which are ensembles of interacting instantons (see T. Schafer and E. Shuryak \cite{Schafer}). For this reason the true QCD vacuum is a superposition of topological vacua $\ket n$ multiplied by phases $e^{in\theta}$ where $\theta$ is so called vacuum angle.

Such system can be modeled in one dimensional quantum mechanics by a periodic potential. According to Bloch theorem, the energy spectrum consists of continuous bands. Each energy in a band is labeled by an angle $\theta$ which appears in energy states in the phase $e^{in\theta}$ multiplying topological vacua. Unlike in the quantum mechanical case, the vacuum angle enters into the Lagrangian of Yang--Mills theory and only one value of $\theta$ is admissible. No energy bands are present and there is usual mass gap between vacuum and the first excited state. The vacuum angle in QCD is responsible for violating CP symmetry. On the other hand, there is no experimental evidence for CP breaking which imposes a limit on the angle $|\theta|<10^{-9}$.

The anharmonic oscillator attracted a continuous attention of physicists sine 1960s \cite{Bender,Caswell,DuncanJones,Meurice}. Its double well version with tunneling effect was extensively studied by J. Zinn--Justin, E.B. Bogomolny and many others in 1970s and 1980s \cite{ZJ,ZJ1,Bogomolny}. It was discovered that there are further corrections to the WKB approximation which can be derived from modified Bohr--Sommerfeld quantization condition \cite{ZJ2}. There are contributions from $n$--instanton molecules (i.e. a classical path in Euclidean spacetime which is composed of instantons that are close to each other) which contribute to the ground energy much less then the traditional instanton. Secondly, each instanton molecule contribution (including single instantons) is multiplied by a series, which is presumably asymptotic. Moreover, as stated recently by M. \"Unsal \cite{Unsal}, interactions between instantons can heal non Borel summability of perturbation series for potentials with degenerate global minima.

One has to keep in mind that instanton considerations neglect perturbative contributions to energies which are much larger. Secondly, there are higher order corrections to instanton contributions which become significant at stronger coupling. It is understandable that there is a need to verify statements concerning instantons and see in what regime of the coupling constant the instanton picture is valid. In quantum mechanics there is a very efficient method, called cut Fock space method (see J. Wosiek \cite{Wosiek:2002}), which we have at hand. It originates from the variational Tamm--Dancoff method. In \cite{Dancoff} S. Dancoff studied the ground energy of fields of two relativistic nucleons. He formulated an eigenvalue problem (which is the time independent Schr\"odinger equation) and proposed a trial wavefunction which would represent collapsing nucleons. Energy which was obtained turned out to be higher than the energy from former adiabatic approximation. It meant that no collapse would take place. It shows that this variational approach needs a good understanding of physics of the system under consideration to propose an adequate trial function. In the cut Fock space method one takes basis states of Fock space $\ket{n}$ with $n$ smaller than a certain cut--off which is supposed to be large. A price for taking so many states rather than a few trial functions is that the calculations have to be performed numerically. On the other hand, it is very efficient at least in three dimensional quantum mechanics. Indeed, convergence of energies with growing cut--off has been observed numerically \cite{Wosiek}. Accuracy of this method is limited only by precision of computations and the size of the Hamiltonian (appearing in Schr\"odinger equation) which is precisely equal to the cut--off. Apart from these limitations, this method is exact and is then a powerful tool for testing WKB approximation.

Both, cut Fock space method and instanton calculus start from the Hamilton operator. Let us then shortly discuss dimensional analysis of the Hamiltonian which will simplify notation. A Hamiltonian $H$ may be given in a form
\begin{align}
\mathbf H=\frac{1}{2m}\mathbf P^2+\mathbf{ \hat V}.
\end{align}
The potential $\mathbf{\hat V}$ may be given in a form
\begin{align}
\mathbf{\hat V}=m\omega^2 \mathbf{a}^2  V(\mathbf X/\mathbf a)
\end{align}
where $V(x)$ is a real function of a dimensionless parameter. The function $V(x)$ can have arbitrary shape as long as it is bounded from below. The parameter $\mathbf a$ is scale of the potential. The semiclassical approximation which will be addressed in this dissertation is valid when $\mathbf a$ is large. In fact, the same limit can be obtained by taking $\hbar$ to be small which is the classical limit. Dimensions of given operators and parameters are:
\begin{align*}
[\mathbf H]&=kg\ cm^2\ s^{-2}, &[\mathbf X]&=cm,\\
[\mathbf P]&=kg\ cm\ s^{-1},\\
[m]&=kg,&[\omega]&=s^{-1},\\
[\mathbf a]&=cm,&[\hbar]&=kg\ cm^2\ s^{-1}.
\end{align*}
The dimensionfull operators can be then substituted with dimensionless ones.
\begin{align}
P&=\frac{1}{\sqrt{\hbar m \omega}}\mathbf P&X&=\sqrt\frac{m\omega}{\hbar}\mathbf X\\
H&=\frac{1}{\hbar \omega}\mathbf H&a&=\sqrt\frac{m\omega}{\hbar}\mathbf a.
\end{align}
Then the dimensionless Hamiltonian is
\begin{align}
H=\frac{1}{2} P^2+\hat V=\frac{1}{2} P^2+ a^2 V(X/a).
\end{align}
We use the symbol $\hat{}$ only do distinguish between scaled and non--scaled potentials.

The idea behind introducing the scale parameter $a$ is that for large value of $a$ minima of the potential are separated by a high barrier and are far away from each other. It means that they interact weakly. It is common for the semiclassical approximation: there is a macroscopic scale $a$ which is big compared to quantum length scale $\sqrt{m\omega/\hbar}$. On the other hand, usually a small coupling constant rather than a large parameter is responsible for weak interactions. For this reason we introduce the coupling constant
\begin{align}
g=\frac{1}{a^2}.
\end{align}
Let us assume that the function $V(x)$ has a minimum at $x=x_0/a$ and $V(x_0/a)=0$. Then
\begin{align}
\hat V=\frac{V''(x_0)}{2}(X-x_0)^2+\mathcal O(\sqrt g)
\end{align}
and the system is a perturbed harmonic oscillator centered at $x_0$. We will be mainly using the coupling constant $g$ whereas the scale parameter $a$ will be used for convenience in instanton calculus.

Plan of the thesis is the following. In chapter \ref{ch:anhramonic_oscillator} we introduce the cut Fock space method on example of anharmonic oscillator with $X^4$ interaction. A typical convergence of the energy levels with growing cut--off is presented. Then results are compared with another numerical technique -- shooting method which employees the standard Runge--Kutta algorithm for solving differential equations. In chapter \ref{ch:double_well} we turn to the double well potential. The standard instanton calculus is presented and then compared with numerical results obtained with the cut Fock space method. As it is well known, the WKB approximation in Euclidean space give relevant information only on the energy splitting and not on energies themselves because it does not include perturbative corrections to energies. For this reason only difference of the two lowest energies is compared with numerical results. In chapter \ref{ch:periodic_potentials} the cosine potential in the weak coupling limit is considered. It is observed that in a periodic space with periodic boundary conditions $\psi(0)=\psi(Ka)$ where there are $K$ minima of the potential within one period of the wavefunction, there are $K$ energies that are split by a nonperturbative quantity. The splittings are obtained in the WKB approximation and with the cut Fock space method for $K=2,3$. The case of $K=\infty$ which is closest to the Yang--Mills theory is also addressed by both techniques. Due to discrete translation symmetry occurring in this case, the cut Fock space method turns out to be very efficient. In the last chapter we study the anharmonic triple well potential. It is bound to have different expansions about different minima. This is why there might be no tunneling between all three minima. In order to have it, the potential has to be fine tuned.

\chapter{Anharmonic oscillator}\label{ch:anhramonic_oscillator}
The aim of this chapter is to introduce the cut Fock space method and to compare it with shooting method.
To do this we deal with the anharmonic oscillator using both approaches.
For the cut Fock space method on needs matrix representation of the Hamiltonian. Its relatively simple structure makes this task feasible when one uses eigenbasis of the occupation number operator. Eigenvalues of the truncated matrix approximate energies of the system. The approximation gets better as the cut--off grows.
Parallelly the shooting method is applied. Symmetries of the system allows one to reduce the Schr\"odinger equation $H\psi=E\psi$ to a differential equation on the interval $[0,\infty)$ (rather than on $(-\infty,+\infty)$) with initial conditions at the origin. Standard numerical techniques can be applied to such equation.
At the end of the chapter agreement of both techniques is checked.  The two methods are placed in different representations of the Hilbert space and are thus completely independent. Agreement of results implies that they are correct.

In this chapter we will be using Hamiltonian in parametrisation
\begin{align}\label{eq:chapter1_hamiltonian}
H=\frac{1}{2}P^2+\frac{\epsilon}{2}X^2+\frac{g}{4}X^4+c.
\end{align}
For $\epsilon>0$ the system is a harmonic oscillator with frequency $\omega=\sqrt \epsilon$ perturbed by a quartic potential $\frac{g}{4}X^4$. For $\epsilon<0$ it is a double well potential which will be studied in detail in following chapter. The constant $c$ is introduced to keep the potential (and then also energies) positive for $\epsilon<0$.
\section{Cut Fock space method}\label{ch:TD_for_anharmonic_oscillator}
This method, which is performed here after \cite{Wosiek}, makes use of the Fock space in which Hamiltonian (\ref{eq:chapter1_hamiltonian}) has a very simple structure.

We will now briefly present construction of the Fock space. Let $a$ and $a^\dag$ be annihilation and creation operators respectively, satisfying canonical commutation relations
\begin{align*}
[a,a^\dag]&=1,&[a,a]=[a^\dag,a^\dag]&=0.
\end{align*}
Then the vacuum state $\ket{0}$ is defined to be such a normalized vector, that $a\ket{0}=0$ is satisfied. Other basis vectors are constructed recursively by relation $\ket{n}=\frac{1}{\sqrt n}a^\dag\ket{n-1}$. The Hilbert space $\mathcal H$ spanned by all $\ket{n}$'s is called the Fock space. From the commutation relations it follows that $a\ket{n}=\sqrt n\ket{n-1}$. In this paper we constrain our calculations to a finite dimensional space $\mathcal H_{M}=\mathrm{lin}\{\ket{n}: n\leq M\}$ called cut Fock space, M being the cut--off.

In order to be able to compute how the Hamiltonian acts on the basis states one has to express it in terms of creation and annihilation operators using $X=\frac{1}{\sqrt 2}(a^\dag+a)$, $P=\frac{i}{\sqrt 2}(a^\dag-a)$ and canonical commutation relations. The Hamiltonian constrained to $\mathcal H_M$ is a sparse matrix $H_M$, which elements can be easily explicitly calculated:
\begin{align}\label{eq:Hamiltonianmatrixelements}
\begin{split}
\braket{m|H|n}&=\Big((n+\frac{1}{2})\frac{1+ \epsilon}{2}+\frac{g}{16}(6n^2+6n+3)+c\Big)\delta_{mn}+ \frac{g(n-\frac{1}{2})-1+\epsilon}{4}\sqrt{n(n-1)}\delta_{m,n-2}\\
&\quad+\frac{g(n+\frac{3}{2})-1+\epsilon}{4}\sqrt{(n+1)(n+2)}\delta_{m,n+2}\\
&\quad+\frac{g}{16}\Big(\sqrt{n(n-1)(n-2)(n-3)}\delta_{m,n-4}+\sqrt{(n+1)(n+2)(n+3)(n+4)}\delta_{m,n+4}\Big).
\end{split}
\end{align}

For $M=\infty$ eigenvalues of the matrix $H_M$ are energies of the system. Crucial question is the rate of convergence of eigenvalues to the spectrum when $M$ grows. It has been observed \cite{Trzetrzelewski:2004nz} that the eigenvalues indeed converge. An eigenvalue $\lambda(M)$ converges to energy $E$ exponentially if $E$ belongs to discrete spectrum and like $1/M$ if $E$ is in continuous spectrum. These conclusions are based on numerical data. Behavior of energies against $M$ is shown in Figure \ref{fig:macierz1}. The value of $M$ which is needed to obtain energies with desired precision highly depends on the coupling constant $g$.

Let us now concentrate on the structure of energies which are presented in Figs \ref{fig:macierz1}, \ref{fig:studnia1}. The figure was obtained for $g=1/98,\ \epsilon=-1/2$ and $c=49/8$. For this choice of parameters, the potential $V(x)=c+\frac{\epsilon}{2}x^2+\frac{g}{4}x^4$ has two global minima at $x=\pm7$ and a maximum at $x=0$. One can clearly see a change of behavior at the value of energy $E\approx 6$. This value is height of the potential barrier, $V(0)=c=6.125$. In a classical system, energies $E>c$ would correspond to states which have enough energy to propagate over the barrier. For $E<c$ it cannot go through the potential barrier and stays in one minimum. Then there are two independent states, one in the left and one in the right minimum with equal energy. This is why energies smaller than $c$ are paired. Because the system is not classical but quantum, each state with energy $E<c$ is a superposition of a wavefunctions localized left and right minima. The lower energy of each pair always corresponds to symmetric superposition and thus is a vector composed of basis vectors with even number of quanta $\ket{2n}$ exclusively. The higher one is the antisymmetric superposition which is a vector containing only basis states with odd number of quanta $\ket{2n+1}$. These states have parity $+1$ and $-1$ respectively. In position representation their wavefunctions are even and odd. Since the distance between minima is big and the barrier is high, there is only weak tunneling between them. In the limit of infinitely separated minima and an infinite potential barrier, energies would be exactly degenerate. The two corresponding wavefunctions could be chosen is a such way that one would be localized in the left minimum and the other in the right. Because it is not precisely the case, the tunneling effect mixes those states causing the true eigenstates to be even and odd with a slight energy splitting. Representations of eigenvectors in the configuration space are presented in Figure \ref{fig:studnia1}.

We will now explain the relation between energy splittings and tunneling. In the anharmonic double well potential, a few lowest are paired into say $\bar E_k-\delta E_k$ and $\bar E_k+\delta E_k$ where $\bar E_k$ is mean energy of the k'th pair and the energy splitting $\delta E_k$ is small. These energies correspond to even and odd states $\ket{\psi_k^{(+)}}$ and $\ket{\psi_k^{(-)}}$ respectively. In configuration space, both states are very similar up to a sign in neighborhood of each minimum (see Fig. \ref{fig:studnia1}). One can than create a state $\ket{L_k}=\frac{1}{\sqrt 2}(\ket{\psi_k^{(+)}}+\ket{\psi_k^{(-)}})$. If phases are chosen as in Fig. \ref{fig:studnia1} then the interference is constructive around the left minimum and destructive around right minimum. If one evolves the state, each component acquires a different phase:
\begin{align}\begin{split}
\ket{L_k(t)}&=\frac{1}{\sqrt 2}e^{i(\bar E_k-\delta E_k)t}\ket{\psi_k^{(+)}}+\frac{1}{\sqrt 2}e^{i(\bar E_k+\delta E_k)t}\ket{\psi_k^{(-)}}\\
&=\frac{1}{\sqrt 2}e^{i(\bar E_k-\delta E_k)t}\left(\ket{\psi_k^{(+)}}+e^{2i\delta E_kt}\ket{\psi_k^{(-)}}\right).
\end{split}\end{align}
At some point in time $t=t_0$, the relative phase becomes $e^{2i\delta E_kt_0}=-1$ and the interference is constructive in the right minimum and destructive in the left one. We say that the state $\ket{L_k}$ tunnels into the right minimum in time $t_0$. When the barrier grows and widens, the energy splitting becomes smaller and tunneling is slower. More detailed analysis of such system, with small coupling constant, will be performed in the following chapter.

Let us make one more observation on convergence of energies. As one can see from Fig. \ref{fig:macierz1}, the two lowest energies converge at first (for $M<30$) linearly rather than exponentially as it was stated before. This is caused by the particular choice of basis. The eigenstates of harmonic oscillator $\ket{n}$ have support growing with $n$. Therefore, for small $M$, the states $\ket{n}$ with $n<M$ do not yet explore minima of the potential $V(x)$. The first state which reaches the minimum is $\ket{25}$. A few more states are needed to form an approximation of a gaussian function centered at $x=7$ and from $M\approx30$ convergence of the energy starts being exponential. For higher states, e.g. the 10th, the decrease is at first exponential, which is related to forming a wave with energy higher than the barrier. When $M$ is large enough and the minima are more explored, excited stated in both minima can be formed. Formation of those states is reflected by linear decrease of the energy. Than it becomes exponential again. For yet higher states, e.g. 18th, the convergence is exponential from the beginning and does not become linear at any $M$ because it converges to a energy greater that $c$.

\begin{multicols}{2}
\begin{figurehere}
\centering
\fbox{\resizebox{.97\columnwidth}{!}{\includegraphics{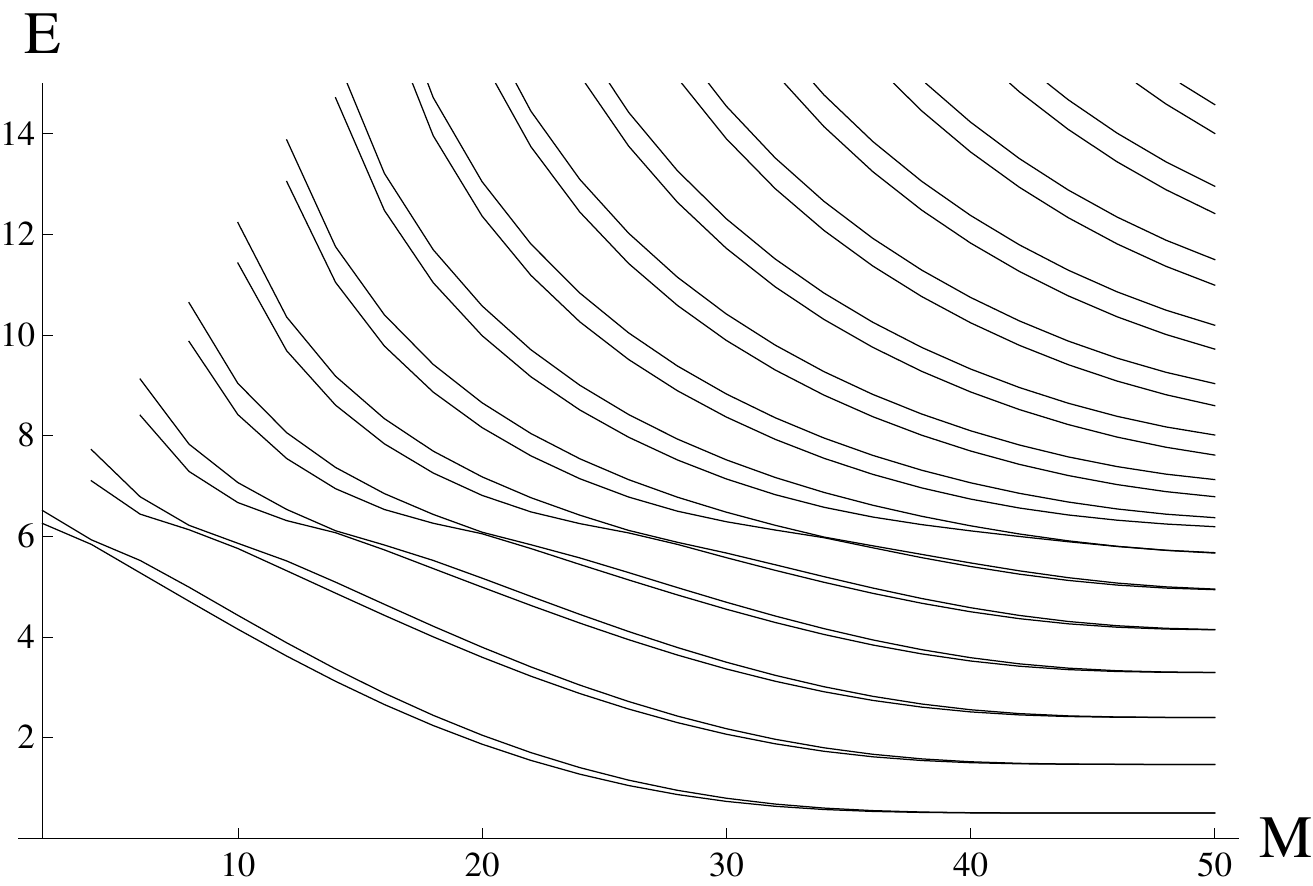}}}
\caption{Energies of Hamiltonian $ H_M$ with parameters $\epsilon=-1/2,\  g=1/98$. A few lowest energies are almost degenerate.}
\label{fig:macierz1}
\end{figurehere}
\begin{figurehere}
\centering
\fbox{\resizebox{.97\columnwidth}{!}{\includegraphics{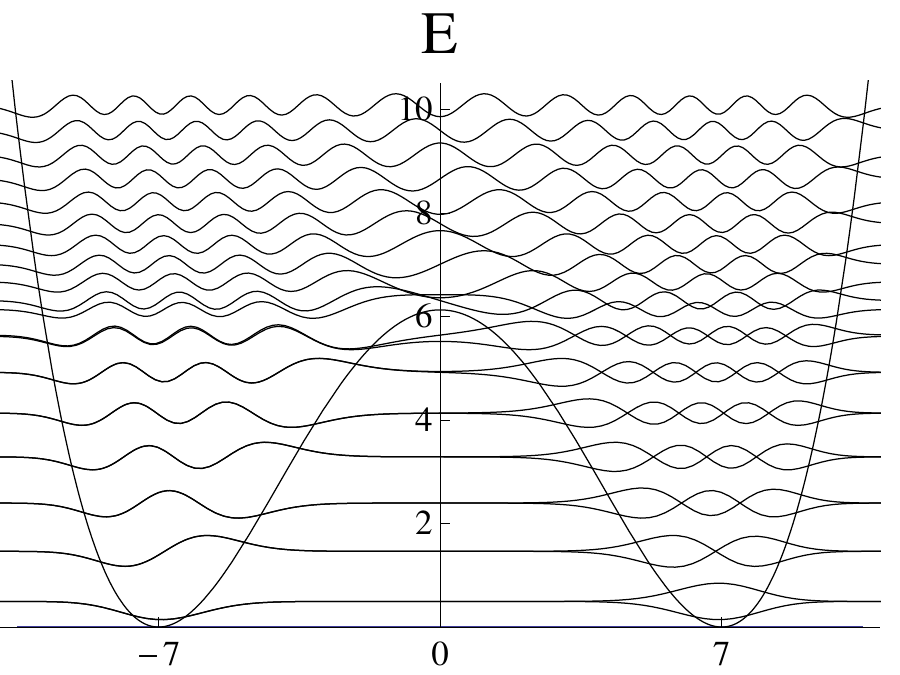}}}
\caption{Eigenstates in position representation for parameters $\epsilon=-1/2,\ g=1/98$. Each function is shifted by energy to which it corresponds. The near degenerate energy pairs can be observed. Note that the phase is chosen to be such that the low energy wavefunctions cover themselves on the left side. Multiplying every second state by $-1$ would reverse the situation.}
\label{fig:studnia1}
\end{figurehere}
\end{multicols}
\section{Shooting method}
Another popular way to determine eigenvalues of $H$ is the shooting method. The Schr\"odinger equation reads:
\begin{align}\label{eq:onedimposition}
-\frac{1}{2}f''(x)+\frac{\epsilon}{2}x^2f(x)+\frac{g}{4}x^4f(x)=E f(x).
\end{align}
Equation (\ref{eq:onedimposition}) has solutions for every value of $E$. However, only for a discrete set $\{E_n\}$ there exists a normalizable solution $f(x)$. This observation is the essence of the shooing method. Therefore, we will fix the initial values at the point $x=0$ and adjust the energy parameter in order to get a solution vanishing at infinity.

Since the equation (\ref{eq:onedimposition}) remains invariant under the transformation $x\rightarrow -x$, function $f(-x)$ is a solution whenever $f(x)$ is one. Let $f_0(x)$ satisfy (\ref{eq:onedimposition}). Then symmetric and antisymmetric parts of $f_0(x)$, namely $f_s(x)=\frac{1}{2}(f_0(x)+f_0(-x))$ and $f_a(x)=\frac{1}{2}(f_0(x)-f_0(-x))$, are also solutions of (\ref{eq:onedimposition}). Moreover, they obey simpler initial conditions
\begin{align}
&\left\{
\begin{array}{rl}
f_s(0)&=f_0(0);\\
f_s'(0)&=0,
\end{array}
\right.
&\left\{
\begin{array}{rl}
f_a(0)&=0;\\
f_a'(0)&=f_0'(0).
\end{array}
\right.
\end{align}
By the virtue of Picard theorem both functions are uniquely determined by these relations. Moreover, if $f_0(x)$ has finite norm then $f_s(x)$ and $f_a(x)$ also have. This shows that it is possible to consider only symmetric or antisymmetric functions at a time. We will constrain search for energies to functions satisfying one of the following initial conditions
\begin{align}
&
\left\{
\begin{array}{rl}
f(0)&=1;\\
f'(0)&=0,
\end{array}
\right.
&
\left\{
\begin{array}{rl}
f(0)&=0;\\
f'(0)&=1.
\end{array}
\right.
\end{align}
which leaves us with $E$ as the only free parameter.

The question that now arises is which values of $E$ shall be considered to be eigenvalues of the Hamiltonian when equation (\ref{eq:onedimposition}) is solved numerically. For a given $E$ one may write the differential equation in the form
\begin{align}
f''(x)=(\epsilon x^2+\frac{g}{2}x^4-2E)f(x).
\end{align}
It follows, that for $x$ such that $\epsilon x^2+\frac{g}{2}x^4-2E<0$ signs of $f''(x)$ and $f(x)$ are different. Thus, the function $f$ is "accelerated" towards the zero value and the equation is stable. However, beyond this limit the equation begins to be unstable, so one cannot expect any good behavior of a solution computed numerically for $x$ greater than some critical value. As an example, a plot of a solution of
\begin{align}\label{eq:1dimshoot}
\begin{split}
-\frac{1}{2}f''(x)+\frac{1}{2}x^2f(x)&=\frac{1}{2} f(x);\\
f(0)&=1;\\
f'(0)&=0
\end{split}
\end{align}

\begin{wrapfigure}{l}{.48\textwidth}
\fbox{
\includegraphics[width=.45\textwidth]{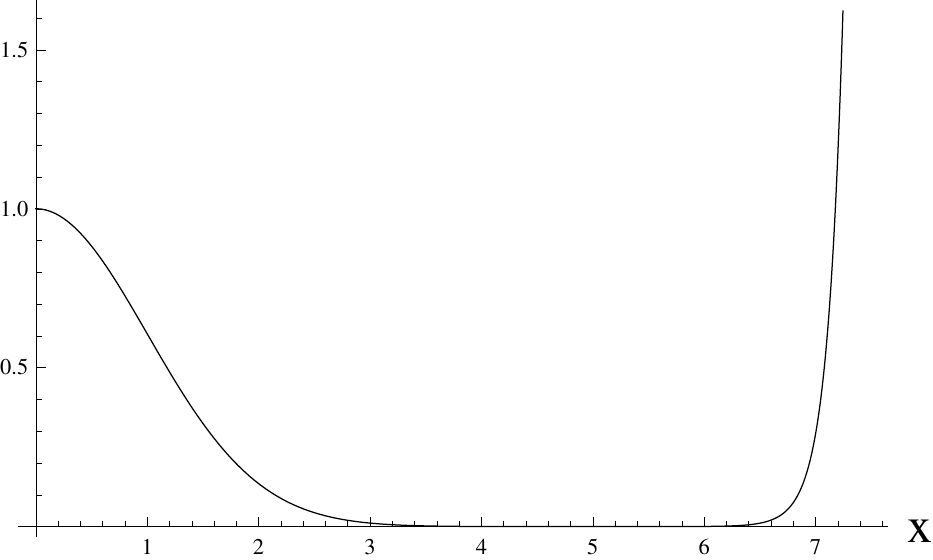}
}

\caption{Numerically computed harmonic oscillator ground state wavefunction. Loss of stability can be seen near $x=6.5$.}
\label{fig:strzaly1}
\vspace{5pt}
\end{wrapfigure}
obtained numerically is presented in Figure \ref{fig:strzaly1}. The solution is supposed to be wavefunction of the ground state of harmonic oscillator. The equation (\ref{eq:1dimshoot}) loses stability at $x=1$ while the solution starts exploding towards $+\infty$ about $x$ value of 6.5.
Similarly, a numerical solution of (\ref{eq:onedimposition}) taken with exact value of energy should converge to zero on a wide interval in the unstable region before it diverges. Clearly, width of the interval depends on precision of computations. We may then aim for functions which get closest to zero together with derivative. It can be done by finding minima of the function
\begin{align}\label{eq:mfunction}
m(E)=\!\min_{0<x<K}\{|f(x)|+|f'(x)|:\ f\text{ satisfies (\ref{eq:1dimshoot})}\}
\end{align}
starting with symmetric or antisymmetric initial conditions and varying the energy. $K$ is the upper boundary above which the function $f(x)$ is known to be large. Let us first note that in the unstable region the function $f(x)$ may tend to zero only if its first derivative has opposite sign than the function itself. Once the sign of the derivative agrees with sign of the function, the function $f(x)$ diverges instantaneously. Thus, the simplest way to choose $K$ is to take the first value of $x$ in the unstable region for which $f(x)f'(x)>0$. This is done during computations. Ex. for the ground energy with $\epsilon=g=1$ size of step equal to 0.01 and 16 digit precision, the equation loses stability at $x\approx0.93$ while $K=3.15$. Of course, value of $K$ grows with precision of computations. Noteworthy, decreasing the stepsize is not necessary do increase $K$, but is needed to obtain more precise values of energy.

\section{Comparison of the results}
Energies obtained with both methods and their relative differences are presented in Table \ref{tab:1dimcomparison}. $M$ was set to $20$. As it can be seen, the relative difference for small energies is negligible and starts to grow when we move to higher energies. Since $M$ is rather small, the error presumably comes from the truncation of the Hamiltonian matrix. Indeed, for $M=40$ we get $(E_F-E_s)/E_s<10^{-7}$ up to the 7th energy. The remaining difference is small and can easily be explained by numerical inaccuracy and finiteness of step size used in Runge-Kutta algorithm used for the shooting method.

\begin{table}[h!]
\centering
\begin{tabular}{r @{.} l r @{.} l r @{$\cdot$} l}
\multicolumn{2}{c}{$E_F$}&\multicolumn{2}{c}{$E_s$}&\multicolumn{2}{c}{$(E_F-E_s)/E_s$}\\
\hline
 0&620927&0&620927&1.34&$10^{-7}$\\
 2&02597&2&02597&2.67&$10^{-7}$\\
 3&69845&3&69845&6.48&$10^{-7}$\\
 5&55758&5&5576&3.44&$10^{-6}$\\
 7&56842&7&56935&1.22&$10^{-4}$\\
 9&70915&9&71146&2.38&$10^{-4}$\\
 11&9645&11&9697&4.32&$10^{-4}$\\
\end{tabular}\caption{Energies of the anharmonic oscillator with parameters $\epsilon=g=1$ obtained by the cut Fock space ($E_F$) and shooting ($E_s$) methods.}
\label{tab:1dimcomparison}
\end{table}

\section{Summary}
Results of this chapter confirm that approximating energy of a system by eigenvalues of a finite matrix is a correct approach. In fact, it is much faster than the shooting method and thus allows to reach higher precision. We have shown that for a double well potential energies are nearly degenerate and explained this fact by the tunneling effect. It is an introduction to the following chapters where we compare the numerical method with analytical results obtained in the semi--classical approximation.

\chapter{Symmetric double well potential}\label{ch:double_well}
The symmetric double well potential, especially anharmonic oscillator with negative quadratic term, is the simplest and most classical example of a system with tunneling. Still, it has some nontrivial properties of advanced physical systems. This is not only splitting of energies due to tunneling. Perturbative series of ground energy is asymptotic and non Borel summable, which is a common feature in field theory.

In this chapter we will address the anharmonic potential
\begin{align}\label{eq:two_minima_potential}
V(x)=\frac{1}{8}(x^2-1)^2.
\end{align}
Recall, that the Hamiltonian is then given by
\begin{align}\label{eq:double_well_hamiltonian}\begin{split}
H&=\frac{1}{2}P^2+a^2V(X/a)=\frac{1}{2}P^2+\frac{1}{8a^2}(X^2-a^2)^2.
\end{split}\end{align}
Note that when one translates the coordinate system by value $a$, so that zero is in the left minimum, $X\to X-a$ and uses the coupling constant $g=a^{-2}$. Then the Hamiltonian
\begin{align}
H&=\frac{1}{2}P^2+\frac{1}{2}X^2-\frac{\sqrt g}{2}X^3+ \frac{g}{8}X^4
\end{align}
can be viewed as a perturbed Hamiltonian of a harmonic oscillator.

\section{WKB approximation}\label{sec:double_well_instantons}
\begin{figure}
\centering
\includegraphics[width=.6\textwidth]{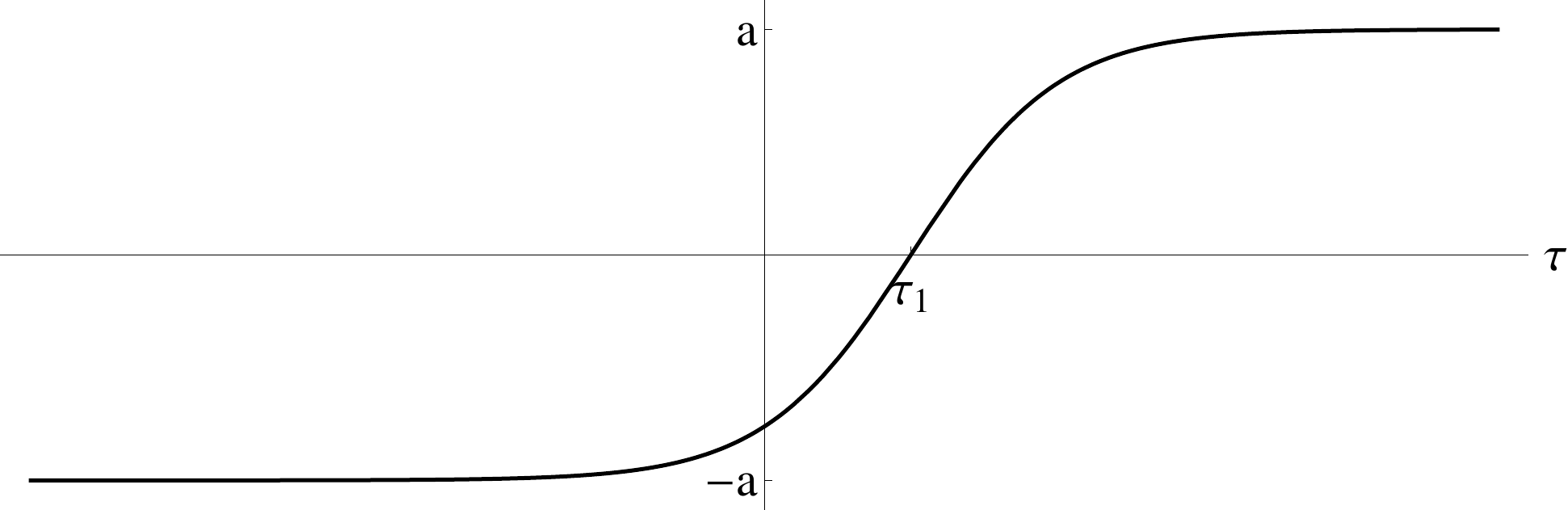}
\caption{A typical shape of an instanton. Position of the instanton is at $\tau=\tau_1$.}
\label{fig:instanton}
\end{figure}

We will now sketch the method for obtaining energies of the Hamiltonian using semiclassical approximation. It is done after S. Coleman \cite{Coleman}. A detailed analysis is performed in Appendix \ref{ch:instanton_calculus}. The key to this approach is calculating amplitudes
\begin{align}\label{eq:ij_amplitude}
\braket{a|e^{-TH}|\pm a}
\end{align}
in the large $T$ limit. Points $x=\pm a$ are minima of the scaled potential $a^2V(x/a)$. First we take the minus sign. One may represent the unity operator in terms of bound states of the system, $I=\sum\ket{E}\bra{E}$, so that
\begin{align}\label{eq:ij_expansion_for_double_well}
\braket{a|e^{-TH}|-a}&=\sum_{E,E'}\braket{a|E}\bra{E}e^{-TH}\ket{E'}\braket{E'|-a}=
\sum_{E}e^{-TE}\braket{a|E}\braket{E|-a}\\
&\approx e^{-TE_0}\braket{a|E_0}\braket{E_0|-a}+e^{-TE_1}\braket{a|E_1}\braket{E_1|-a}.
\end{align}
We shall now calculate the amplitude (\ref{eq:ij_amplitude}) using path integrals. It can be written as
\begin{align}
\braket{a|e^{-TH}|-a}&=\mathcal N\int \mathcal D[x(\tau)]e^{-S_E[x(\tau)]}
\end{align}
where $x(\tau)$ satisfies the boundary conditions $x(-T/2)=-a$ and $x(T/2)=a$. The Euclidean action $S_E[x(\tau)]$ is given by the formula
\begin{align}
S_E[x(\tau)]=\int_{-T/2}^{T/2}d\tau\left(\frac{1}{2}\dot x(\tau)^2+a^2V(x(\tau)/a)\right).
\end{align}
We calculate the integral using saddle point approximation. $S_E[x(\tau)]$ is maximized by $x(\tau)$ satisfying $\frac{\delta S_E[x(\tau)]}{\delta x(\tau)}=0$ and the boundary conditions:
\begin{align}\label{eq:instanton_equations}\begin{split}
-\ddot{\bar x}(\tau)+aV'(\bar x(\tau)/a)&=0,\\
\bar x(-T/2)&=-a,\\
\bar x(T/2)&=a.
\end{split}\end{align}
Expanding $aV'(x/a)$ about the point $x=-a$ gives an equation with the solution $\bar x(\tau)= -a+ce^{\tau}$ with an arbitrary constant $c$. It means that for large negative values of $\tau$ the solution remains exponentially close to $-a$, then is suddenly jumps to neighborhood of $a$ and approaches it again exponentially fast. For this reason $\bar x(\tau)$ is called an instanton. We will also use term classical solution because it is a solution of classical equations of motion in Euclidean space. A typical shape of an instanton is presented in Fig. \ref{fig:instanton}. For finite $T$ the classical solution crosses $0$ at $\tau=0$ because of symmetric boundary conditions. However, for infinite $T$ the problem possesses translational symmetry. We may choose any time $\tau_1$ at which $\bar x(\tau)$ passes through zero. $\tau_1$  is called position of the instanton.

There is a very convenient way to express Euclidean action of a instanton as a function of the potential only. It is possible since the energy $\mathcal E=\frac{1}{2}\dot x^2-V(x)$ is conserved and is zero for the instanton solution. The following formula is valid in the large $T$ limit:
\begin{align}
S_E[\bar x(\tau)]\xrightarrow[T\to\infty]{} S_0=\int_{-a}^adx\sqrt{2a^2V(x/a)}=a^2\int_{-1}^1dx\sqrt{2V(x)}.
\end{align}
Expanding $S_E[x(\tau)]$ around $\bar x(\tau)$ yields
\begin{align}\begin{split}\label{eq:operator_for_eigenvalues}
S_E[x(\tau)]&\approx S_0+\frac{1}{2}\int d\tau' d\tau''\left.\frac{\delta^2S_E[x(\tau)]}{\delta x(\tau')\delta x(\tau'')}\right|_{\delta x(\tau)=0}\delta x(\tau')\delta x(\tau'')\\
&=S_0+\frac{1}{2}\int d\tau\delta x(\tau)\left(-\frac{d^2}{d\tau^2}+V''(\bar x(\tau)/a)\right)\delta x(\tau),
\end{split}\end{align}
where $\delta x(\tau)=x(\tau)-\bar x(\tau)$.
Let us denote by $\{\lambda_n\}$ eigenvalues of the operator in (\ref{eq:operator_for_eigenvalues}):
\begin{align}
\left(-\frac{d^2}{d\tau^2}+V''(\bar x(\tau)/a)\right)x_n(\tau)&=\lambda_n x_n(\tau),&x_n(\pm T/2)=0.
\end{align}
Eigenfunctions $x_n$ form a complete set of functions. Expanding $\delta x(\tau)$ in this basis $\delta x(\tau)=\sum c_n x_n(\tau)$ gives
\begin{align}\begin{split}
S_E[x(\tau)]&\approx S_0+\frac{1}{2}\sum_{n,m=0}^\infty\lambda_n c_n^2.
\end{split}\end{align}
One may change variables from integration over all paths $x(\tau)$ to integration over coefficients $\{c_n\}$. Additional constant coming from change of variables is absorbed by $\mathcal N$.
\begin{align}\begin{split}
\braket{a|e^{-TH}|-a}
&=e^{-S_0}\mathcal N\prod_{n=0}^\infty\int \frac{d c_n}{\sqrt{2\pi}}e^{-\frac{1}{2}\lambda_n c_n^2}\\
&=e^{-S_0}\mathcal N\prod_{n=0}^\infty \frac{1}{\sqrt\lambda_n}\\
&=e^{-S_0}\mathcal N\det{}^{-\frac{1}{2}}\left[-\frac{d^2}{d\tau^2}+V''(\bar x(\tau)/a)\right].
\end{split}\end{align}
Determinant is by definition a product of all eigenvalues of relevant operator. For large $T$ the action does not change significantly if one changes position of the instanton $\tau_1$. It means that the system has a zero mode. It is reflected by the fact that the lowest eigenvalue $\lambda_0\to0$ when $T\to\infty$. We have then to integrate out the zero mode explicitly in order to avoid a divergence in the determinant.
Proper normalization of the zero mode yields $c_0=\tau_1\sqrt{S_0}$. Then,
\begin{align}
\int\frac{d c_0}{\sqrt{2\pi}}e^{-\frac{1}{2}\lambda_0c_0^2}&\approx\int\frac{d c_0}{\sqrt{2\pi}} 1=\int_{-T/2}^{T/2}\sqrt{\frac{S_0}{2\pi}}d\tau_1=\sqrt{\frac{S_0}{2\pi}}T.
\end{align}
Let $\det'$ denote the determinant with the lowest eigenvalue omitted. Then,
\begin{align}\begin{split}
\braket{a|e^{-TH}|-a}&=e^{-S_0}\sqrt{\frac{S_0}{2\pi}}T\mathcal N\left(\det{}'\left[-\frac{d^2}{d\tau^2}+V''(\bar x(\tau)/a)\right]\right){}^{-1/2}.
\end{split}\end{align}

\begin{figure}
\centering
\includegraphics[width=.6\textwidth]{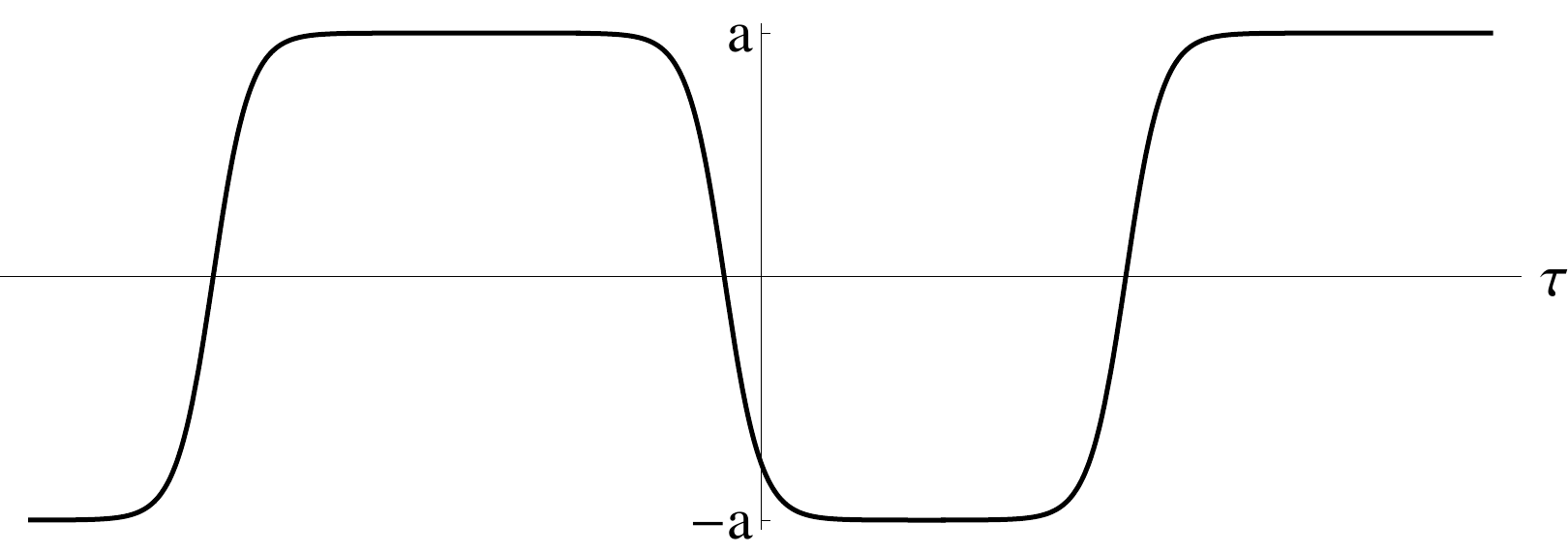}
\caption{A 3--instanton.}\label{fig:3instanton}
\end{figure}

There are other approximate classical solutions which are called multi--instantons. If individual components of such solutions are widely separated then they can be constructed from one--instanton solution by gluing instantons and antiintstantons (instanton reversed in time) in a sequence. Instantons and antiinstantons must be glued anternatively so that they form a continuous function. A plot of 3--instanton is shown in Fig. \ref{fig:3instanton}
Let us denote an $n$--instanton solution by $\bar x_n(\tau)$. The Euclidean action for $\bar x_n(\tau)$ is simply $n$ times larger than for a single instanton: $S_E[\bar x_n(\tau)]\approx n S_0$. Because of the boundary conditions (\ref{eq:instanton_equations}) the number of instantons $n$ has to be odd. In the other case, (\ref{eq:ij_amplitude}) taken with the plus sign, $n$ would have to be even. An $n$--instanton has $n$ zero modes due to translational symmetry of each instanton separately. We shall then integrate them out as previously. Note that we cannot change order of instantons.
\begin{align}\label{eq:integrating_positions}
\int_{-T/2}^{T/2}\sqrt{\frac{S_0}{2\pi}}d\tau_1\int_{\tau_1}^{T/2}\sqrt{\frac{S_0}{2\pi}}d\tau_2\ldots\int_{\tau_{n-1}}^{T/2}\sqrt{\frac{S_0}{2\pi }}d\tau_n=\left(\frac{S_0}{2\pi}\right)^{n/2}\frac{T^n}{n!}
\end{align}

Calculation of the determinant is technical and is presented in the Appendix \ref{ch:instanton_calculus}. The final formula is
\begin{align}\label{eq:n_determinant}
\mathcal N \left(\det{}'\left[-\frac{d^2}{d\tau^2}+V''(\bar x_n(\tau)/a)\right]\right)^{-1/2}&=e^{-\frac{T}{2}}\frac{1}{\sqrt\pi}\left(\sqrt{\frac{2}{S_0}}aA\right)^n
\end{align}
The constant $A$ is defined as follows. Let $\bar x(t)$ be the one--instanton solution:
\begin{align}\begin{split}
\ddot{\bar x}(t)&=aV'(\bar x(t)/a),\\
\bar x(-\infty)&=-a,\\
\bar x(\infty)&=a.
\end{split}\end{align}
Then asymptotic behavior of $\bar x(\tau)$ is
\begin{align}\label{eq:Apm}
\dot {\bar x}(t)&\approx a A_\pm e^{-|\tau|}&\tau\to\pm\infty,
\end{align}
and the constant $A$ is defined to be $A=\sqrt{A_+A_-}$. Contributions from all $n$--instantons have now to be summed. One has then an arbitrary many instantons which are separated from each other. It is known as the dilute instanton gas approximation. It assumes that there can be arbitrary many instantons as long as they are separated by a time interval much larger than the size of a single instanton. This condition is not taken into account in the integral (\ref{eq:integrating_positions}). However, corrections are of order $T^{-1}$. Every classical solution has to begin at $-a$ and end at $a$ so only functions with odd number of instantons contribute:
\begin{align}\begin{split}
\braket{a|e^{-TH}|-a}&=e^{-\frac{T}{2}}\frac{1}{\sqrt\pi}\sum_{n=0}^\infty\frac{1}{(2n+1)!}\left(e^{-S_0}\frac{aA}{\sqrt\pi} T\right)^{2n+1}\\
&=e^{-\frac{T}{2}}\frac{1}{\sqrt\pi}\sinh\left(e^{-S_0}\frac{aA}{\sqrt\pi}T\right).
\end{split}\end{align}
The other amplitude, $\braket{a|e^{-TH}|a}$ is calculated in the same way. The only difference is that now only even numbers of instantons contribute:
\begin{align}\begin{split}
\braket{a|e^{-TH}|a}&=e^{-\frac{T}{2}}\frac{1}{\sqrt\pi}\sum_{n=0}^\infty\frac{1}{(2n)!}\left(e^{-S_0}\frac{aA}{\sqrt\pi}T\right)^{2n}\\
&=e^{-\frac{T}{2}}\frac{1}{\sqrt\pi}\cosh\left(e^{-S_0}\frac{aA}{\sqrt\pi}T\right)^{2n+1}.
\end{split}\end{align}

All previous considerations were valid for any potential $V(x)$ with minima at $x=\pm1$. For the potential (\ref{eq:two_minima_potential}) we have
\begin{align}
V(x)&=\frac{1}{8}(x^2-1)^2,\\
S_0&=a^2\int_{-1}^1dx\sqrt{2V(x)}=a^2\frac{1}{2}\int_{-1}^1dx (1-x^2)=\frac{2}{3}a^2,\\
\bar x(\tau)&=a\tanh(\tau/2).
\end{align}
It is simple to observe that $A=2$. Finally,
\begin{align}\begin{split}
\braket{-a|e^{-TH}|-a}&=e^{-\frac{T}{2}}\frac{1}{\sqrt\pi}\cosh\left(e^{-\frac{2}{3}a^2}\frac{2a}{\sqrt\pi}T\right),\\
\braket{a|e^{-TH}|-a}&=e^{-\frac{T}{2}}\frac{1}{\sqrt\pi}\sinh\left(e^{-\frac{2}{3}a^2}\frac{2a}{\sqrt\pi}T\right).
\end{split}\end{align}
Using (\ref{eq:ij_expansion_for_double_well}) we read off the two lowest energies and amplitudes of corresponding eigenstates at minima. Expressed with the parameter $g=a^{-2}$ they take the form
\begin{align}\begin{aligned}\label{eq:double_well_energies}
E_0&=\frac{1}{2}-\frac{2}{\sqrt{g\pi}} e^{-2/3g},\makebox[3cm][l]{}&\braket{a|E_0}&=\braket{-a|E_0}=\left(4\pi\right)^{-1/4},\\
E_1&=\frac{1}{2}+\frac{2}{\sqrt{g\pi}} e^{-2/3g},\makebox[3cm][l]{}&\braket{a|E_1}&=-\braket{-a|E_1}=\left(4\pi\right)^{-1/4}.
\end{aligned}\end{align}
Phases were chosen such that $\braket{a|E_i}>0$.

One has to remember that the WKB approximation does not include perturbative corrections to energies, which are much greater than the nonperturbative terms in (\ref{eq:double_well_energies}). These perturbative contributions are identical for $E_0$ and $E_1$. It follows that the instanton calculus provides us relevant information about the energy difference
\begin{align}\label{eq:double_well_WKB_splitting}
\Delta E_{WKB}&=\frac{4}{\sqrt{g\pi}} e^{-2/3g}
\end{align}
rather than about energies themselves.
\section{Comparison with the Tamm-Dancoff method}\label{sec:double_well_numerics}
It is instructive to compare the semiclassical approximation with a numerical method which we know that is essentially exact. This allows us to estimate accuracy of the WKB approximation and indicate in which region of $g$ it is valid.
To this end we write the Hamiltonian (\ref{eq:double_well_hamiltonian}) in the form
\begin{align}\label{eq:double_well_hamiltonian_expanded}
H=\frac{1}{2}P^2-\frac{1}{4} X^2+\frac{g}{8}X^4+\frac{1}{8g}.
\end{align}
Energies of the system can be computed in the way that was introduced in the preceding chapter, by constructing a finite matrix that will approximate the Hamiltonian. The formula (\ref{eq:Hamiltonianmatrixelements}) for matrix elements of Hamiltonian may be used with substitutions
\begin{align}
\epsilon&\to-\frac{1}{2},&
g&\to\frac{g}{2}.
\end{align}
The constant $\frac{1}{8g}$ which appears at the end of the formula (\ref{eq:double_well_hamiltonian_expanded}) and was not present earlier has to be added to diagonal elements. Then the matrix elements are
\begin{align}
\begin{split}
\braket{m|H|n}&=\Big(\frac{1}{4}(n+\frac{1}{2})+\frac{g}{32}(6n^2+6n+3)+\frac{1}{8g}\Big)\delta_{mn}+ \left(\frac{(n-\frac{1}{2})g}{8}-\frac{3}{8}\right)\sqrt{n(n-1)}\delta_{m,n-2}\\
&\quad+\left(\frac{(n+\frac{3}{2})g}{8}-\frac{3}{8}\right)\sqrt{(n+1)(n+2)}\delta_{m,n+2}\\
&\quad+\frac{g}{32}\Big(\sqrt{n(n-1)(n-2)(n-3)}\delta_{m,n-4}+\sqrt{(n+1)(n+2)(n+3)(n+4)}\delta_{m,n+4}\Big).
\end{split}
\end{align}

\begin{wrapfigure}{l}{.5\textwidth}
\fbox{
\includegraphics[width=.45\textwidth]{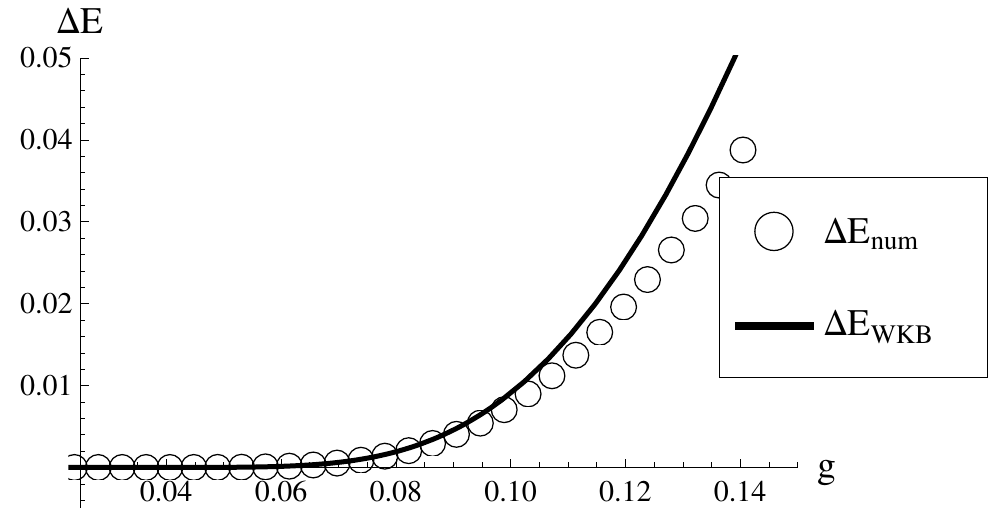}
}
\caption{Comparison of numerical and semiclassical energy splitting.}
\label{fig:wkb_duze}
\end{wrapfigure}
In the first step it is checked for which $g$ the WKB approximation result $\Delta E_{WKB}$ agrees with numerical outcome $\Delta E_{num}$. It is shown in Figure \ref{fig:wkb_duze} that $\Delta E_{num}$ and $\Delta E_{WKB}$ coincide for $g\approx 0.08$.
More precisely, for $g=0.04$ the relative difference $\delta=(\Delta E_{WKB}-\Delta E_{num})/\Delta E_{WKB}$  is around $7\%$ and decreases when $g$ becomes smaller. The numerical procedure allows to go to much smaller values of $g$.

Since the energy splitting $\Delta E_{num}$ is small, high precision computations have to be performed. We succeeded to reach $g=8\cdot 10^{-5}$ which requires $3600$ digits to be taken into account in all computations. As we expect $\Delta E_{num}$ to be of the same order as $\Delta E_{WKB}$, one shall consider at least a few more digits than $\log_{10}\Delta E_{WKB}$. This precision can be easily obtained eg. in \emph{Mathematica} by setting precision of input to desired number of digits. To keep cut--off effects insignificant one has to take $M$ large enough so that $\Delta E_{num}$ does not depend significantly on it. Needed cut--off appears to be approximately $M=1.6/g$, i.e. for $g=8\cdot 10^{-5}$ we took $M=20000$.

In Fig. \ref{fig:wkb_energies} we present dependence of the energy splitting $\Delta E_{num}$ and $\Delta E_{WKB}$ (given in (\ref{eq:double_well_WKB_splitting})) on $g$. It is seen that both approaches are in agreement in that region. However, in order to conclude consistency of semi-classical approximation with cut Fock space method, one has to perform a more detailed analysis. More sensitive way to present our results is to plot above $\delta$ as a function of $g$ which is done in Fig. \ref{fig:relative_difference}. From this it can be seen, that numerical values agree with WKB approximation as $g\to0$. Still, $\delta$ is nonzero for finite $g$. The numerical method is essentially exact and $\Delta E_{WKB}$ was calculated up to a coefficient which is $1+\mathcal O(g)$. Thus our numerical data on $\delta(g)$ can be used to determine further corrections to $\Delta E_{WKB}$.  The dashed line is a function $\Delta E_{num}=\Delta E_{WKB}(1-\alpha g-\beta g^2-\gamma g^3)$ fitted to five points of data corresponding to smallest values of $g$. The fitted parameters are
\begin{align*}
\alpha&=1.47916667\pm6.8\cdot 10^{-7}\\
\beta&=1.36693\pm7.8\cdot 10^{-4}\\
\gamma&=4.10\pm0.18
\end{align*}
The fit is consistent with the result of \cite{ZJ} where author proposes complete structure of the non-perturbative contribution. According to that paper coefficients $\alpha,\ \beta$ are obtained from calculating higher order perturbations around one instanton contribution to the energy. They are \cite{ZJ2}
\begin{align*}
\alpha&=\frac{71}{48}\approx   1.479166667\\
\beta&=\frac{6299}{4608}\approx 1.3669705
\end{align*}
Coefficient $\gamma$ is not given. Coefficient $\alpha$ is in perfect agreement, $\beta$ is away form the value given in \cite{ZJ2} by $3\sigma$. This is because of higher order terms contributions which limit the accuracy of the fit. Of course better approximation of $\alpha,\beta,\gamma$ and determining higher order terms would be possible if we were able to reach smaller values of $g$.

\begin{multicols}{2}
\begin{figurehere}
\centering
\fbox{\resizebox{.97\columnwidth}{!}{\includegraphics{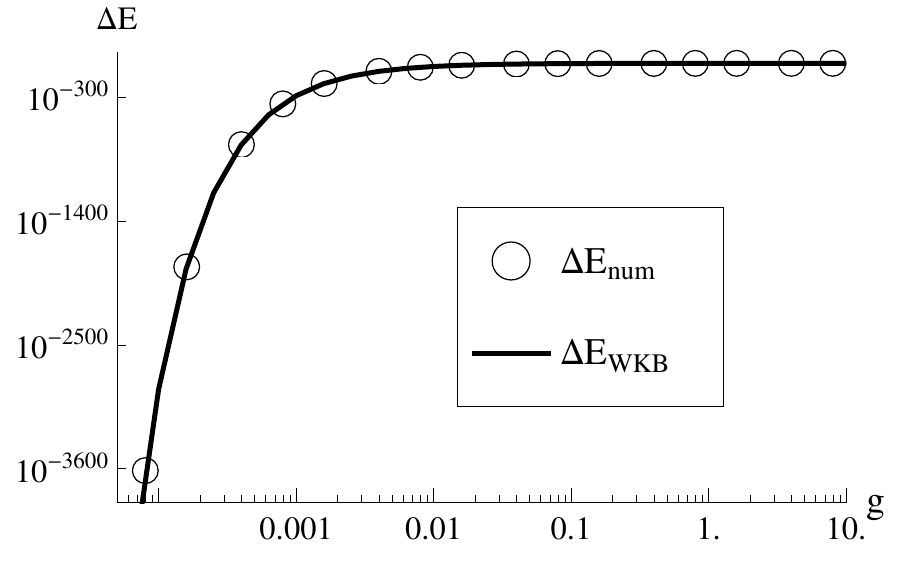}}}
\caption{Numerical and theoretical energy splitting for small parameter $g$.}
\label{fig:wkb_energies}
\end{figurehere}
\begin{figurehere}
\centering
\fbox{\resizebox{.97\columnwidth}{!}{\includegraphics{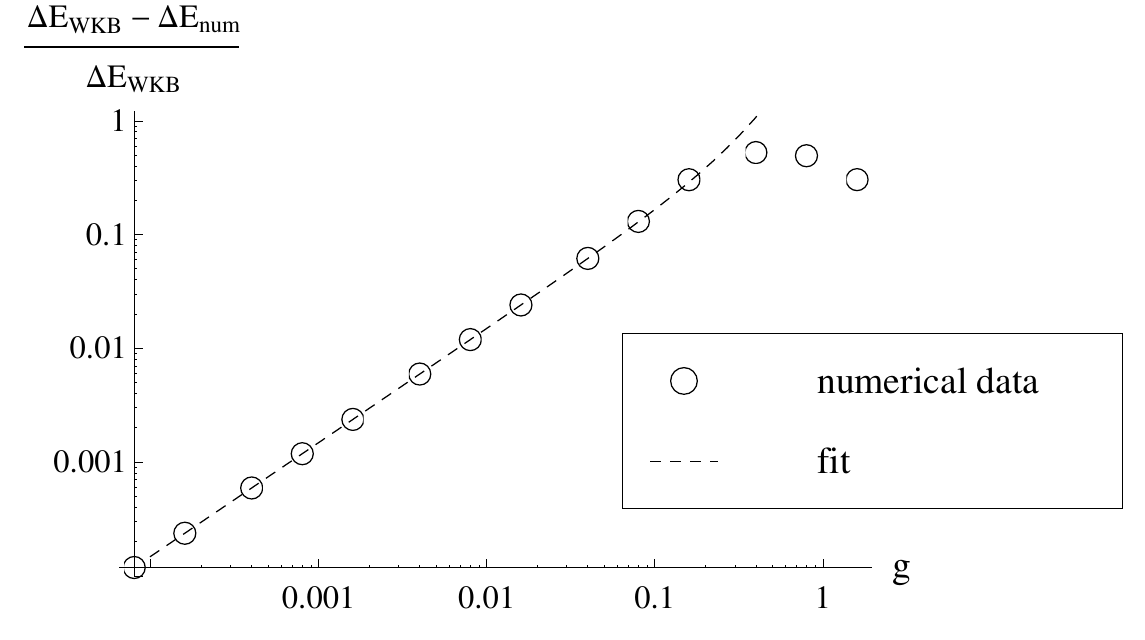}}}
\caption{Relative difference between numerical and theoretical energy splitting.}
\label{fig:relative_difference}
\end{figurehere}
\end{multicols}

As we have seen, the energy splitting obtained by the WKB approximation and cut Fock space approach agree for $g\leq 0.04$. For smaller values of $g$ the agreement improves. Computations were done in such high precision and so small values of $g$ that further corrections were easy to determine. We have given their fittes values and observed that they are in perfect agreement with the results available in the literature.

\chapter{Periodic potentials}\label{ch:periodic_potentials}
This part of the thesis will be devoted to periodic potentials. As it will be shown in the last chapter, it is much easier to study tunneling when minima of potential are equivalent, i.e. Taylor expansion of the potential about each minimum is identical up to reflection. Periodicity of potential guarantees such equivalence. The most natural candidate for a periodic potential is cosine function. Such potential has infinite number of minima. One can takle with this problem by imposing periodic boundary conditions on the wavefunction. Then, the number of minima can be chosen arbitrarily. In fact, one can choose to have only one minimum and study contribution of instantons to the ground energy. However, no splitting of energy will take place. In the first section, we discuss potentials with two and three minima. In such potential multiinstanton trajectories can go around and get to the same minimum. We say that such trajectories have nonzero winding number.

In second section we address the problem of a periodic potential in infinite space. Even though it is more complicated, the semi--classical approximation can be performed analytically. On the other hand, the cut Fock space method is much more challenging.

\section{Tunneling in periodic space}
In this section we will be interested in cosine potential in a periodic space with two and three minima. On one hand, there is no reason why the instanton calculus should not work in such space. Indeed, the reasoning is the same as in previous cases and only counting number of instantons gets more complicated. The cut Fock space analysis will be performed in a slightly different manner. One cannot start from eigenbasis of the harmonic oscillator, because there is no such system in the periodic space. On the other hand, there is a natural basis of plane waves which one can use to express the Hamiltonian as a matrix.

Parametrization of the potential is as follows:
\begin{align}
\hat V=g^{-1}V(\sqrt gx)&=\frac{1}{4g\pi^2}\left(1-\cos\left(2\pi\sqrt g x\right)\right).
\end{align}
The potential has minima at $x=ng^{-1/2}$ for $n\in\mathbf Z$. We will be interested in two cases, $x\in(0,2g^{-1/2})$ and $x\in(0,3 g^{-1/2})$ and will make some remarks on more general case $x\in (0,K g^{-1/2})$ for $K$ minima.
\subsection{Instanton calculus}
The Euclidean action and one--instanton solution are independent of $K$. They can be calculated explicitly. Recall that customarily for calculations in the semiclassical approximation we use the scale of the potential $a=g^{-1/2}$ rather than the coupling constant $g$.
\begin{align}\label{eq:periodic_action}
S_0&=a^2\int_0^1 dx\sqrt{2V(x)}=\frac{a^2}{\sqrt 2\pi}\int_0^1dx\sqrt{1-\cos(2\pi x)}=\frac{2}{\pi^2}a^2.
\end{align}
The one--instanton solution satisfies the equation
\begin{align}\begin{split}
-\ddot{\bar x}(\tau)+\frac{a}{2\pi}\sin\left(2\pi \bar x(\tau)/a\right)&=0,\\
\bar x(-\infty)&=0,\\
\bar x(+\infty)&=a.
\end{split}\end{align}
It is a simplified version of the well known Sine--Gordon equation. Its solution is
\begin{align}
\bar x(\tau)&=\frac{2a}{\pi}\arctan\left(e^{\tau}\right),
\end{align}
up to a shift in $\tau$. The constants $A_\pm$, which are defined in \ref{eq:Apm} as
\begin{align}
\dot {\bar x}(t)&\approx a A_\pm e^{-|\tau|},&\tau\to\pm\infty,
\end{align}
are in this case
\begin{align}\label{eq:periodic_A}
A_\pm&=\lim_{\tau\to\pm\infty}e^{|\tau|}\dot{\bar x}(\tau)/a=\frac{2}{\pi}\lim_{\tau\to\pm\infty}e^{|\tau|}\frac{e^{\tau}}{1+e^{2\tau}}=\frac{2}{\pi}
\end{align}
and $A=\sqrt{A_+A_-}=\frac{2}{\pi}$.

The number of paths of $n$--instanton $N_n$ depends on periodicity of the space and the two cases have to be addressed separately. Let us first consider the simpler case, $x\in(0,2a)$. Then there are two minima at zero and $a$. Similarly as for the double well potential, there are two amplitudes to be calculated.

Let us first calculate $\braket{a|e^{-TH}|0}$. According to formulas (\ref{eq:integrating_positions}) and (\ref{eq:n_determinant}) the contribution of $n$--instanton solution integrated over zero modes is
\begin{align}\label{eq:more_general_n_instanton_contribution}
e^{-\frac{T}{2}}\frac{1}{\sqrt\pi}\frac{1}{n!}\left(e^{-S_0}\frac{aA}{\sqrt\pi} T\right)^n.
\end{align}
It does not take into account the number of topologically different paths of $n$--instanton solution.  Each instanton of an $n$--instanton trajectory changes the minimum in which a state is localized. Therefore, only odd number of instantons contributes to this amplitude. Because of the periodic boundary conditions, an instanton starting at $0$ can go either left or right and it ends at $a$. There are two admissible paths for each instanton, so $2^{2n+1}$ possible paths for a $2n+1$ instanton solution.

Taking sum over $n$ one obtains
\begin{align}\begin{split}
\braket{a|e^{-TH}|0}
&\approx e^{-\frac{T}{2}}\frac{1}{\sqrt\pi}\sum_{n=0}^\infty \frac{2^{2n+1}}{(2n+1)!}\left(e^{-\frac{2}{\pi^2}a^2}\frac{2}{\pi^{3/2}}a T\right)^{2n+1}\\
&=e^{-\frac{T}{2}}\frac{1}{\sqrt\pi}\sinh\left(e^{-\frac{2}{\pi^2}a^2}\frac{4}{\pi^{3/2}}a T\right).
\end{split}\end{align}
Conversely, only even instanton solutions contribute to $\braket{0|e^{-TH}|0}$. The number of $2n$--instantons starting and ending at $0$ is given by an analogical formula, $N_{2n}=2^{2n}$ where $n\geq0$.
\begin{align}\begin{split}
\braket{0|e^{-TH}|0}
&\approx e^{-\frac{T}{2}}\frac{1}{\sqrt\pi}\sum_{n=0}^\infty \frac{2^{2n}}{(2n)!}\left(e^{-\frac{2}{\pi^2}a^2}\frac{2}{\pi^{3/2}}a T\right)^{2n}\\
&=e^{-\frac{T}{2}}\frac{1}{\sqrt\pi}\cosh\left(e^{-\frac{2}{\pi^2}a^2}\frac{4}{\pi^{3/2}}a T\right)
\end{split}\end{align}
We now read off the energies and amplitudes. In terms of $g=1/a^2$ they are
\begin{align}\label{eq:periodic_2_energies}
E_0&=\frac{1}{2}-e^{-2/\pi^2g}\frac{4}{\sqrt{g}\pi^{3/2}},\\
E_1&=\frac{1}{2}+e^{-2/\pi^2g}\frac{4}{\sqrt{g}\pi^{3/2}}.
\end{align}
\begin{align}
\braket{E_0|0}&=\frac{1}{\sqrt 2}\pi^{-1/4},&
\braket{E_0|a}&=\frac{1}{\sqrt 2}\pi^{-1/4},\nonumber\\
\braket{E_1|0}&=\frac{1}{\sqrt 2}\pi^{-1/4},&
\braket{E_1|a}&=-\frac{1}{\sqrt 2}\pi^{-1/4}.
\end{align}

\begin{figure}[ht!]
\centering
\includegraphics[width=.8\textwidth]{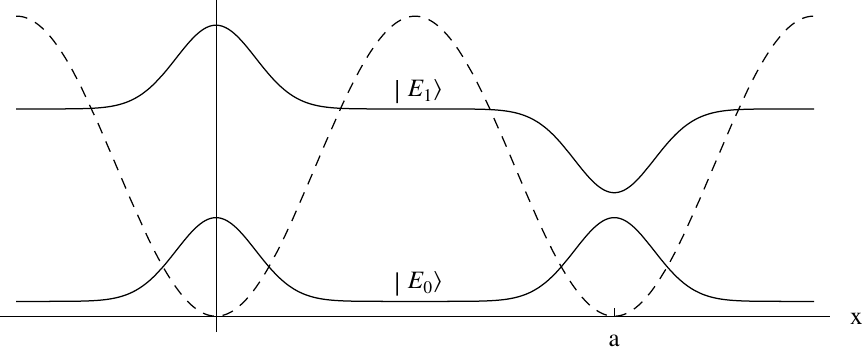}
\caption{Wavefunctions of the two lowest energy states (solid) for potential (dashed) with two minima for $a=10$. Amplitudes in each minimum are calculated in the semiclassical approximation. Presented wavefunctions are gaussians in neighborhood of each minimum which is zeroth approximation in $g\to0$ limit.}\label{fig:periodic_2_amplitudes}
\end{figure}

Let us now discuss the three minima case. Because only the topology changes and not shape of the potential, action of the classical trajectory $S_0$ and the constant $A$ are given by formulas (\ref{eq:periodic_action}) and (\ref{eq:periodic_A}). The remaining part is determination of numer of paths $N_n$.  Consider a triangle with vertices denoted by $A,B,C$. Each instanton moves a particle along one of the edges. Let us denote by $N_n(v_1,v_2)$ the number of $n$--instanton paths starting at vertex $v_1$ and ending at vertex $v_2$. Due to rotation and reflection symmetry, following equalities hold:
\begin{align}
c_n^{(0)}&\equiv N_n(A,A)=N_n(B,B)=N_n(C,C)\\
c_n^{(1)}&\equiv N_n(A,B)=N_n(B,C)=N_n(C,A)=N_n(B,A)=N_n(A,C)=N_n(C,B)
\end{align}
One can easily observe that $c_0^{(0)}=1$ and $c_0^{(1)}=0$. Consider now an $n$--instanton path which starts at $A$ and ends at $B$. In the first step it moves either to $B$ or $C$ and then there are $n-1$ steps left. This fact is represented by equation $N_n(A,B)=N_{n-1}(B,B)+N_{n-1}(C,B)$. Similarily, an $n$--instanton trajectory starting and ending at $A$ moves to $B$ or $C$ in the first step and then it has additional $n-1$ steps to return. It follows that $N_n(A,A)=N_{n-1}(B,A)+N_{n-1}(C,A)$. These two equations may be expressed as
\begin{align}
c_{n}^{(1)}&=c_{n-1}^{(0)}+c_{n-1}^{(1)},\label{eq:recursion1}\\
c_{n}^{(0)}&=2c_{n-1}^{(1)}.\label{eq:recursion2}
\end{align}
Inserting (\ref{eq:recursion2}) to (\ref{eq:recursion1}) gives
\begin{align}\label{eq:periodic_cn}
c_{n}^{(1)}=2c_{n-2}^{(1)}+c_{n-1}^{(1)}.
\end{align}
Taking $c_{n}^{(1)}=r^n$ one obtains a quadratic equation for $r$ which solutions are $r=2$ and $r=-1$. Then, the general solution of (\ref{eq:periodic_cn}) is $c_{n}^{(1)}=\alpha 2^n+\beta (-1)^n$. Using the initial conditions,
\begin{align}\begin{split}
c_{n}^{(1)}=\frac{1}{3}2^n-\frac{1}{3}(-1)^n.
\end{split}\end{align}
The other series yields
\begin{align}\begin{split}
c_{n}^{(0)}=\frac{1}{3}2^n+\frac{2}{3}(-1)^n.
\end{split}\end{align}
Relation $N_n=c_n^{(0)}$ or $N_n=c_n^{(1)}$ can be used to calculate amplitudes $\braket{q_2|e^{-TH}|q_1}$ with $q_1=q_2$ and $q_1\neq q_2$ respectively. From formula (\ref{eq:more_general_n_instanton_contribution}) multiplied by $c_n^{(0)}$ and summed over $n$ one gets the final result
\begin{align}\begin{split}
\braket{0|e^{-TH}|0}&=\braket{a|e^{-TH}|a}=\braket{-a|e^{-TH}|-a}=e^{-\frac{T}{2}}\frac{1}{\sqrt\pi}\sum_{n=0}^\infty \frac{c_n^{(0)}}{n!}\left(e^{-S_0}\frac{aA}{\sqrt\pi}T\right)^n\\
&=\frac{1}{\sqrt\pi}\left(\frac{1}{3}\exp\left(-\frac{T}{2}+2e^{-\frac{2}{\pi^2}a^2}\frac{2aT}{\pi^{3/2}}\right)+\frac{2}{3}\exp\left(-\frac{T}{2}-e^{-\frac{2}{\pi^2}a^2}\frac{2aT}{\pi^{3/2}}\right)\right),
\end{split}\end{align}
\begin{align}\begin{split}
\braket{a|e^{-TH}|0}&=\braket{a|e^{-TH}|-a}=\braket{-a|e^{-TH}|0}=e^{-\frac{T}{2}}\frac{1}{\sqrt\pi}\sum_{n=0}^\infty\frac{c_n^{(1)}}{n!}\left(e^{-S_0}\frac{aA}{\sqrt\pi}T\right)^n\\
&=\frac{1}{\sqrt \pi}\left(\frac{1}{3}\exp\left(-\frac{T}{2}+2 e^{-\frac{2}{\pi^2}a^2}\frac{2a T}{\pi^{3/2}}\right)-\frac{1}{3}\exp\left(-\frac{T}{2}-e^{-\frac{2}{\pi^2}a^2}\frac{2a T}{\pi^{3/2}}\right)\right).
\end{split}\end{align}
From the expansion $\braket{q_2|e^{-TH}|q_1}=\sum_E e^{-TE}\braket{q_2|E}\braket{E|q_1}$ one can easily read off the energies:
\begin{align}\label{eq:periodic_3_energies}\begin{split}
E_0&=\frac{1}{2}-e^{-2/\pi^2g}\frac{4}{\pi^{3/2}\sqrt{g}},\\
E_1&=\frac{1}{2}+e^{-2/\pi^2g}\frac{2}{\pi^{3/2}\sqrt{g}}
\end{split}\end{align}
in terms of $g=1/a^2$ and solve equations for amplitudes of the ground state $\ket{E_0}$:
\begin{align}
\braket{0|E_0}=\braket{-a|E_0}=\braket{a|E_0}=\frac{1}{\sqrt3},
\end{align}
which are determined up to a phase. The equations for amplitudes of the wavefunction corresponding to $E_1$ are
\begin{align}\label{eq:nondegenerate_states_amplitude_equations}
|\braket{0|E_1}|^2=|\braket{-a|E_1}|^2=|\braket{a|E_1}|^2&=\frac{2}{3},\\
\label{eq:2nd_for_amplitudes}\braket{0|E_1}\braket{E_1|a}=\braket{a|E_1}\braket{E_1|-a}=\braket{-a|E_1}\braket{E_1|0}&=-\frac{1}{3}.
\end{align}
Multiplying three expressions on the left hand side of (\ref{eq:2nd_for_amplitudes}) gives
\begin{align}
|\braket{0|E_1}\braket{a|E_1}\braket{-a|E_1}|^2&=-\frac{1}{27}
\end{align}
which is clearly a contradiction. Indeed, we made a wrong assumption that there is only one state of energy $E_1$ while there are actually two such states, i.e. the energy $E_1$ is degenerate. Let then $\ket{E_1^{(1)}}$  and $\ket{E_1^{(2)}}$ be the two orthogonal states corresponding to energy $E_1$. Instead of (\ref{eq:nondegenerate_states_amplitude_equations}) and (\ref{eq:2nd_for_amplitudes}) we obtain
\begin{align}\begin{split}
|\braket{0|E_1^{(1)}}|^2+|\braket{0|E_1^{(2)}}|^2&=|\braket{a|E_1^{(1)}}|^2+|\braket{a|E_1^{(2)}}|^2\\
&=|\braket{-a|E_1^{(1)}}|^2+|\braket{-a|E_1^{(2)}}|^2=\frac{2}{3},\\
\braket{0|E_1^{(1)}}\braket{E_1^{(1)}|a}+\braket{0|E_1^{(2)}}\braket{E_1^{(2)}|a}&=\braket{a|E_1^{(1)}}\braket{E_1^{(1)}|-a}+\braket{a|E_1^{(2)}}\braket{E_1^{(2)}|-a}\\
&=\braket{-a|E_1^{(1)}}\braket{E_1^{(1)}|0}+\braket{-a|E_1^{(2)}}\braket{E_1^{(2)}|0}=-\frac{1}{3}.
\end{split}\end{align}
One solution to these equations is
\begin{align}
\braket{-a|E_1^{(1)}}&=-\frac{1}{\sqrt 6}&\braket{0|E_1^{(1)}}&=\sqrt\frac{2}{3}&\braket{a|E_1^{(1)}}&=-\frac{1}{\sqrt 6},\\
\braket{-a|E_1^{(2)}}&=-\frac{1}{\sqrt 2}&\braket{0|E_1^{(2)}}&=0&\braket{a|E_1^{(2)}}&=\frac{1}{\sqrt 2}.
\end{align}
It is not unique since there is an infinite number of possibilities in which one can choose a basis of the two--dimensional eigenspace corresponding to the energy $E_1$. In our choice $\ket{E_1^{(1)}}$ is symmetric and $\ket{E_1^{(2)}}$ is antisymmetric.

\begin{figure}[ht!]
\centering
\includegraphics[width=.8\textwidth]{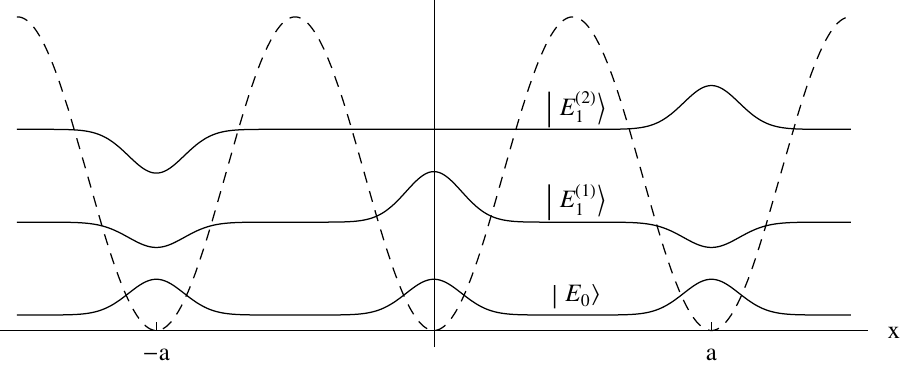}
\caption{Wavefunctions of the three lowest energy states (solid) for potential with three minima (dashed) for $a=10$. Amplitudes in each minimum are calculated in the semiclassical approximation. Presented wavefunctions are gaussians in neighborhood of each minimum which is zeroth approximation in $g\to0$ limit.}\label{fig:periodic_3_amplitudes}
\end{figure}

\subsection{Numerical computations of energy levels}
In this section we will show that energies of the Hamiltonian may be computed in a space with periodic boundary conditions even though we may no longer use creation and annihilation  operators as in the case of infinite space. Construction of the matrix will be presented for general number of minima $K$, but computations will be done only for $K=2,3$ for which we have derived energy levels in semiclassical approximation.

The most natural choice of basis for finite space are plane waves. We label them as $\ket{n}$:
\begin{align}
\braket{x|n}&=\frac{g^{1/4}}{\sqrt{K}}\exp\left(\frac{2n\pi i \sqrt gx}{K}\right),&n\in\mathbb Z.
\end{align}
The Hamiltonian is
\begin{align}
H=\frac{1}{2}P^2+\hat V=\frac{1}{2}P^2+ g^{-1}V(\sqrt gX).
\end{align}
Action of momentum operator and the potential is easy to calculate in this basis:
\begin{align}
P^2\ket{n}&=g\left(\frac{2\pi n}{K}\right)^2\ket{n},\\
\hat V\ket{n}&=\frac{1}{4\pi^2g}\ket{n}-\frac{1}{8\pi^2g}\ket{n+K}-\frac{1}{8\pi^2g}\ket{n-K}.
\end{align}

We shall note that the Hamiltonian has a translation symmetry $x\to T_ax\equiv x+a=x+1/\sqrt g$. This is $\mathbb Z_K$ symmetry. The basis states transform under this symmetry as follows:
\begin{align}
\braket{x|T_a|n}=\frac{g^{1/4}}{\sqrt{K}}\exp\left(\frac{2n\pi i \sqrt g(x+1/\sqrt g)}{K}\right)=\exp\left(\frac{2n\pi i}{K}\right)\braket{x|n}.
\end{align}
There are $K$ different values of $\exp\left(\frac{2n\pi i}{K}\right)$ so the Hilbert space can be divided into $K$ sectors
\begin{align}
\mathcal S_k&=\{\ket{k+nK},n\in\mathbf Z\},&k=0,\ldots K-1.
\end{align}
The Hamiltonian can be diagonalized in each sector independently.
One can see that the potential part of the matrix of Hamiltonian does not depend on sector $\mathcal S_k$, i.e.
\begin{align}
\braket{k+mK|\hat V|k+nK}=\frac{1}{8\pi^2g}\times\left\{\begin{array}{rl}2&n=m\\-1&n=m\pm1\\0&\text{otherwise}\end{array}\right.
\end{align}
On the other hand, the kinetic part does:
\begin{align}
\braket{k+mK|\frac{1}{2} P^2|k+nK}=g\left(\frac{2\pi n}{K}\right)^2\delta_{nm}.
\end{align}
The momentum operator depends only on the value of $n^2$. Therefore, matrices of Hamiltonian are identical in sectors $\mathcal S_k$ and $\mathcal S_{-k}\equiv\mathcal S_{K-k}$. Energies are doubly degenerate in each sector apart from $S_{0}$ and $S_{K/2}$.

The Hamiltonian has also parity symmetry. However, it cannot be used together with the $\mathbb Z_K$ symmetry since sectors $\mathcal S_k$ are not invariant under parity transformation but transform $\mathcal S_k\to\mathcal S_{-k}$. Thus, the only two parity invariant sectors are $S_0$ and $S_{K/2}$ (unless $K$ is odd). Parity symmetry can be used to reduce sizes of these two sectors into $\mathcal S_k^+$ and $\mathcal S_k^-$ ($k=0,K/2$) formed by cosine and sine functions respectively. It computations more effective in these sectors.

For $K=2$ there are two cosine and two sine sectors. We note that the two states of lowest energies lie in cosine sectors $\mathcal S_0^+$ and $\mathcal S_1^+$. This observation is based on analysis of amplitudes obtained by the WKB approximation and on making computations for some small $g$. For $K=3$ there are three cosine and three sine sectors. There are three energies of order $1/2$. One of them corresponds to function in $\mathcal S_0^+$ and the other two to functions in sectors $\mathcal S_1$ and $\mathcal S_{-1}$.

As before, we take a finite number of basis states and construct matrices of the Hamiltonian in each sector separately. Precision of computations is chosen to be such that energy splitting estimated by the semiclassical approximation is seen. The cut--off for states has to be chosen experimentally.

\subsection{Comparison of the results}
\begin{figure}[h!]
  \centering
  \subfloat[Two minima]{\label{fig:periodic_2_energies}\includegraphics[width=0.5\textwidth]{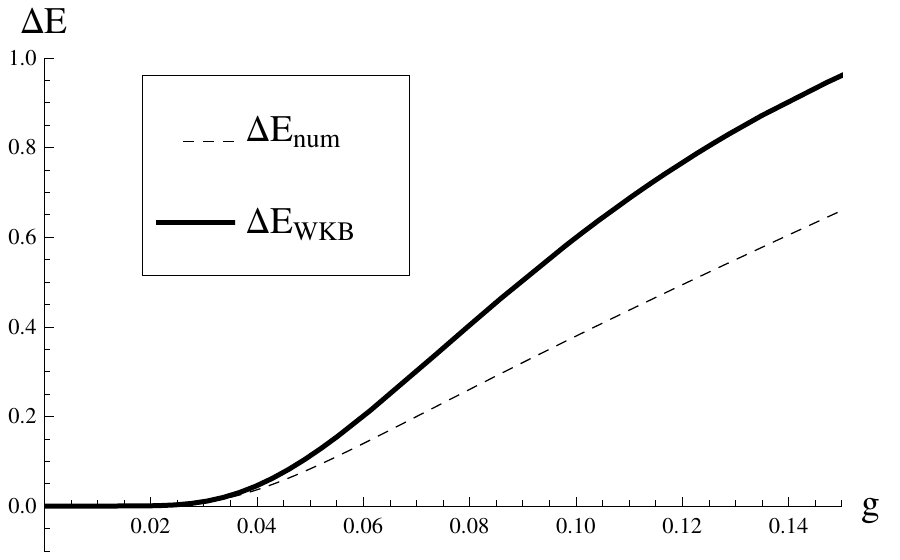}}
  \subfloat[Three minima]{\label{fig:periodic_3_energies}\includegraphics[width=0.5\textwidth]{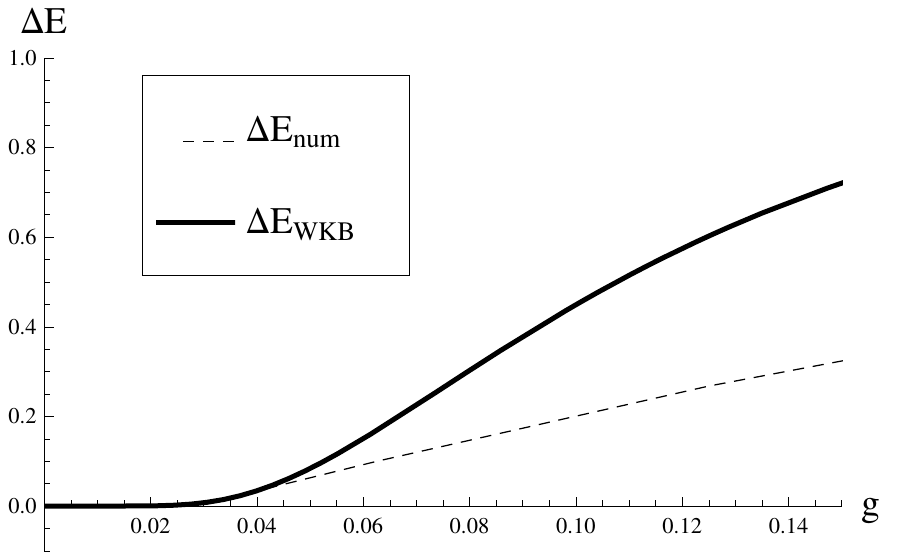}}
  \caption{Energy splitting obtained with WKB and numerical methods for small values of parameter $g$. Agreement of analytical and numerical approaches starts at $g\approx 0.04$.}
  \label{fig:periodic_energies}
\end{figure}

\begin{figure}[ht!]
  \centering
  \subfloat[Two minima]{\label{fig:periodic_2_splitting}\includegraphics[width=0.5\textwidth]{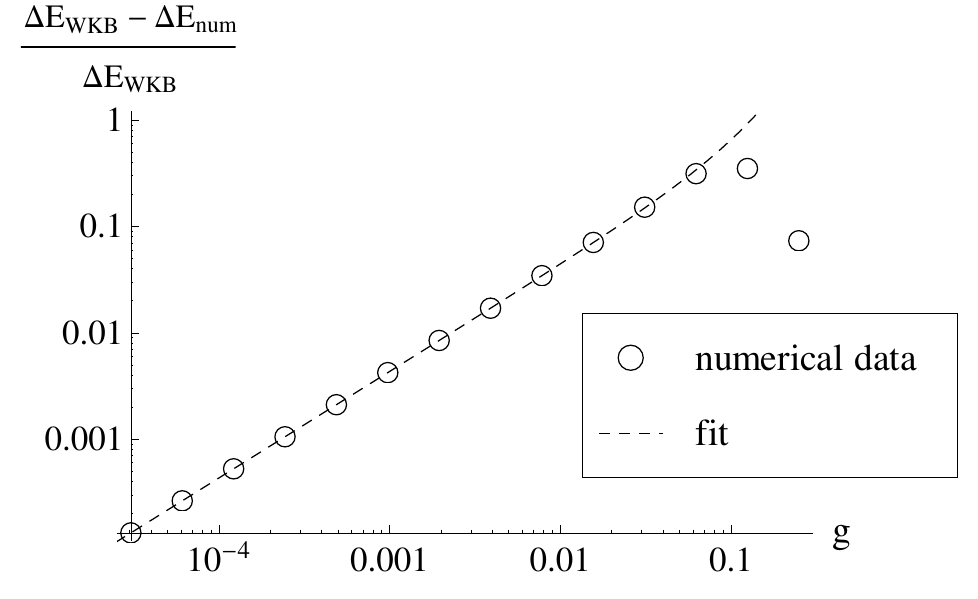}}
  \subfloat[Three minima]{\label{fig:periodic_3_splitting}\includegraphics[width=0.5\textwidth]{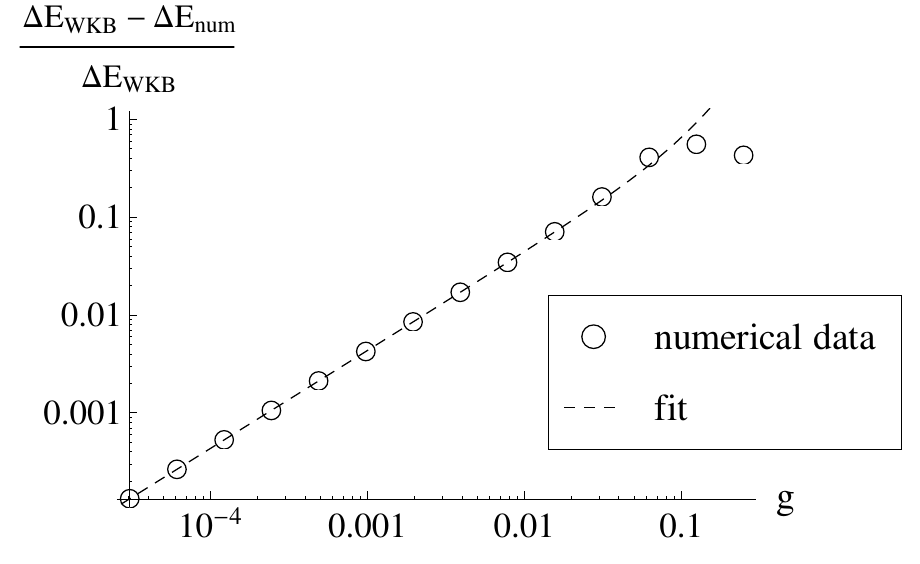}}
  \caption{Plot of relative difference of energy splitting obtained with analytical and numerical method. Points are values obtained from numerical computations and WKB approximation. The dashed line is a fit.}
  \label{fig:periodic_relative_difference}
\end{figure}

Finally, we compare results obtained with semiclassical approximation with exact, numerical data. In Figure \ref{fig:periodic_energies} we display energy splitting for two and three minima.
As one can see, energies start to agree around $g=0.04$ similarly as for two minima. To be more precise, in both cases relative error at $g=0.04$ is around 20\%. For $g=0.011$ it is already only 5\%. As the energy splitting converges to $0$ very fast, it is more instructive to look at relative difference of energy splitting obtained with both methods. In Fig. \ref{fig:periodic_relative_difference} there are presented plots of $(\Delta E_{WKB}-\Delta E_{num})/\Delta E_{WKB}$ for cases with two and three minima.
The relative error tends to zero. It is a proof that the semiclassical approximation and Fock space method agree. For small $g$ the points align on a straight line. It suggests that there are further corrections of order $g$ to $\Delta E_{WKB}$. Indeed, one can fit a polynomial to the relative difference of energy splitting. We assume that $\Delta E_{num}=\Delta E_{WKB}(1-\alpha g-\beta g^2-\gamma g^3)$. Results of fitting values $\alpha,\ \beta,\ \gamma$ are presented below.
\begin{align*}
&\text{two minima}&&\text{three minima}\\
\alpha&=4.31795193\pm9\times 10^{-7}&\alpha&=4.31795193\pm9\times 10^{-7}\\
\beta&=11.2246\pm3.8\times 10^{-3}&\beta&=11.2246\pm3.8\times 10^{-3}\\
\gamma&=106.4\pm3.5&\gamma&=106.4\pm3.5
\end{align*}
We have thus determined further corrections to the WKB energy splitting. However, there are no results in the literature to which we can compare these corrections. A very interesting fact is that the corrections are exactly the same. This is because perturbation about an instanton does not depend on topology of the space. It is an indication of the fact, that the only difference between instanton calculus for two and three minima, even in higher orders of $g$, is counting of instantons, i.e. the coefficient $N_n$. Coefficients $N_n$ are purely topological.

\begin{figure}
\centering
\includegraphics[width=.8\textwidth]{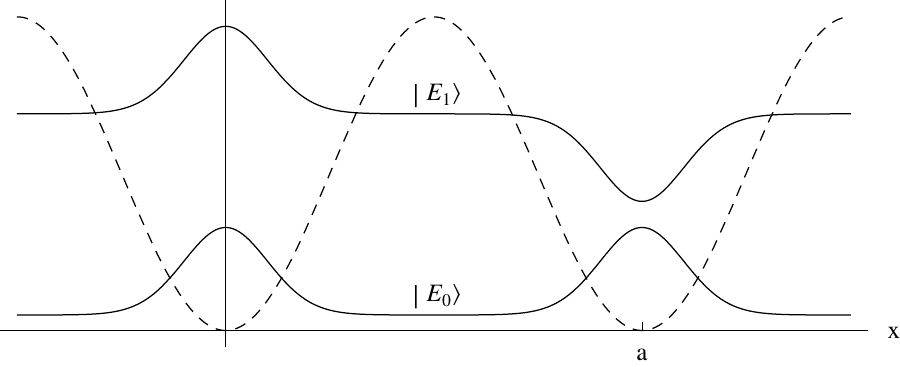}
\caption{Wavefunctions of the two lowest energy states obtained with the cut Fock space method for $a=10$ in the case with two minima. They are very similar to wavefunctions presented in Fig. \ref{fig:periodic_2_amplitudes}.}\label{fig:periodic_2_amplitudes_numerics}
\end{figure}

We have shown wavefunctions of the lowest energy states in the case of two minima obtained with numerical approach for $g=0.01$ in Figure \ref{fig:periodic_2_amplitudes_numerics}. They are in agreement with the expected shapes of wavefunctions obtained by semiclassical approximation.

\begin{figure}
\centering
\includegraphics[width=.8\textwidth]{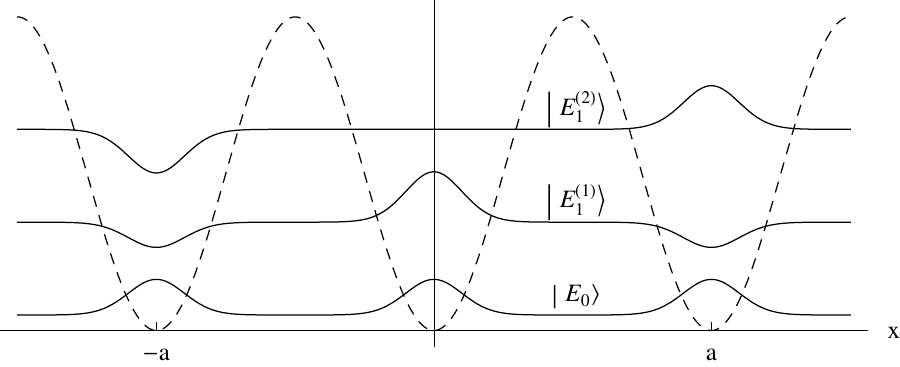}
\caption{Wavefunctions of the two lowest energy states (solid line) obtained with the cut Fock space method for $g=0.01$ in the case with three minima. They agree with the functions obtained with the semiclassical method presented in Fig. \ref{fig:periodic_3_amplitudes}. The potential is drawn with dashed line.}\label{fig:periodic_3_amplitudes_numerics}
\end{figure}

Note that in the case of three minima amplitudes of states achieved with WKB approximation were chosen symmetric or antisymmetric. In the cut Fock space approach the three lowest energies lie in sectors $\mathcal S_0^{+}$, $\mathcal S_1$ and $\mathcal S_{-1}$. Functions from the $\mathcal S_0^{+}$ sector are symmetric, but the two sectors $\mathcal S_1,\ \mathcal S_{-1}$ contain no symmetric or antisymmetric functions. In order to obtain comparable results we have to construct wavefunctions with positive of negative parity. Let $\ket{\psi_1}$ be the state with energy $E_1$ in the sector $\mathcal S_1$ (which is the lowest energy in this sector but the first excited energy in the whole Hilbert space). The parity transformation $P$ maps the sector $\mathcal S_1$ into $\mathcal S_{-1}$. The Hamiltonian commutes with the parity operator. Therefore, the state $P\ket{\psi_{1}}$ has the same energy as $\ket{\psi_1}$ and belongs to $\mathcal S_{-1}$. We now construct two new states:
\begin{align}\begin{split}
\ket{E_1^{(1)}}&=\frac{1}{\sqrt2}\left(\ket{\psi_1}+P\ket{\psi_1}\right),\\
\ket{E_1^{(2)}}&=\frac{-i}{\sqrt2}\left(\ket{\psi_1}-P\ket{\psi_1}\right)
\end{split}\end{align}
which are both eigenstates of parity with $P=+1$ and $P=-1$ respectively and are eigenstates of the Hamiltonian corresponding to the same energy $E_1$. Phases can be chosen such that the wavefunctions are real. The wavefunctions are plotted in Fig. \ref{fig:periodic_3_amplitudes_numerics}. As one can see amplitudes in each minimum agree with the result of WKB.

Concluding, numerical results agree with semiclassical approximation for small $g$. We are able to identify further corrections to the instanton contributions. They are identical in both cases, potential with two and three minima.

\section{Cosine potential in infinite space}
In this section we study the cosine potential in infinite space (on $\mathbb R$ without boundary conditions). One can view it as $K\to\infty$ limit of cosine potential in periodic space. The potential $V(x)$ is given by
\begin{align}
V(x)&=\frac{1}{(2\pi)^2}(1-\cos(2\pi x)),&\quad x\in\mathbb R.
\end{align}
Recall that the potential term in Hamiltonian is $\hat V=g^{-1}V(\sqrt gX)$. The feature that distinguishes this system from the former ones is the fact that there is infinite number of states with the energy close to $E=1/2$. These energies form a continuous band. This is a general property of systems with infinite number of minima as discussed by Bloch \cite{Bloch}. Also it can be seen directly in WKB calculation. In numerical computations we take finite $K$ and study behavior of energies with growing $K$. We observe that there is an interval $(E_0-\Delta E/2,E_0+\Delta E/2)$ with $K$ energies. In the limit $K\to\infty$ the interval is filled -- it is the continuous energy band. The middle of the interval $E_0$ is perturbative value of the ground energy and $E_0\to \frac{1}{2}$ as $g\to0$. Width of the interval is a nonperturbative function of $g$ and vanishes for $g=0$. Note that the ground energy of the system is $E_0-\Delta E/2$, not $E_0$.
\subsection{Semiclassical approximation}
The cosine potential has an infinite number of minima. We will be interested in obtaining all possible amplitudes $\braket{(m+k) a|\exp(-TH)|ma}$. Because of the discrete symmetry of translation by $a$
\begin{align}\label{eq:k_th_scalar_prod}
\braket{(m+k) a|e^{-TH}|ma}&=\braket{ka|e^{-TH}|0}.
\end{align}
The Euclidean action and the one--instanton solution may be obtained precisely as in the previous section,
\begin{align}
S_0&=a^2\int_0^1dx\sqrt{\frac{2}{(2\pi)^2}(1-\cos(2\pi x))}=\frac{2}{\pi^2}a^2,\\
\bar x(\tau)&=\frac{2a}{\pi}\arctan(e^\tau).
\end{align}
Then $A=\frac{2}{\pi}$.
The number of all $n$--instanton paths contributing to (\ref{eq:k_th_scalar_prod}) can be estimated in the same way one counts probability for the simple random walk on $\mathbb Z$. One has to make even number of steps to move by an even number of minima. The number of all possible paths moving $2k$ sites to the right in $2n$ or $2n+1$ steps is given by a Newton's binomial
\begin{align}\label{eq:no_of_paths_on_R}
N_{2n}^{(2k)}&=\binom{2n}{n+k},\\
N_{2n+1}^{(2k)}&=0.
\end{align}
We modify formulas for transition amplitudes from periodic case by including proper number of paths (\ref{eq:no_of_paths_on_R}). For (\ref{eq:k_th_scalar_prod}) we find
\begin{align}\begin{split}
\braket{2ka|e^{-TH}|0}&=e^{-\frac{T}{2}}\frac{1}{\sqrt\pi}\sum_{n=|k|}^\infty\frac{1}{(2n)!}\binom{2n}{n+|k|}\left(e^{-\frac{2}{\pi^2}a^2}\frac{2 a}{\pi^{3/2}} T\right)^{2n}\\
&=e^{-\frac{T}{2}}\frac{1}{\sqrt\pi}I_{2|k|}\left(e^{-\frac{2}{\pi^2}a^2}\frac{4a}{\pi^{3/2}}T\right)
\end{split}\end{align}
where $I_\alpha(x)$ is the modified Bessel function of the first kind.
For odd number of minima one gets
\begin{align}
N_{2n}^{(2k+1)}&=0,\\
N_{2n+1}^{(2k+1)}&=\binom{2n+1}{n+k+1},
\end{align}
which results in
\begin{align}\begin{split}
\braket{(2k+1)a|e^{-TH}|0}&=e^{-\frac{T}{2}}\frac{1}{\sqrt\pi}\sum_{n=k_0(k)}^\infty\frac{1}{(2n+1)!}\binom{2n+1}{n+k+1}\left(e^{-\frac{2}{\pi^2} a^2}\frac{2a}{\pi^{3/2}}T\right)^{2n+1}\\
&=e^{-\frac{T}{2}}\frac{1}{\sqrt\pi}I_{|2k+1|}\left(e^{-\frac{2}{\pi^2}a^2}\frac{4 a}{\pi^{3/2}}T\right),
\end{split}\end{align}
where $k_0(k)=k$ for $k$ nonnegative and $k_0(k)=-k-1$ for $k$ negative.
Summing up, we have
\begin{align}\label{eq:almost_general}
\braket{ka|e^{-TH}|0}
&=e^{-\frac{T}{2}}\frac{1}{\sqrt\pi}I_{|k|}\left(e^{-\frac{2}{\pi^2}a^2}\frac{4a}{\pi^{3/2}}T\right).
\end{align}
It is a known fact that for integer $k$
\begin{align}\label{eq:for_Bessel}
I_{k}(x)&=\frac{1}{\pi}\int_0^\pi d\theta e^{x\cos(\theta)}\cos(k\theta).
\end{align}
After combining (\ref{eq:almost_general}) and (\ref{eq:for_Bessel}) one gets
\begin{align}
\braket{ka|e^{-TH}|0}
&=\frac{1}{\pi^{3/2}}\int_0^\pi d\theta \cos(k\theta)\exp\left(-T\left(\frac{1}{2}-\cos(\theta) e^{-\frac{2}{\pi^2} a^2}\frac{4a}{\pi^{3/2}}\right)\right),
\end{align}
or in the $g=a^{-2}$ variable:
\begin{align}\label{eq:cosine_expansion_integral}
\braket{ka|e^{-TH}|0}
&=\frac{1}{\pi^{3/2}}\int_0^\pi d\theta \cos(k\theta)\exp\left(-T\left(\frac{1}{2}-\cos(\theta) e^{-2/\pi^2g}\frac{4}{\pi^{3/2}\sqrt g}\right)\right).
\end{align}
The energies form a continuous band, so one should use expansion with integral over energies rather than sum
\begin{align}
\braket{ka|e^{-TH}|0}&=\int dE\braket{ka|E}\braket{E|0}e^{-TE}.
\end{align}
Energies are parameterized by an angle $\theta$:
\begin{align}
E(\theta)&=\frac{1}{2}-\cos(\theta) e^{-2/\pi^2g}\frac{4}{\pi^{3/2}\sqrt g}&\theta\in(0,\pi)
\end{align}
and they form a band which has width
\begin{align}\label{eq:theta_energy}
\Delta E&=e^{-2/\pi^2g}\frac{8}{\pi^{3/2}\sqrt g}.
\end{align}
Energy states are improper (not normalizable) which is a typical situation for continuous spectrum. Their amplitudes at each minimum can be read off from (\ref{eq:cosine_expansion_integral}):
\begin{align}\label{eq:theta_amplitude}
\braket{ka|E(\theta)}=\cos(k\theta).
\end{align}
We know that in neighborhood of each minimum the wavefunction $\braket{x|E(\theta)}$ is approximately a gaussian of width 1. We can write a general expression
\begin{align}
\braket{x|E(\theta)}=\cos(x\theta/a)\phi(x)
\end{align}
where $\phi(x)$ is a periodic function with period $a$ and
\begin{align}
\cos(x\theta/a)\phi(x)&\approx e^{-x^2/2}&&\text{for }|x|\ll a/2.
\end{align}
If $\theta/\pi a$ is not a rational number then one can find such integer numbers $l,l'$ that the equation $l\theta/a=2l'\pi-\pi/2$ is satisfied with arbitrarily precision. Then
\begin{align}
\braket{x+la|E(\theta)}=\cos(x\theta/a+l\theta/a)\phi(x)\approx\sin(x\theta/a)\phi(x).
\end{align}
Due to the discrete translational symmetry there is another eigenstate of the Hamiltonian
\begin{align}
\braket{x|E(\theta)}_1=\sin(x\theta/a)\phi(x).
\end{align}
Because $\ket{E(\theta)}$ and $\ket{E(\theta)}_1$ correspond to the same energy, their superposition
\begin{align}\label{eq:Bloch_state}
\braket{x|E(\theta)}_B=e^{\pm ix\theta/a}\phi(x).
\end{align}
is also an eigenstate corresponding to the same energy. In fact, all eigenstates of the Hamiltonian -- also for rational $\theta/\pi a$ can be given in the form (\ref{eq:Bloch_state}). The wavefunction (\ref{eq:Bloch_state}) is also an eigenfunction of lattice translation $x\to x+a$ corresponding to the eigenvalue $e^{\pm i \theta}$.
This is in agreement with the Bloch theorem which states that eigenstates of the Hamiltonian with periodic potential can be always given as lattice translation eigenstates. They are a plane waves $e^{ikx}$ modulated by a periodic function $\phi(x)$. The lattice momentum $k$ is inside the Brillouin zone $k\in(-\pi/a,\pi/a)$ and energy of such state is $E=E_0-\Delta \cos(\theta)$. In our case $k=\pm \theta /a$ and $\Delta =\frac{1}{2}\Delta E$.

Concluding, the Hamiltonian $H$ can be diagonalized simultaneously with the translation operator $T_a x\to x+a$. Energy $E(\theta)$ corresponds to two points in spectrum of $T_a$, namely $e^{i\theta}$ and $e^{-i\theta}$.
\begin{figure}[h]
\centering
\includegraphics[width=.5\textwidth]{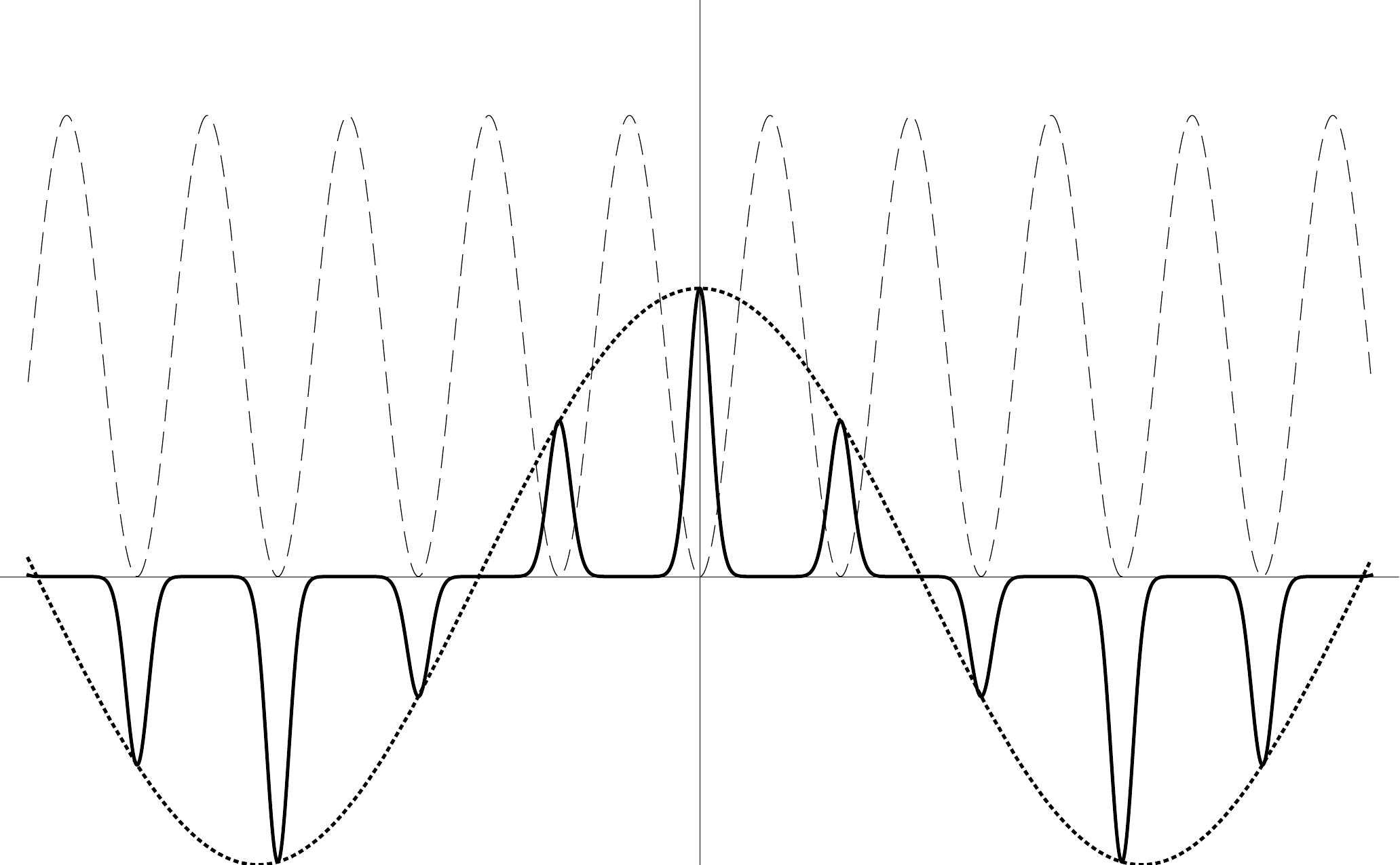}
\caption{A typical wavefunction (solid) for the cosine potential (dashed). An envelope of the gaussians is shown (dotted). For this function we chose $\theta=1$.}
\end{figure}
\subsection{Comparison with the Tamm-Dancoff method}
In this section we will show how to obtain energies for the cosine potential with the cut Fock space method. One can do it straightforwardly as it was done in Section \ref{sec:double_well_numerics}, i.e. write the Hamiltonian as a matrix in basis of the harmonic oscillator and then diagonalize it. This approach has two major disadvantages. Firstly, matrix of the Hamiltonian is no longer sparse. Secondly, we do not know explicit formulas for matrix elements so we have to expand the cosine in power series and then sum it as far as it is possible which is very time consuming. For this reasons we look for another approach. The crucial point is to properly exploit the translational symmetry $x\to x+1/\sqrt g$. In the infinite space the translation operator $T_{g^{-1/2}}x=x+1/\sqrt g$ has a continuous spectrum $\{e^{i\phi}:\phi\in(0,2\pi)\}$. It does not have proper eigenspaces which could be used to diagonalize the Hamiltonian. A remedy to this problem is to impose periodic boundary conditions. It means that we have to take the same Hamiltonian as in the previous section but with large $K$. Matrix elements of the Hamiltonian in sector $\mathcal S_k$ are given by formulas:
\begin{align}\begin{split}
(H_k)_{mn}&=\frac{1}{2}\braket{k+mK|P^2|k+nK}+\braket{k+mK|\hat V|k+nK},\quad m,n\in\mathbb Z\\
\braket{k+mK|P^2|k+nK}&=g\left(\frac{2\pi (k+nK)}{K}\right)^2\delta_{mn}\\
\braket{k+mK|\hat V|k+nK}&=\frac{1}{8\pi^2g}\times\left\{\begin{array}{rl}2&n=m\\-1&n=m\pm1\\0&\text{otherwise}\end{array}\right.
\end{split}\label{eq:cosine_expansion_matrices}\end{align}
Independently of the number of minima, matrix of the Hamiltonian $(H_k)_{mn}$ is tridiagonal in each sector $\mathcal S_k$ and one can use a fast algorithm for sparse matrices like Ardnoldi's iteration in order to find lowest eigenvalues. Moreover, if one increases the number of minima $K$ and the cutoff simultaneously, so that the cutoff is constant in each sector $\mathcal S_k$, then the time of computations grows linearly and needed amount of computer memory is constant! It proves that it is possible to reach high number of minima with low cost of computational time. Additionally, we know already that energies in sector $\mathcal S_k$ are the same as in the sector $\mathcal S_{-k}$ so it suffices to find only half of the eigenvalues. One shall remember that this approach is so good mostly because of the very special form of the potential. For any periodic potential which is not a polynomial in sines and cosines, matrix of the Hamiltonian will not be sparse. Still, using this approach one would always benefit from the $\mathbb Z_K$ symmetry.

\begin{figure}
\centering
\includegraphics[width=0.8\textwidth]{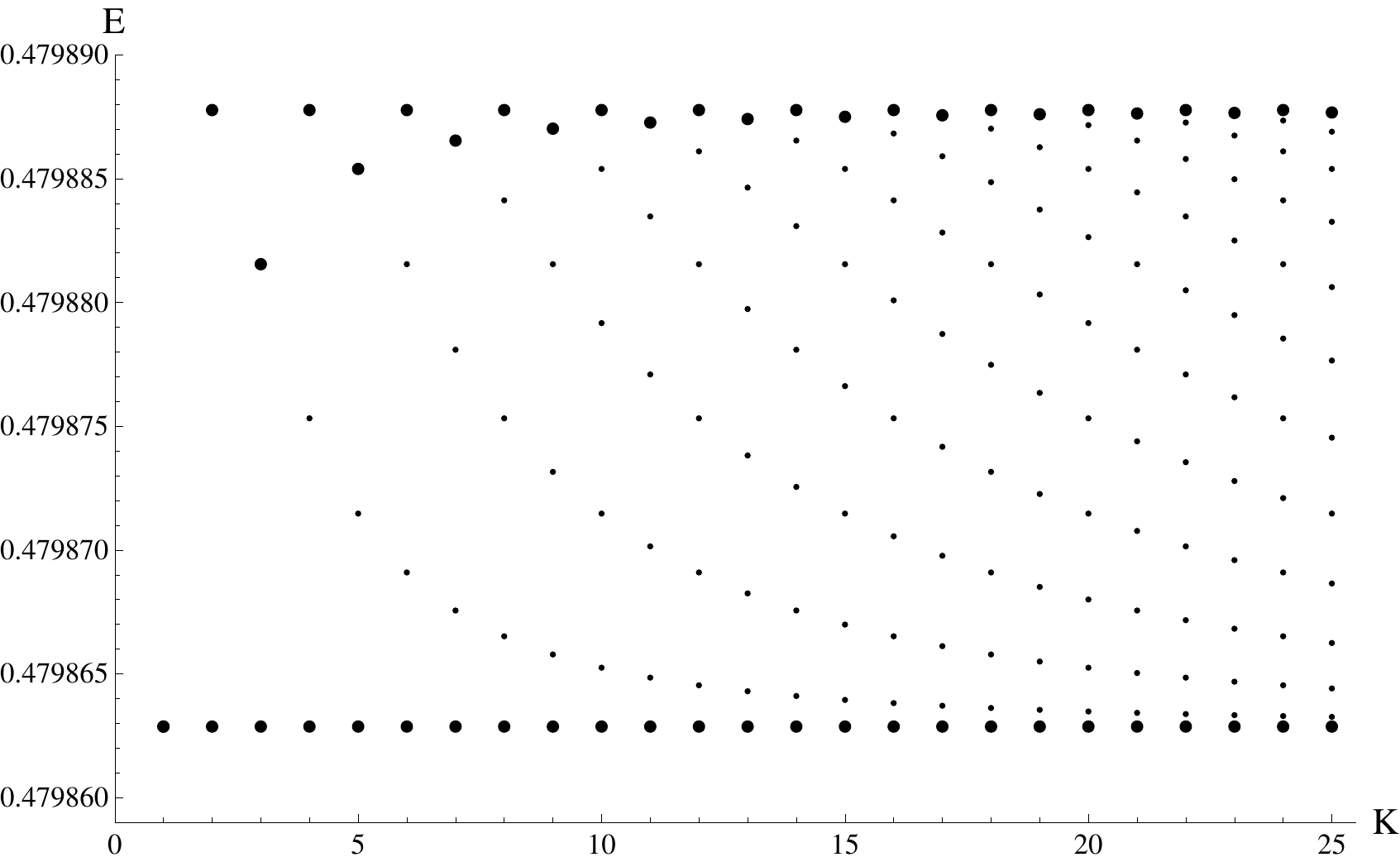}
\caption{Energies of the Hamiltonian plotted against growing $K$. Coupling constant is $g=0.00024$. The lowest and highest energies for each $K$ are marked.}\label{fig:cosine_expansion_growing_K}
\end{figure}

We will study two aspects of the spectrum. One is width of the lowest energy band and the other is distribution of energies in this band. In Fig. \ref{fig:cosine_expansion_growing_K} we plot eigenvalues of the Hamiltonian against $K$. One can see that the lowest energy $E_{min}$ is the same for each $K$ while the highest energy $E_{max}$ is constant only for even $K$. For odd values of $K$ the highest energy is lower than for even $K$'s but converges to the value for even $K$ as $K\to\infty$. This fact can be understood on both grounds, the semiclassical approximation and cut Fock space approach.

Let us consider formulas (\ref{eq:theta_energy}) and (\ref{eq:theta_amplitude}). We see that the lowest energy $E(\theta)$ is for $\theta=0$. Amplitude of the wavefunction is the same in each minimum: $\braket{ka|E(0)}\propto \cos(k\cdot 0)=1$. This can be realized in space with periodic boundary conditions with arbitrary number of minima. In contrary, the higher energy if for $\theta=\pi$. Then, the amplitude of the wavefunction alters: $\braket{ka|E(\pi)}\propto(-1)^k$. It is possible in space with even, but not with odd number of minima. Eg. for three minima we get:
\begin{align}
\braket{E(\pi)|0}=-\braket{E(\pi)|a}=\braket{E(\pi)|2a}=-\braket{E(\pi)|3a}\equiv-\braket{E(\pi)|0}
\end{align}
This is why for odd number of minima the highest energy in the band is lower than in the case of even number of minima.

In computations for the cut Fock space method we observe that the lowest energy in sectors $\mathcal S_k$ grows with $k$ for $k\in(0,K/2)$ and then decreases as $k$ goes from $K/2$ to $K-1$. From the explicit formula (\ref{eq:cosine_expansion_matrices}) one can see that the matrix of the potential part does not depend neither on the sector nor on number of minima. The part that depends on these values is the momentum part. However, in the $\mathcal S_0$ sector
\begin{align}
(P^2)_{mn}=\braket{mK|P^2|nK}&=g\left(\frac{2\pi (nK)}{K}\right)^2\delta_{mn}=g\left(2\pi n\right)^2\delta_{mn}
\end{align}
and in sector $\mathcal S_{K/2}$
\begin{align}
(P^2)_{mn}=\braket{K/2+mK|P^2|K/2+nK}&=g\left(\frac{2\pi (K/2+nK)}{K}\right)^2\delta_{mn}=g\left(2\pi(n+1/2)\right)^2\delta_{mn}
\end{align}
so we see that they do not depend on number of minima. This is why the lowest and highest energies in the band are constant for even $K$. For odd $K$ there is no sector $\mathcal S_{K/2}$ and for this reason the highest energy is lower than for even $K$.

In order to compare width of the energy band it is enough to study energies in sectors $\mathcal S_{0}$ and $\mathcal S_{K/2}$ which are independent of $K$ as long as it is even. Therefore, they are the same as in the previously considered case $K=2$. Semiclassical prediction of the width of energy band is
\begin{align}
\Delta E&=e^{-\frac{2}{\pi^2}a^2}\frac{8a}{\pi^{3/2}}.
\end{align}
is also exactly the same as the energy splitting in space with two minima. Then comparison of the semiclassical prediction with numerical computations is presented in Fig. \ref{fig:periodic_2_splitting} which was obtained for potential with two minima.

There is yet another property of the semiclassical approximation which we can compare with numerical results, namely energy dependence on the angle $\theta$. Consider the formula (\ref{eq:cosine_expansion_integral}) for $k=0$:
\begin{align}\begin{split}
\braket{0|e^{-TH}|0}&=\frac{1}{\pi^{3/2}}\int_0^\pi d\theta e^{-TE(\theta)}
\end{split}\end{align}

\begin{figure}
\centering
\includegraphics[width=.8\textwidth]{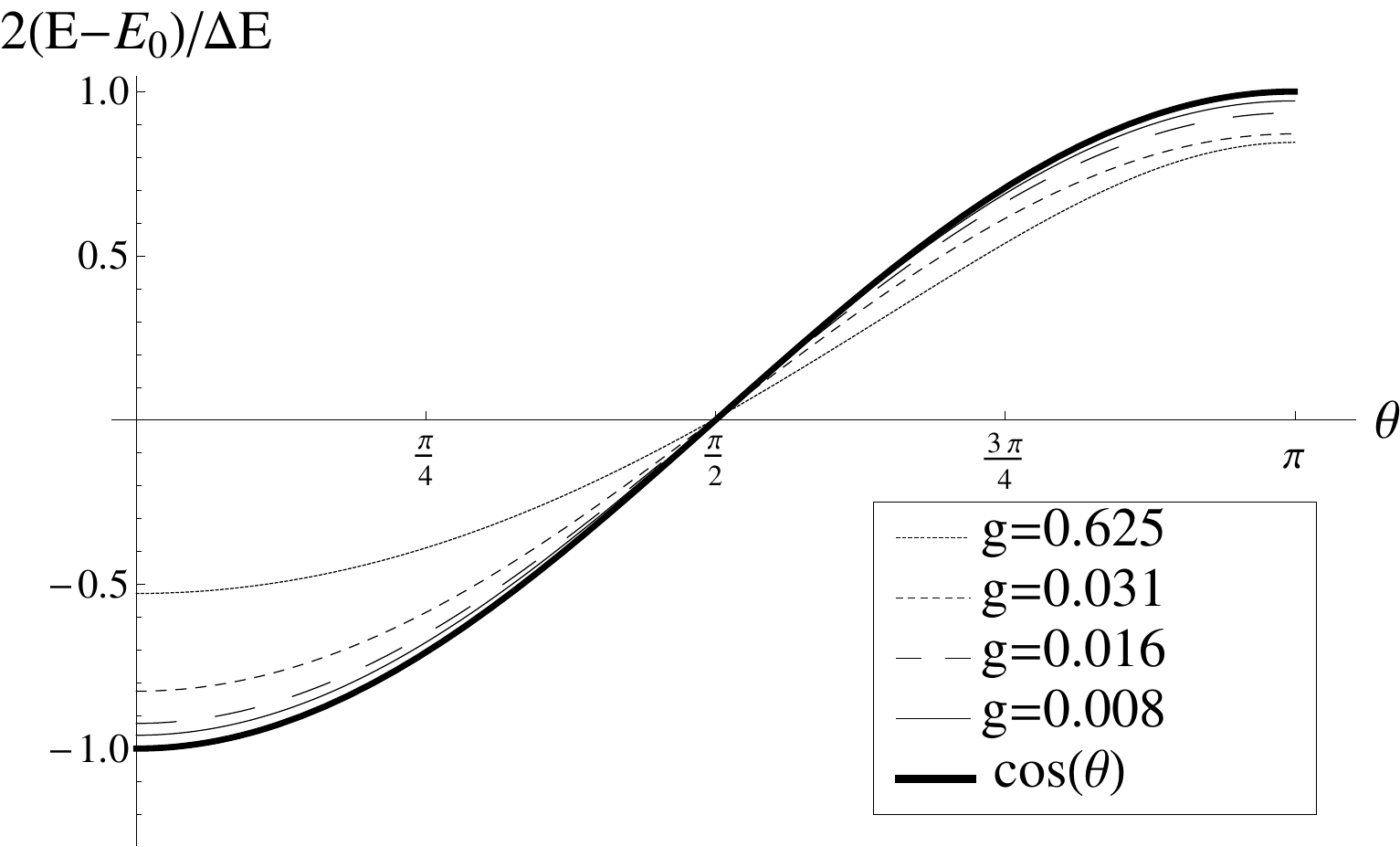}
\caption{According to the WKB approximation, dependence of the energies in the lowest energy band on the angle $\theta$ yields $2(E(\theta)-E_0)/\Delta E=\cos(\theta)$. This plot shows convergence to the cosine function for small $g$. We put $E_0=E(\pi/2).$ Value of $E_0$ is determined numerically and $\Delta E$ is the result of semiclassical approximation.}\label{fig:cosine_expansion_smalling_g}
\end{figure}

Recall that the energy state $\ket{E(\theta)}$ may be decomposed into eigenstates of translation operator $T_a$ corresponding to eigenvalues $e^{i\theta}$ and $e^{-i\theta}$. Basis states of a sector $\mathcal S_k$ are also eigenstates of $T_a$ corresponding to eigenvalue $e^{i\theta_k}=e^{2\pi ik/K}$. For large $K$ points $e^{i\theta_k}$ cover the unit circle uniformly. In each sector $\mathcal S_k$ there is precisely one energy contained in the lowest energy band. Therefore, picking some large $K$ and finding the lowest energy in each sector $\mathcal S_k$ enables us to find the relation between energy and angle. Remember that the instanton calculus does not take into account perturbative correction, so the energy density is centered at value $E_0\neq 1/2$. Now we can compare energies obtained with the cut Fock space method with as functions of $\theta$ with the WKB formula $E(\theta)=E_0-\cos(\theta) \Delta E/2$. To do this, we plotted the energy for $g=0.625,\ 0.031,\ 0.016,\ 0.08$ and $K=1000$ in Fig. \ref{fig:cosine_expansion_smalling_g}. One can clearly see that the energies come closer to the WKB (or Bloch) formula as $g\to0$.

Concluding, using the semiclassical approximation we have shown that energies which are of order $E=1/2$ form a continuous band and calculated its width. We have shown that eigenstates of each energy may be written as eigenstates of the lattice translation operator. All above facts are in agreement with the Bloch theorem. Next we confirmed width of the band numerically and showed agreement in energy dependence on the angle $\theta$.

\section{Summary}
In this chapter we were comparing results of the semiclassical approximation with the cut Fock space method for periodic potential $V(x)=\frac{1}{4\pi^2}(1-\cos(2\pi x))$ in the weak coupling limit. We studied two cases with imposed periodic boundary conditions with $x\in(0,K g^{-1/2})$ with $K=2,3$ and a separate case $x\in\mathbb R$.

It was shown that the instanton calculus for the cosine potential is very similar to the case of double well. The main difference is calculating the number of topologically different $n$--instanton solutions with fixed values at infinities. The classical one--instanton solution as well as its Euclidean action were found analytically. In both cases, two and three minima, we found two energies in the lowest energy regime. However, for $K=3$ one of them is degenerate which we were able to prove within the instanton calculus. Explicit formulas for splitting of energies were given in (\ref{eq:periodic_2_energies}) and (\ref{eq:periodic_3_energies}).

On the basis of these results and observations made in Section \ref{ch:TD_for_anharmonic_oscillator} we state that for any finite $K$, energies which are below the potential barrier join into groups of $K$ energies including degeneracies. Energy splittings in each group are nonperturbative. However, for $K>3$ similar calculations are more difficult because recursive relations for number of $n$--instanton solutions become more complex.

In the case of infinite space, there is an energy band which also has a nonperturbative width given in (\ref{eq:theta_energy}). We also remarked that the results agree qualitatively with Bloch theorem. Let $\psi(x)$ be a wavefunction of energy state. One may introduce an angle $\theta$ being a relative phase of $\psi(x)$ in subsequent minima, i.e. $\psi(x+a)=e^{i\theta}\psi(x)$. Then energy in a band corresponding to $\psi$ is given by the formula $E=E_0-\frac{\Delta E}{2}\cos(\theta)$.

Thanks to the translational $\mathbb Z_K$ symmetry we were able to construct a version of Tamm--Dancoff method which turned out to be especially efficient for large $K$. Using numerical analysis we were able to establish agreement with energy splitting for $K=2,3$ as well as width of energy band for $K=\infty$. We also checked dependence of the energy on angle $\theta$.

\chapter{Anharmonic triple well potential}
In this chapter we will analyze a triple well potential. This is the most natural generalisation of the double well potential. The cut Fock space method is the same as in previous chapters and the instanton calculus is very similar. Because of the fact that there are three minima, one expects three energies to be almost degenerate. Although splitting of energies is indeed small for $g\to0$, we will show that it is nonperturbative only in very special cases. This fact indicates that one has to be careful with the WKB approximation and know weather it applies to certain system or not.

A relevant potential will be now constructed. We impose conditions on the potential so it has minima at $x=0$ and $x=\pm1$ and the second derivative is equal in all minima. They are
\begin{align}\label{eq:triple_well_first_conditions}\begin{split}
V(0)=V(\pm 1)&=0,\\
V'(0)=V'(\pm 1)&=0,\\
V''(0)=V''(\pm 1)&=1.
\end{split}\end{align}
The least order polynomial satisfying these constraints is $8th$ order. However, then the potential is unbounded from below. In order to deal with this problem we impose additional requirements which are
\begin{align}\label{eq:triple_well_add_conditions}
V(\pm 1/2)&=\frac{1}{2\pi^2}.
\end{align}
This condition is inspired by the function $\frac{1}{4\pi^2}\left(1-\cos(2\pi x)\right)$ which satisfies both, (\ref{eq:triple_well_first_conditions}) and (\ref{eq:triple_well_add_conditions}) and is nonnegative. The least order polynomial satisfying all conditions is $10th$ order. It is positive and has three global minima which are $x=0,\pm1$.
\begin{align}\begin{split}
V(x)&=\frac{1}{2}x^2+\left(-\frac{85}{24}+\frac{512}{27\pi^2}\right)x^4+\left(\frac{31}{4}-\frac{512}{9\pi^2}\right)x^6
+\left(-\frac{55}{8}+\frac{512}{9\pi^2}\right)x^8
+\left(\frac{13}{6}-\frac{512}{27\pi^2}\right)x^{10}.
\end{split}\end{align}
Recall that the Hamiltonian is
\begin{align}
H=\frac{1}{2}P^2+\frac{1}{g}V(\sqrt gX).
\end{align}
\section{Cut Fock space method}\label{ch:TD_for_three_minima}
In this section we perform numerical computations to obtain energies of the system.  We apply the finite matrix method in the same way as is was done for the double well potential. Since the potential involves higher powers of $X$, the expression for matrix elements is more complex but still amenable.
\begin{align}
\braket{m|H|n}&=m\delta_{mn}+\frac{1}{2}+\left(-\frac{85}{24}+\frac{512}{27\pi^2}\right)\braket{m|X^4|n}+\left(\frac{31}{4}-\frac{512}{9\pi^2}\right)\braket{m|X^6|n}\\
&\quad+\left(-\frac{55}{8}+\frac{512}{9\pi^2}\right)\braket{m|X^8|n}+\left(\frac{13}{6}-\frac{512}{27\pi^2}\right)\braket{m|X^{10}|n}.
\end{align}
If is convenient to express each power of $X$ as in a form
\begin{align}
X^k=q_{k,0}(N)+\sum_{j=1}^k a^j q_{k,-j}(N)+(a^\dagger)^jq_{k,j}(N)
\end{align}
where $N=a^\dagger a$ is the operator of number of quanta. For $k=4$ it is
\begin{align}
X^4=\frac{3}{4}+\frac{3}{2}N+\frac{3}{2}N^2+a^2\left(-\frac{1}{2}+N\right)+(a^\dagger)^2\left(\frac{3}{2}+N\right)+\frac{1}{4}a^4+\frac{1}{4}(a^\dagger)^2.
\end{align}
Then
\begin{align}
\braket{m|X^k|n}=q_{k,0}(n)\delta_{m,n}+\sum_{j=1}^k q_{k,-j}(n)\delta_{m,n-j}\sqrt\frac{n!}{m!}+q_{k,j}(n)\delta_{m,n+j}\sqrt\frac{m!}{n!}.
\end{align}
One can easily find recursive relations for functions $q_{k,j}(n)$ and thus find explicit expressions for arbitrary (but finite) number of amplitudes of the form $\braket{m|X^k|n}$.

\begin{figure}[h]
\centering
\includegraphics[width=.7\textwidth]{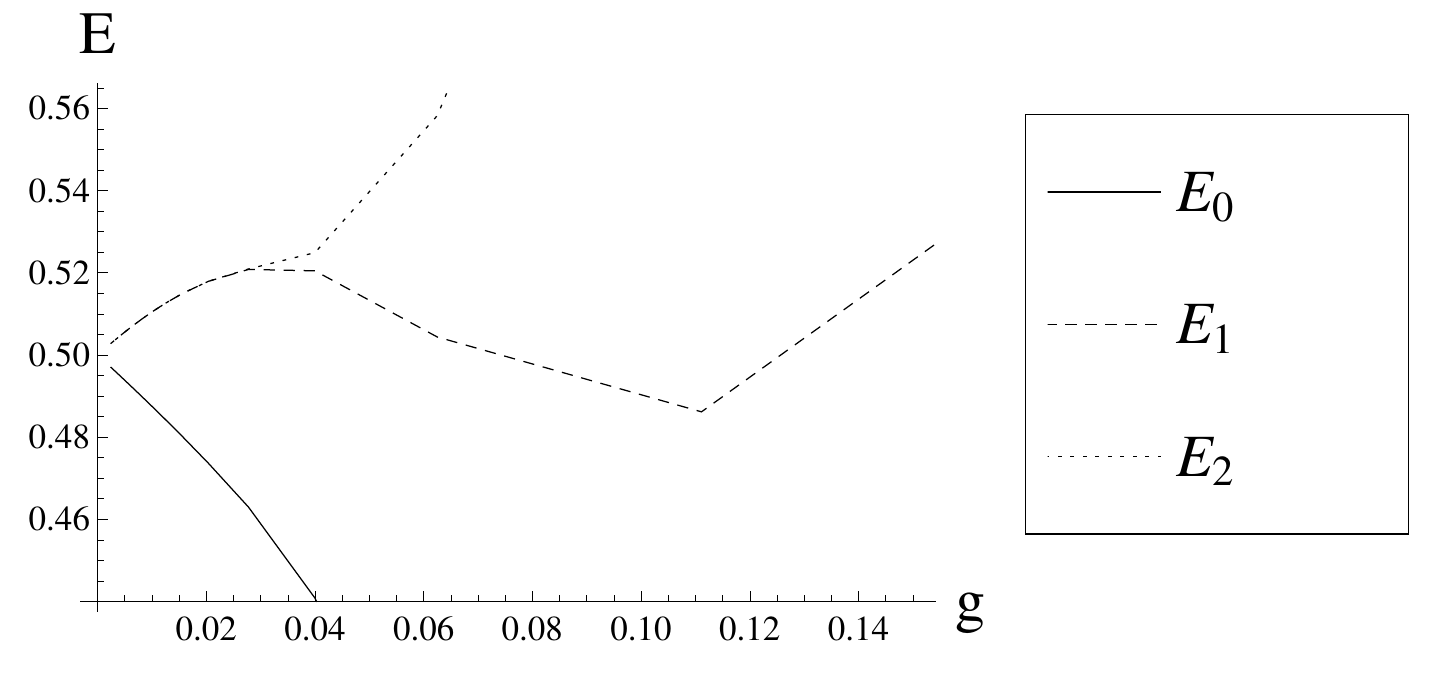}
\caption{The three lowest energies of the Hamiltonian. It can be seen that the lowest energies is lower than the other two by a quantity of order $\mathcal O(g)$}
\label{fig:energies_for _3_minima}
\end{figure}

\begin{figure}
\centering
\includegraphics[width=\textwidth]{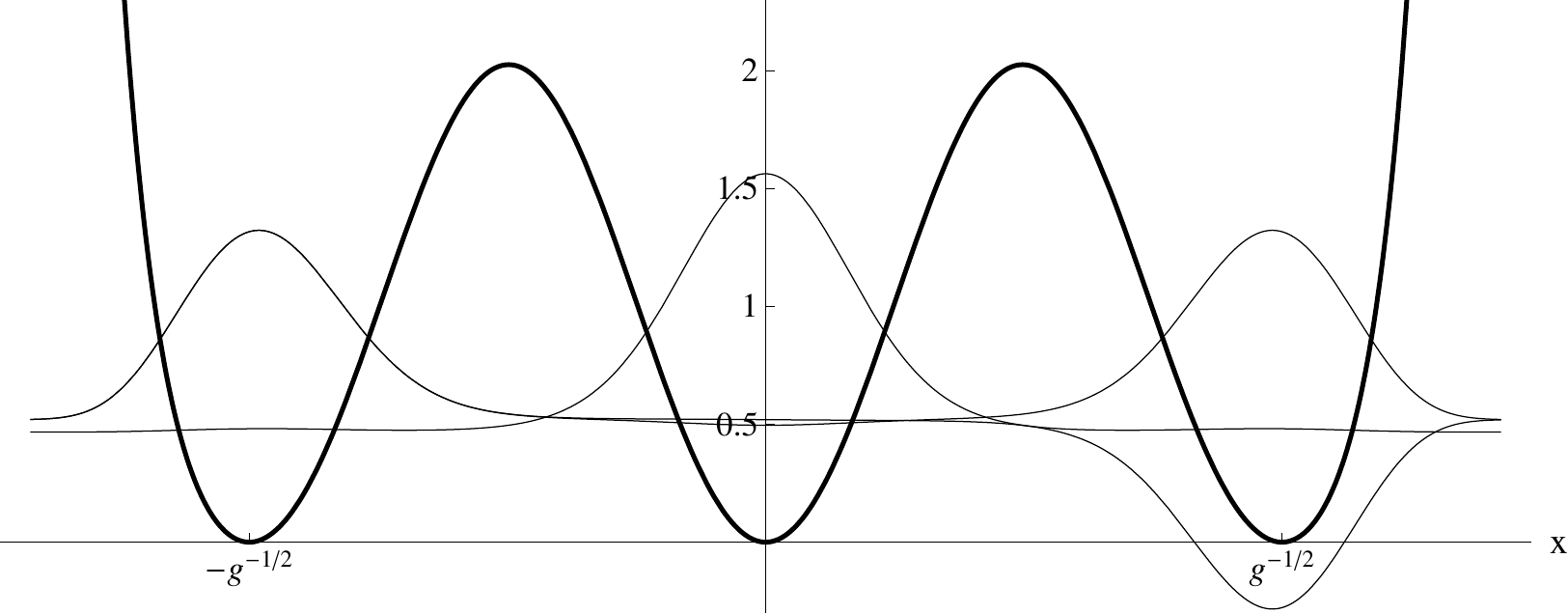}
\caption{Wavefunctions corresponding to the three lowest energies for $g=0.025$. Wavefunctions are lifted by values of energies they correspond to. The lowest energy correspond to wavefunction which has support in the central minimum only. The other two have support in the two side minima.}\label{fig:wavefunctions_at_delta_0}
\end{figure}

We construct matrix of the Hamiltonian and compute its eigenvalues. The three lowest energies for $g\in(0.001,0.1)$ are presented in Fig. \ref{fig:energies_for _3_minima}. As one can see, the ground energy $E_0$ differs from the two higher energies $E_1$ and $E_2$ by a quantity of order $\mathcal O(g)$ while the difference $E_2-E_1$ is nonperturbative. Wavefunctions corresponding to the three lowest energies are shown in Fig. \ref{fig:wavefunctions_at_delta_0} . The wavefunction $\psi_0(x)$ corresponding to the lowest energy has support in neighborhood of the central minimum only. It means that a state situated in the middle minimum does not tunnel to any other minimum. The other two wavefunctions $\psi_{1,2}(x)$ are approximately symmetric and antisymmetric combinations of gaussians localized in left and right minima. A wavefunction $\psi(x)=\frac{1}{\sqrt 2}(\psi_1(x)+\psi_2(x))$ is located in the right minimum. Due to energy splitting it tunnels to the left minimum in a finite time. We conclude that only nonperturbative energy differences are responsible for the tunneling effect.

We may ask if the energy difference $E_2-E_1$ is the same as in the double well potential. This question is reasonable since tunneling ignores the central minimum. According to calculations for double well potential the splitting of energies is
\begin{align}\label{eq:3minima_delta_zero_splitting}
E_2-E_1=\frac{C}{\sqrt g}e^{-S_0}
\end{align}
where $S_0$ is the Euclidean action. As it was shown in the instanton calculus for double well potential, the Euclidean action $S_0$ can be expressed as a simple integral and thus can be obtained numerically: $S_0=\frac{1}{g}\int_{-1}^1dx\sqrt{2V(x)}\approx 0.4036/g$.
The coefficient $C$ is not possible to be determined because there is no classical trajectory like there was in the case of double well potential. A trajectory satisfying $\bar x(\pm\infty)=\pm1$ and equations of motion in the Euclidean space needs an infinite time to get through the middle minimum. For this reason it is not possible to perform a calculation for energy splitting in WKB approximation. Nevertheless we check if the formula (\ref{eq:3minima_delta_zero_splitting}) holds.
\begin{figure}[ht]
\centering
\includegraphics[width=.7\textwidth]{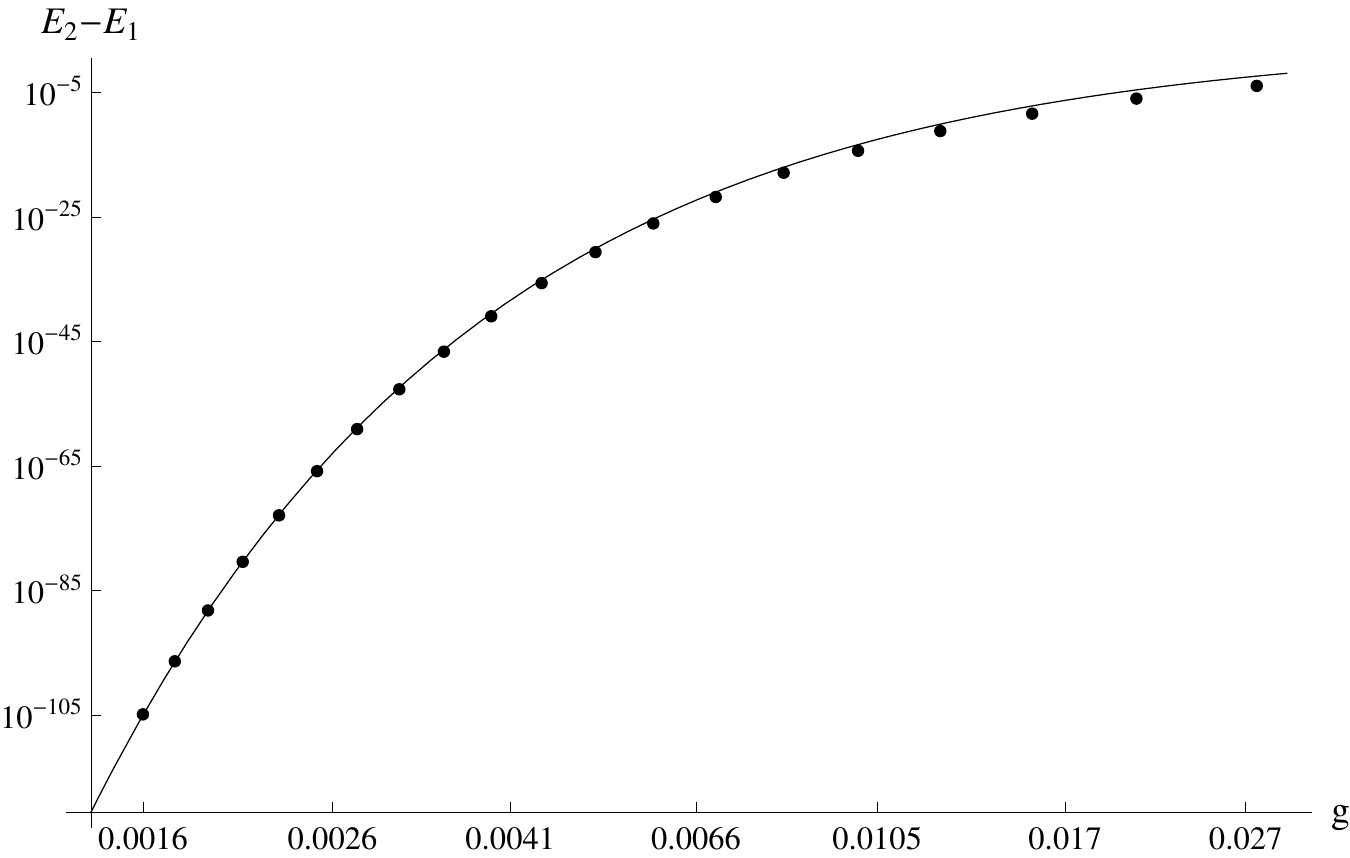}
\caption{Energy splitting between the second and third energy in triple well potential. Points represent numerical data and the solid line is a fit. The discrepancy for larger values of the coupling constant $g$ is an effect of higher order corrections}\label{fig:3minima_splitting_delta_zero}
\end{figure}

Numerical data of the energy difference together with a function given in (\ref{eq:3minima_delta_zero_splitting}) is presented in Fig. \ref{fig:3minima_splitting_delta_zero}. The coefficient $C$ was fitted to several points corresponding to lowest values of coupling constant $g$. Points are very well aligned on the curve which indicates that the formula (\ref{eq:3minima_delta_zero_splitting}) has correct form. Little discrepancy visible for larger values of $g$ is caused by higher order corrections.

Let us now think if there is a possibility to alter the potential in such way that there is tunneling between all three minima. Let us consider potential with an additional parameter $\delta$:
\begin{align}\begin{split}
V_\delta(x)&=\frac{1 + \delta}{2}x^2+\left(-\frac{85}{24}+\frac{512}{27\pi^2}-\frac{7\delta}{2}\right)x^4+\left(\frac{31}{4}-\frac{512}{9\pi^2}+\frac{15\delta}{2}\right)x^6\\
&\quad+\left(-\frac{55}{8}+\frac{512}{9\pi^2}-\frac{13\delta}{2}\right)x^8
+\left(\frac{13}{6}-\frac{512}{27\pi^2}+2\delta\right)x^{10}
\end{split}\end{align}
For $\delta =0$ it is the previous case. Moreover, it is symmetric under the parity transformation so we expect tunneling between right and left minimum to be preserved. Conditions (\ref{eq:triple_well_first_conditions}) remain unchanged apart from the second derivative of the potential at the central minimum which is now larger:
\begin{align}
V''(0)&=1+\delta,&V''(\pm1)&=1.
\end{align}
It follows that in the zero order perturbation expansion energy in the middle minimum is $\frac{1}{2}(1+\delta)$ and in side minima it is $\frac{1}{2}$. On the other hand, from computations done earlier we know that for $\delta=0$ exact energy in the central minimum is lower than in the side ones. It follows that if we increase $\delta$ then the energy $E_0$ grows faster than $E_1$ and they will eventually cross. On the other hand, according to Wigner non--crossing theorem \cite{Wigner} it cannot happen because corresponding wavefunctions $\psi_0(x)$ and $\psi_1(x)$ have the same parity.

Let us consider energies $E_0(\delta)$ and $E_{*}(\delta)$ as functions of $\delta$ with $E_0(\delta)$ being the ground energy and $E_{*}(\delta)$ the first excited energy in the symmetric (parity $+1$) sector. Let then $\psi_0(\delta,x)$ and $\psi_*(\delta,x)$ be corresponding wavefunctions. Let $\phi_{0,*}(x)$ stand for energy eigenfunctions in nonperturbed potential, i.e. $\phi_i(x)=\psi_i(\delta=0,x)$. $\phi_0(x)$ is a wavefunction localized in the central minimum while the other function, $\phi_*(x)$ is a symmetric combination of states localized in left and right minima.

From what was said, if the parameter $\delta$ is large enough then the energy in the central minimum is larger than in left and right minima. Therefore, for $\delta$ large enough $\psi_0(\delta,x)\approx\phi_*(x)$, i.e. the wavefunction corresponding to the lowest energy is localized in left and right minimum. The first excited energy corresponds to a state in the middle minimum so $\psi_*(\delta,x)\approx\phi_0(x)$. We infer that $\psi_0(\delta,x)\approx\alpha(\delta) \phi_0(x)+\beta^*(\delta) \phi_*(x)$ and $\psi_*(\delta,x)\approx\beta(\delta) \phi_0(x)-\alpha(\delta)^* \phi_*(x)$ where $|\alpha|^2+|\beta|^2=1$, $\alpha(0)=1$ and $\beta(\delta)=1$ for $\delta$ large enough. Because the transition is smooth, there is some point $\delta_c$ at which $|\alpha(\delta_c)|=|\beta(\delta_c)|=\frac{1}{\sqrt 2}$. It turns out that both coefficients have the same phase, i.e. we can choose $\alpha(\delta_c)=\beta(\delta_c)=\frac{1}{\sqrt 2}$. Then a wavefunction localized in the middle minimum, i.e. described by function $\phi_0(x)\approx\frac{1}{\sqrt 2}(\psi_0(\delta_c,x)+\psi_*(\delta_c,x))$ will eventually tunnel to left and right minimum. We see that there is tunneling between all three minima.

The question that now arises is how to find the critical value $\delta_c$ of the parameter $\delta$. At $\delta=\delta_c$ both $E_0$ and $E_{*}$ correspond to states which are superpositions of states in all three minima. Then the mean energy of a state in the middle minimum is the same as in side minima but neither of them is an eigenvalue of the Hamiltonian. Difference between energies $E_0$ and $E_{*}$ comes from the tunneling effect only. As we know, the energy splitting has to be small for small $g$. Thus, we may obtain $\delta_c$ by minimizing the energy splitting $E_{*}-E_0$. We shall remember that we neglected the lowest energy in parity $-1$ sector. Numerical computations show that it is between $E_0$ and $E_{*}$. For $\delta=\delta_c$ the energy $E_{*}=E_2$ is the second excited energy and the first excited energy $E_1$ is in parity $-1$ sector. It means that for some $\delta<\delta_c$ there is degeneracy of energy (first excited energy in even sector is equal to lowest energy in odd sector). It is not forbidden by the Wigner's theorem because the energies belong to sectors with different parity.

\begin{figure}
\centering
\includegraphics[width=.6\textwidth]{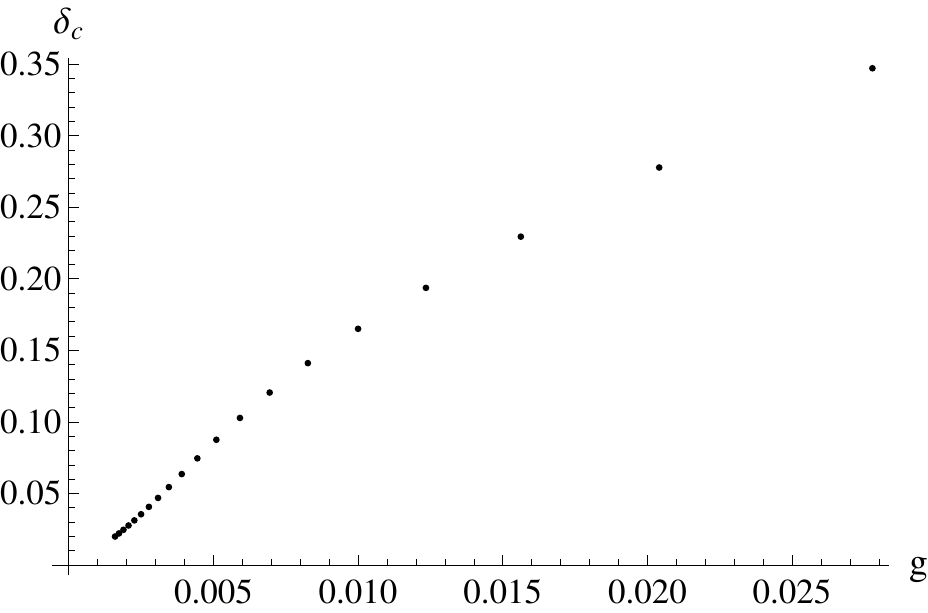}
\caption{Plot of the  value of parameter $\delta_c$ as a function of $g$ obtained from numerical data.}\label{fig:3minima_critical_deltas}
\end{figure}

Minimizing the energy difference is computationally demanding and we cannot reach very small values of parameter $g$. We managed to get down to $g=0.0016$. Certainly, the value of $\delta_c$ depends on the coupling constant. Plot of $\delta_c$ is shown in Fig. \ref{fig:3minima_critical_deltas}. For $\delta=0$ both energies of states in central and in side minima converge to $\frac{1}{2}$ as $g\to0$. It is clear that $\delta_c$ converges to $0$.
For smaller $g$ one has to lift energy of the state in the middle minimum by smaller value in order to exceed energy of a state in left or right minimum.

\begin{figure}
\centering
\includegraphics[width=\textwidth]{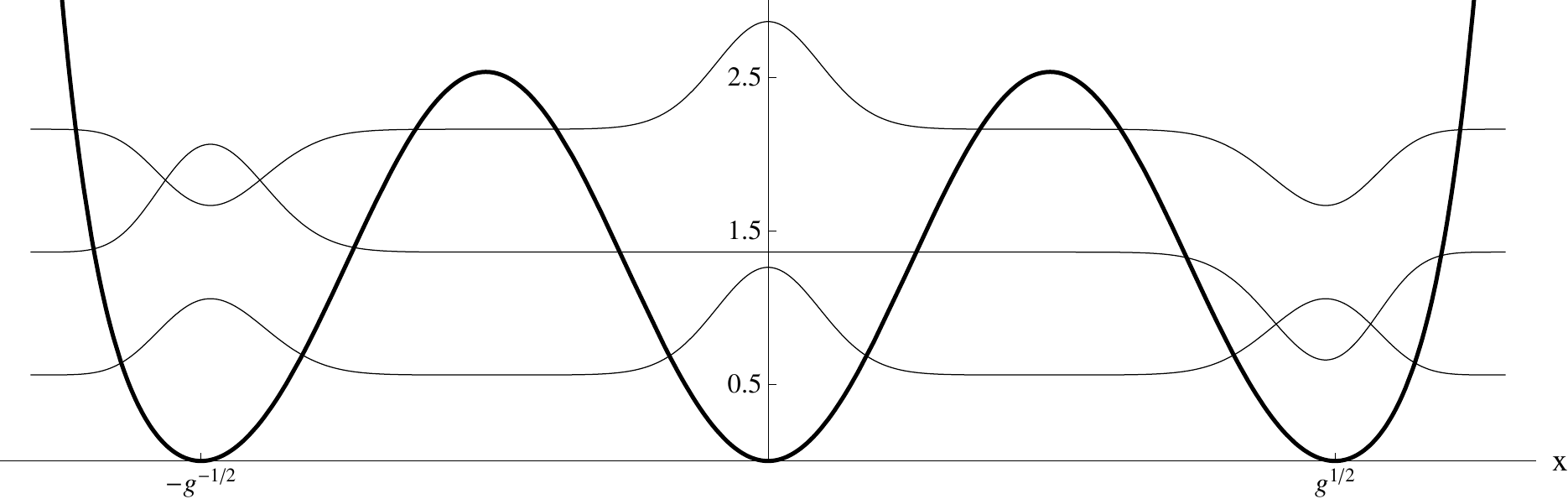}
\caption{Plots of wavefunctions corresponding to the three lowest energies. The lowest wavefunctions is lifted by the energy it corresponds to. The other are lifted by a greater value so that they can be distinguished.}\label{fig:3minima_wavefunctions_delta}
\end{figure}

Wavefunctions are presented in Fig. \ref{fig:3minima_wavefunctions_delta}. One can see that it is possible to construct a wavefunction is localized only in one minimum. The state $\ket{C}=\frac{1}{\sqrt 2}(\ket{E_0}+\ket{E_2})$  represents a state in the central minimum and $\ket{L}=\frac{1}{2}(\ket{E_0}+\sqrt 2\ket{E_1}-\ket{E_2})$ represents a state in the left minimum. As one can see none of them is energy state. It follows that after certain time each of those states will tunnel to other minima.
\section{Instanton calculus}
The main idea of instanton calculus is very similar to cases already considered. However, there are some differences since $V''(0)\neq 1$.
We will be again interested in calculating amplitudes
\begin{align}\label{eq:general_amplitude}
\braket{q_2|e^{-TH}|q_1}
\end{align}
with $q_1,q_2=0,\pm a$ where we put $a=1/\sqrt g$ for convenience. Because of symmetry of the potential, some choices of $q_i$'s yield the same result. There are four different amplitudes:
\begin{align}
\braket{0|e^{-TH}|-a}\label{eq:triple_well_amplitude_left_middle}\\
\braket{a|e^{-TH}|-a}\label{eq:triple_well_amplitude_left_right}\\
\braket{-a|e^{-TH}|-a}\label{eq:triple_well_amplitude_left_left}\\
\braket{0|e^{-TH}|0}\label{eq:triple_well_amplitude_middle_middle}
\end{align}

A single amplitude is a sum of $n$--instanton contributions. The number of possible topologically different paths of an $n$--instanton trajectory $N_n(q_1,q_2)$ has to be taken into account:
\begin{align}
\braket{q_2|e^{-TH}|q_1}=\sum_nN_n(q_1,q_2)\braket{q_2|e^{-TH}|q_1}_n
\end{align}
where $\braket{q_2|e^{-TH}|q_1}_n$ is a contribution from a single $n$--instanton trajectory. By the same argument as for the double well we have
\begin{align}
\braket{q_2|e^{-TH}|q_1}_n&=e^{-S_E[\bar x_n(\tau)]}\mathcal N\det{}^{-\frac{1}{2}}\left[-\frac{d^2}{d\tau^2}+V''(\bar x_n(\tau)/a)\right]
\end{align}
where $\bar x_n(\tau)$ is an $n$--instanton solution starting at $q_1$ and ending at $q_2$.
Calculation of the determinant is more onerous than for the case of double well potential. The full reasoning is given in Appendix \ref{sec:triple_well_appendix}. Let us denote $\omega_i^2=V''(q_i)$. The result is the formula \ref{eq:appendix_tripple_well_determinant}:
\begin{align}\label{eq:tripple_well_determinant}\begin{split}
\mathcal N\left(\det{}'\left[-\frac{d^2}{d\tau^2}+V''(\bar x_n(\tau)/a)\right]\right)^{-\frac{1}{2}}&=\frac{(\omega_1\omega_2)^{1/4}}{\sqrt \pi}
\left(\omega^{1/4}\sqrt\frac{2}{1+\omega}\frac{aA}{\sqrt {S_0}}\right)^ne^{-\frac{T}{2}\frac{\omega_1+\omega_2}{2}}\\
&\quad\times
\left\{\begin{array}{ll}(2\omega(1+\omega))^{n/4}&\omega_1=\omega_2\\
(2\omega(1+\omega))^{(n-1)/4}\sqrt{1+\omega}&\omega_1=\omega,\omega_2=1\\
(2\omega(1+\omega))^{(n-1)/4}\sqrt{2\omega}&\omega_1=1,\omega_2=\omega\\
\end{array}\right.
\end{split}\end{align}
Note that we excluded zero modes which give an additional coefficient
\begin{align}
\frac{T^n}{n!}\left(S_0/2\pi\right)^{n/2}.
\end{align}
Due to reflection symmetry of the potential, action of the instanton solution does not depend on starting and ending point of the instanton. The Euclidean classical action is
\begin{align}
S_0= a^2\int_{-1}^0dx\sqrt{2V(x)}=a^2\int_{0}^1dx\sqrt{2V(x)}.
\end{align}
This integral cannot be calculated analytically, but it is easy to do it numerically with arbitrarily high precision. One has to remember that the potential depends on parameter $\delta$ and the action has to be computed for each value of $\delta$ separately. The highest value of $\delta$ for our range of coupling constant $g$ is $\delta=0.214$. Computations yield
\begin{align}
S_0(\delta=0)/a^2&=0.2019,&S_0(\delta=0.214)/a^2&=0.1987.
\end{align}
As one can see, there is only small dependence of $S_0$ on $\delta$. On the other hand, it is multiplied by a large number $a^2$ and exponentiated. For parameters in our computations including the dependence of $S_0$ on $\delta$ gives a correction from $13\%$ for the smallest coupling up to $22\%$ for largest coupling $g$.

\begin{figure}[h]
  \centering
  \subfloat{\includegraphics[width=0.4\textwidth]{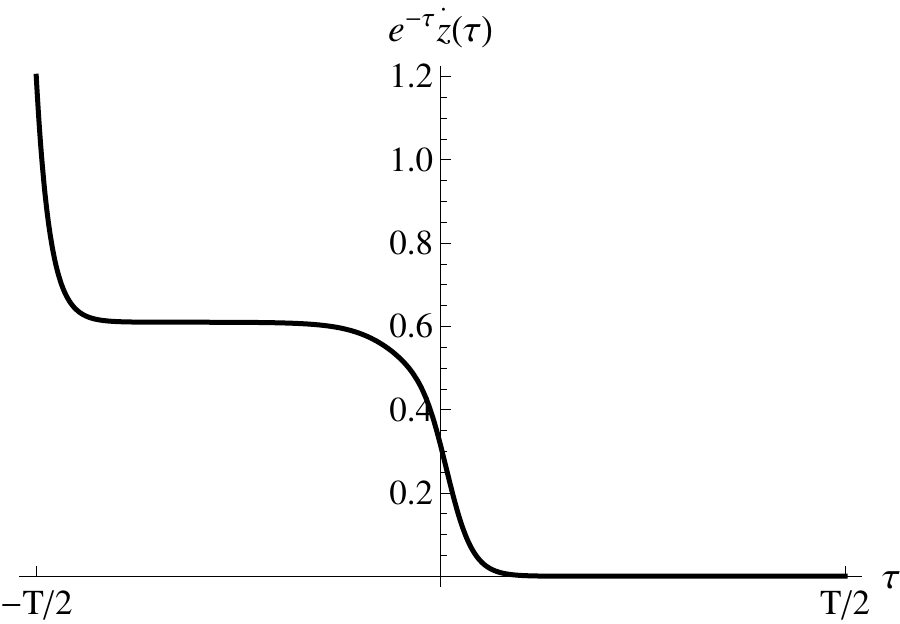}}
  \subfloat{\includegraphics[width=0.4\textwidth]{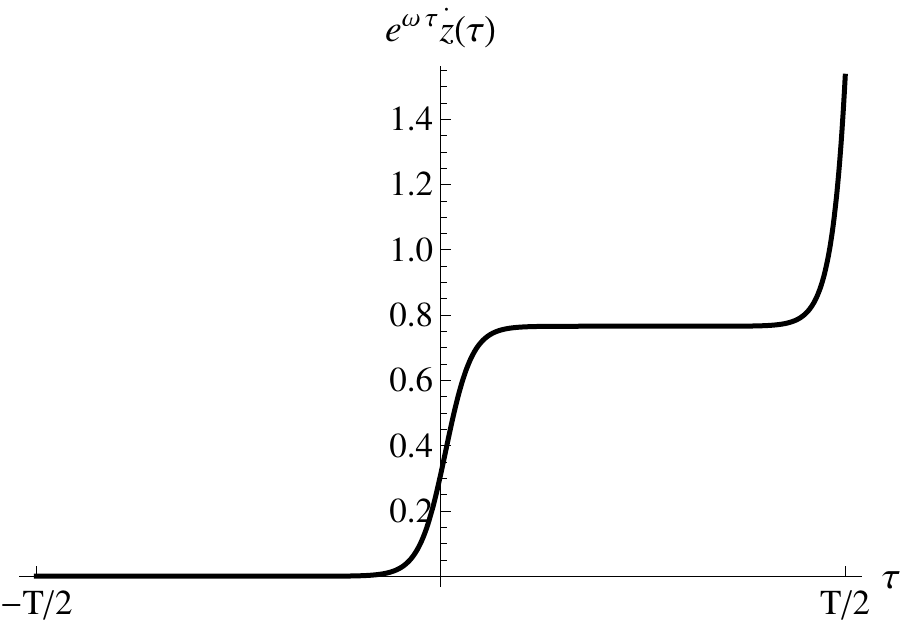}}
  \caption{Behavior of the one--instanton solution near boundaries. Values of $A_\pm$ are determined from height of the plateaux.}
  \label{fig:numerical_insatntons}
\end{figure}

The constant $A$ is defined as a parameter in asymptotic form of the instanton solution. Let $\bar x(\tau)=az(\tau)$ be a one--instanton trajectory going from $-a$ to $0$. Then $z(\tau)$ is a solution of the equation
\begin{align}\label{eq:infinite_boundary_conditions}\begin{split}
\ddot{z}(\tau)&=V'(z(\tau)),\\
z(-\infty)&=-1,\\
z(\infty)&=0.
\end{split}\end{align}
For large $\tau$ there is $\ddot{z}(\tau)=V'(z(\tau))\approx\omega^2 z(\tau)$. From this is follows that $\dot z(\tau)\approx A_+e^{-\omega \tau}$ with some constant $A_+$. For large negative $\tau$ expansion of the potential is different and $\dot z(\tau)\approx A_-e^{\tau}$. By translation in variable $\tau$ one can make these two constants $A_+$ and $A_-$ equal to
\begin{align}
A=A_+^{1/(1+\omega)}A_-^{\omega/(1+\omega)}.
\end{align}
The constant $A$ cannot by determined analytically and numerical estimation has to be performed. Problem (\ref{eq:infinite_boundary_conditions}) cannot be solved numerically because boundary conditions are set at infinity. Thus, we impose boundary conditions at finite time $T$,
\begin{align}\label{eq:finite_boundary_conditions}\begin{split}
\ddot{z}(\tau)&=-V'(z(\tau)),\\
z(-T/2)&=-1,\\
z(+T/2)&=0,
\end{split}\end{align}
with large $T$. Apart from the neighborhood of the boundaries, the solution of (\ref{eq:finite_boundary_conditions}) should behave in the same manner as solution of (\ref{eq:infinite_boundary_conditions}). We expect that functions $e^{\omega\tau}\dot{z}(\tau)$ and $e^{-\tau}\dot{z}(\tau)$ approach nonzero constant values for $t\to T/2$ and $t\to -T/2$ respectively. Numerical solutions of (\ref{eq:finite_boundary_conditions}) confirm this supposition outside vicinity of the boundaries (see Fig. \ref{fig:numerical_insatntons}). $A_\pm$ are obtained by fitting constants to plateaux visible on the plots.
The parameter $A$ also depends on $\delta$. Our computations yield
\begin{align}
A(\delta=0) &\approx 0.4284,&A(\delta=0.214)=0.9119.
\end{align}
One can see that the correction coming from nonzero $\delta$ is significant. Moreover, it accumulates with correction of $S_0$.

We now consider the amplitude $\braket{0|e^{-TH}|-a}$. A classical trajectory that contributes to this expression has to begin at $-a$ and end at $0$. It has to consist of an odd number of instantons, say $2n+1$, first of which goes from $-a$ to $0$. Then there follow $n$ two-instanton pairs. Each pair starts and ends at $0$. However, the first instanton of each pair may go through either $-a$ or $+a$. This freedom of choice gives $2^n$ such classical solutions, so $N_{2n+1}=2^n$. In formula (\ref{eq:tripple_well_determinant}) we put $\omega_1=1,\ \omega_2=\omega$ and change $n\to2n+1$. In the final formula we also include zero modes and exponent of the classical action $\exp(-nS_0)$. At last, we use the $g$ variable instead of $a$.
\begin{align}\label{eq:tripple_well_sineh}\begin{split}
\braket{0|e^{-TH}|-a}&=\sum_{n=0}^\infty 2^ne^{-(2n+1)S_0}\frac{1}{(2n+1)!}\left(\sqrt\frac{S_0}{2\pi}T\right)^{2n+1}\\
&\quad\times\frac{\omega^{1/4}}{\sqrt \pi}\left(\omega^{1/4}\sqrt\frac{2}{1+\omega}\frac{A}{\sqrt{gS_0}}\right)^{2n+1}e^{-\frac{T}{2}\frac{1+\omega}{2}}(2\omega(1+\omega))^{n/2}\sqrt{2\omega}\\
&=\left(\frac{\omega^2}{2\pi^2(1+\omega)}\right)^{1/4}e^{-\frac{T}{2}\frac{1+\omega}{2}}\sinh\left(2^{3/4}e^{-S_0}\sqrt\frac{\omega}{\pi}(1+\omega)^{-1/4}AT/\sqrt g\right)
\end{split}\end{align}
In this amplitude there are seen two energies:
\begin{align}\begin{split}
E_0&=\frac{1+\omega}{4}-2^{3/4}e^{-S_0}\sqrt\frac{\omega}{\pi}(1+\omega)^{-1/4}A/\sqrt g,\\
E_2&=\frac{1+\omega}{4}+2^{3/4}e^{-S_0}\sqrt\frac{\omega}{\pi}(1+\omega)^{-1/4}A/\sqrt g.
\end{split}\end{align}

Let us now consider $\braket{a|e^{-TH}|-a}$. Each contributing approximate classical solution consists of even number of instantons, say $2n+2$. The first instanton starts at $-a$ so it has to end at $0$ while the last instanton ends at $a$ and therefore it has to start from $0$. In between there are $n$ pairs of instantons which start and end at $0$. Again, there is a constant $2^n$ coming from the freedom of choice whether the two-instantons pairs go right or left, i.e. $N_{2n+2}=2^n$. In the formula (\ref{eq:tripple_well_determinant}) we put $\omega_1=\omega_2=1$ and change $n\to2n+2$.
\begin{align}\label{eq:tripple_well_cosh_minus_one}\begin{split}
\braket{a|e^{-TH}|-a}
&=\frac{1}{2\sqrt\pi}e^{-\frac{T}{2}}\left(\cosh\left(2^{3/4}e^{-S_0}\sqrt\frac{\omega}{\pi}(1+\omega)^{-1/4}AT/\sqrt g\right)-1\right)
\end{split}\end{align}
Now there are three energies seen:
\begin{align}\label{eq:triple_well_leftright_energies}\begin{split}
E_0&=\frac{1}{2}-2^{3/4}e^{-S_0}\sqrt\frac{\omega}{\pi}(1+\omega)^{-1/4}A/\sqrt g,\\
E_1&=\frac{1}{2},\\
E_2&=\frac{1}{2}+2^{3/4}e^{-S_0}\sqrt\frac{\omega}{\pi}(1+\omega)^{-1/4}A/\sqrt g.
\end{split}\end{align}
For the amplitude $\braket{-a|e^{-TH}|-a}$ the calculation is the same apart from the fact that there is a trivial (0--instanton) constant solution which gives $N_0=1$. Then,
\begin{align}\label{eq:tripple_well_cosh_plus_one}\begin{split}
\braket{-a|e^{-TH}|-a}
&=\frac{1}{2\sqrt\pi}e^{-\frac{T}{2}}\left(\cosh\left(2^{3/4}e^{-S_0}\sqrt\frac{\omega}{\pi}(1+\omega)^{-1/4}AT/\sqrt g\right)+1\right).
\end{split}\end{align}
It gives energies identical to (\ref{eq:triple_well_leftright_energies}). The last possibility, $\braket{0|e^{-TH}|0}$ gives again two energies, but with mean value $\frac{\omega}{2}$. As we can see, various amplitudes give different values of mean of the three energies. Nevertheless, they all yield the same energy splitting. It shows that only the nonperturbative correction is relevant in semiclassical approximation in Euclidean space while the constant term is irrelevant. Still, one can read off amplitudes of energy states at minima $\braket{q_i|E_j}$:
\begin{align}
\braket{-a|E_0}&=\frac{1}{2}\pi^{-1/4}&\braket{0|E_0}&=\left(\frac{2\omega^2}{1+\omega}\right)^{1/4}\frac{1}{\sqrt 2}\pi^{-1/4}&\braket{a|E_0}&=\frac{1}{2}\pi^{-1/4}\nonumber\\
\braket{-a|E_1}&=\frac{1}{\sqrt 2}\pi^{-1/4}&\braket{0|E_1}&=0&\braket{a|E_1}&=-\frac{1}{\sqrt 2}\pi^{-1/4}\\
\braket{-a|E_2}&=-\frac{1}{2}\pi^{-1/4}&\braket{0|E_2}&=\left(\frac{2\omega^2}{1+\omega}\right)^{1/4}\frac{1}{\sqrt 2}\pi^{-1/4}&\braket{a|E_2}&=-\frac{1}{2}\pi^{-1/4}\nonumber
\end{align}
These amplitudes agree with plots of wavefunctions of energy states presented in Fig. \ref{fig:3minima_wavefunctions_delta}.

\section{Comparison of the results}
As stated in the preceding paragraph, it is necessary to introduce a parameter $\delta$ to rise the lowest energy so that the corresponding state mixes with the two higher energy states and there is tunneling between all three minima. Value of $\delta$ was obtained numerically by finding minimum of energy splitting of the two lowest energies in symmetric sector. Numerical computations show that for these values of parameter $\delta$ there is indeed tunneling between all three minima as expected.

\begin{figure}[h!]
\centering
\includegraphics[width=.8\textwidth]{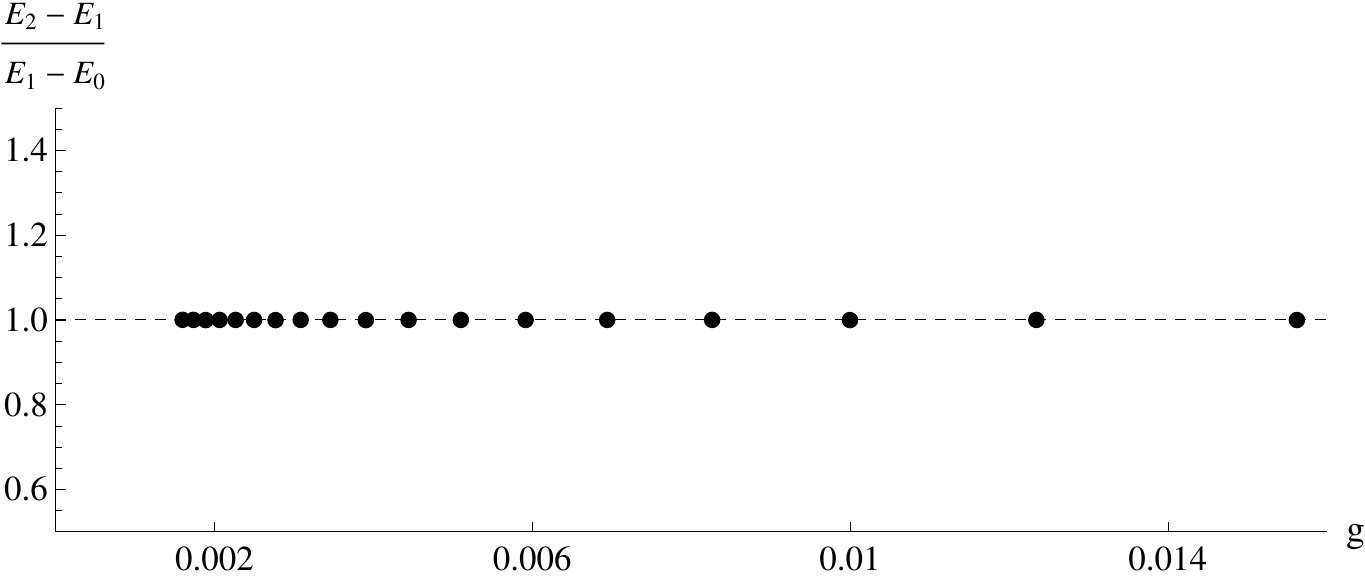}
\caption{Energy splitting of the second and third energy levels compared to the splitting of the two lowest states. As predicted by the semiclassical approximation, the splitting is the same.}\label{fig:3isitanaverage}
\end{figure}

\begin{figure}[h!]
\centering
\includegraphics[width=.8\textwidth]{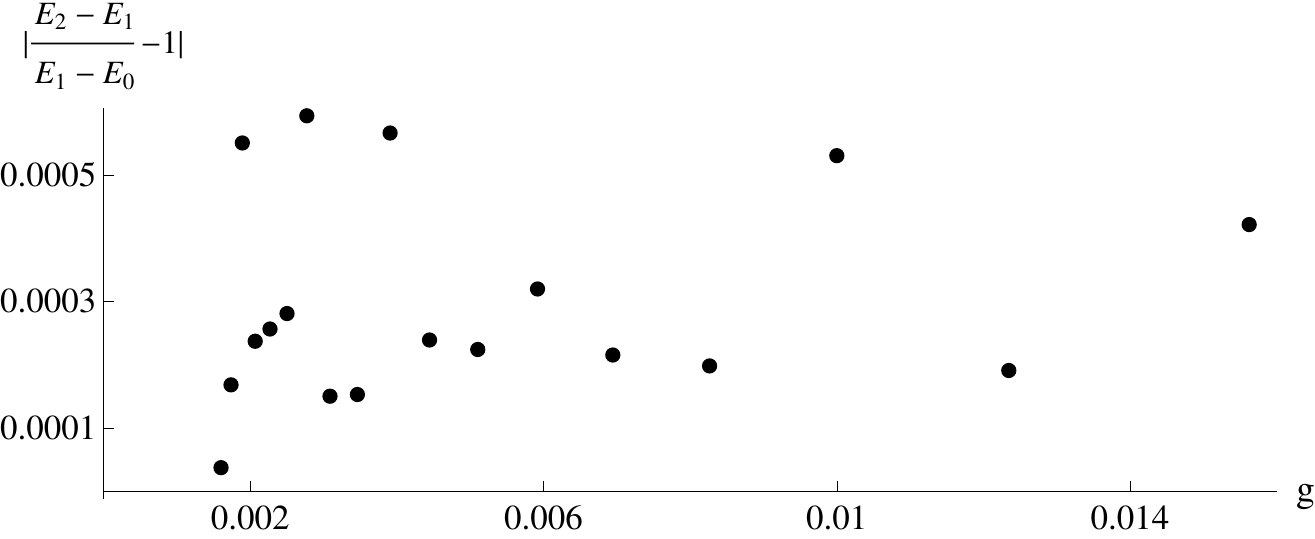}
\caption{Ratio of the two splittings in higher resolution. Deviation of $(E_2-E_1)/(E_1-E_0)$ from 1 is not bigger than $0.0006$ and does not depend on $g$. Such difference can be explained by numerical errors in computing eigenvalues and by accuracy of determining the parameter $\delta$.}\label{fig:3isitanaverage2}
\end{figure}

\begin{figure}[ht!]
\centering
\includegraphics[width=.8\textwidth]{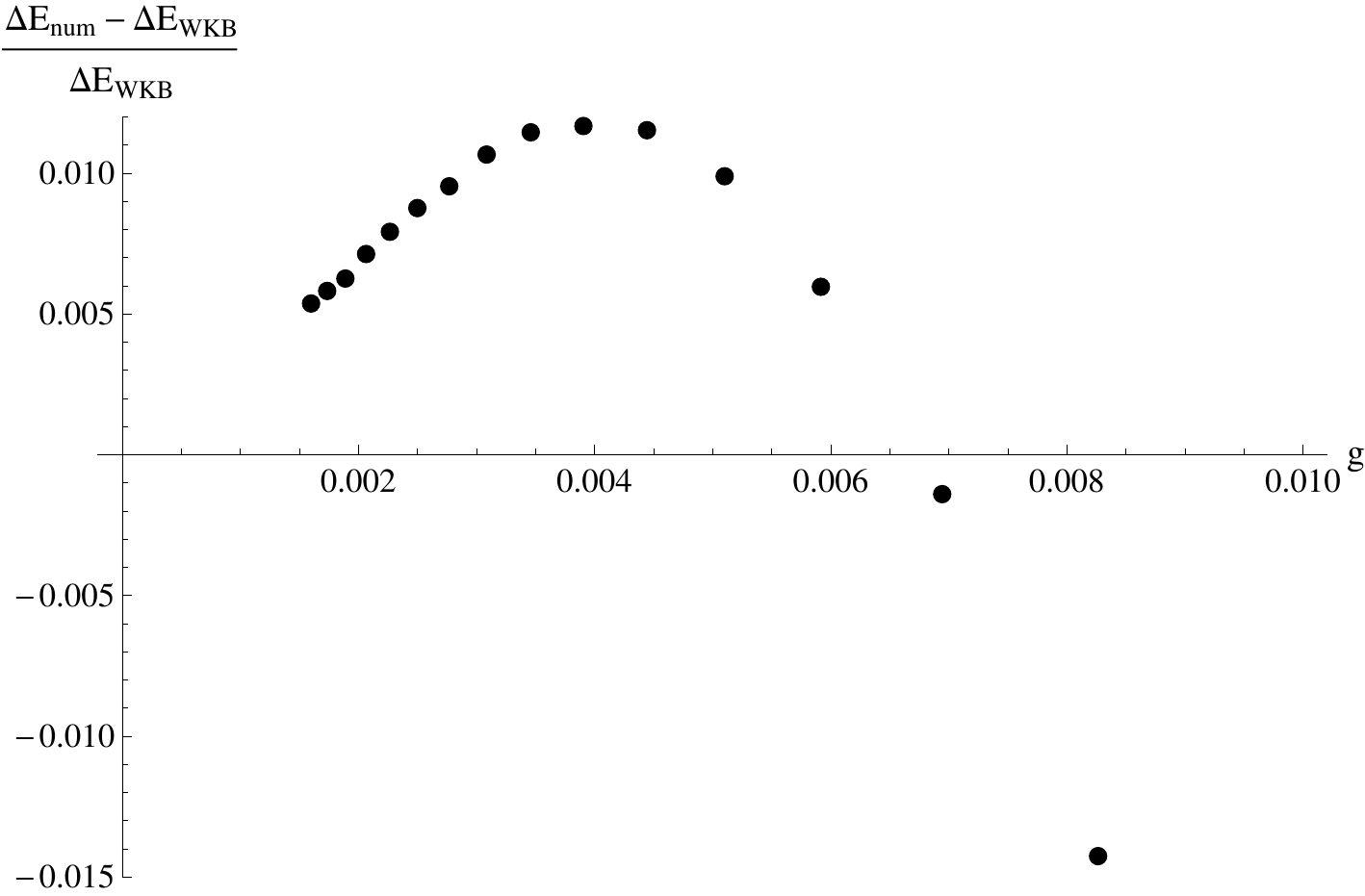}
\caption{Relative difference of energy splitting of the two lowest energy states.}\label{fig:3relative_difference}
\end{figure}

Basic result of the semiclassical approximation is that there is identical splitting between successive energies, i.e. $E_2-E_1=E_1-E_0$. Ratio of these two splittings is presented in Fig. \ref{fig:3isitanaverage}. It can be seen that the ratio is 1 with great precision. A plot in higher resolution is shown in Fig. \ref{fig:3isitanaverage2}. Random behavior of ratio of the splittings around $1$ is most likely a numerical artefact.
To this end we will be considering only splitting of the two lowest energies $\Delta E=E_1-E_0$. The most interesting quantity is the relative difference of $\Delta E$ obtained by both, numerical and semiclassical methods. Relevant plot is presented in Fig. \ref{fig:3relative_difference}. As on can see, the relative difference $(\Delta E_{num}-\Delta E_{WKB})/\Delta E_{WKB}$ decreases for $g\to0$ although the indication that if vanishes for $g=0$ is not as strong as in the cases of double well or cosine potential. This is due to difficulties in determining value of $\delta$ for small couplings $g$. Note that unlike in previous cases this plot was made in linear scale because the range of the coupling constant $g$ is limited. Definitely, the semiclassical approximation does not apply for $g>0.004$ where higher order corrections take over.

Summing up, we observed that for the potential with three minima, in which the minima are not equivalent, the naive supposition that tunneling takes place between all three minima fails. There is nonperturbative energy splitting only between the second and third energy and corresponding wavefunctions are localized in left and right minima. Therefore, there is tunneling only between side minima. Still, for such case WKB approximation can be done (which is putting $\delta=0$ in our calculations) and predicts tunneling between all three minima. It follows that the WKB approximation cannot distinguish whether there is a tunneling effect but only determine how big it is under the assumption that tunneling takes place. For nonzero parameter $\delta$, when it was confirmed numerically that there is tunneling between all three minima we checked agreement of energy splitting obtained with both, cut Fock space and instanton methods. Agreement of both approaches was confirmed in the available range of data.

\chapter{Summary}
In this thesis we were dealing with one dimensional quantum mechanical systems with multiple minima. Such systems exhibit the well know phenomenon of tunneling. Energy states are superpositions of wavefunctions localized at each minimum. Tunneling is responsible for splitting energies corresponding to different superpositions. A wavefunction which is localized in one minimum is composed of several eigenfunctions of the Hamiltonian and has nontrivial evolution which leads to tunneling into different minima after large but finite time. It is well known that in the limit of weak coupling constant the WKB approximation gives quantitative results for energy splittings. The toy model for investigating the tunneling effect is the double well potential in one dimensional quantum mechanics. Despite its simplicity it cannot be solved exactly and only approximate solutions are available.

In Yang--Mills part of QCD the gluon fields $A_\mu^a$ have minimal energy when they are pure gauge. However, pure gauge fields may have different topological properties and thus cannot be transformed continuously one into another. Therefore, each two topologically inequivalent pure gauge fields represent two different topological vacua. These vacua are in fact denumerable and are labeled by a Pontryagin index. This situation can be modeled by a periodic potential in quantum mechanics which is also considered in this thesis.

The energy splitting in a wide variety of double well potentials was extensively studied by many authors. An elegant derivation for the most classical, symmetric anharmonic potential can be found in \cite{Coleman}. In 1980 E. Bogomolny observed \cite{Bogomolny} that there are higher order corrections which come from instanton attractive interactions. After that, in 1981 J. Zinn--Justin proposed \cite{ZJ,ZJ1} a formula for multi--instanton contributions to energy. Still, it was based on observations of only few instanton interactions.
With the advent of the increase of computer speed it became possible to verify instanton calculations numerically. Because in the WKB approximation it is assumed that coupling constant is small, the first question to ask is in what range of the coupling constant $g$ it is still a good approximation. We have also clarified what is the rate of convergence of the energy splitting obtained numerically to the splitting estimated in WKB approximation. Another thing which was checked is how the semiclassical approximation works for other potentials with multiple minima. The most interesting one is a periodic potential which is the simplest possible model for vacuum in Yang--Mills theory.

In the first part of the dissertation we introduced the cut Fock space method which is essentially extension of the Tamm--Dancoff approximation. It was compared with shooting method -- a classical numerical technique for solving eigenequations. It turned out that in the case of anharmonic potential the Fock space method gives satisfactory results already for small cut--offs. Moreover, it is much faster than the shooting method and thus is a good candidate for investigating more challenging examples. For the double well potential we presented a detailed derivation of the energy splitting in the semiclassical approximation. The result was compared with numerical computations for couplings spanning from $0.00008$ to $1.6$. To this end we performed computations in very high precisions ranging up to 3600 digits. For $g=0.06$ the difference between energy splitting obtained from both methods was already smaller than $10\%$ to become $1\%$ for $g=0.006$. It was shown in \cite{ZJ} that the instanton contribution to energies is multiplied by a perturbative series in $g$. In the WKB approximation in Euclidean space one can find only zeroth term of this series. Results from cut Fock space confirm that the next correction is of order $\mathcal O(g)$.  Further coefficients may be found in calculations based on Bohr--Sommerfeld quantization condition and several are given in \cite{ZJ2}. Fit to numerical data confirms there analytical result.

The next step was to perform analogical analysis for periodic potential. The instanton calculus was done for the cosine potential on $\mathbb R$ and in periodic space where the potential had 2 or 3 minima. In the two latter cases there are two and three states respectively with corresponding energies split by a nonperturbative quantity. For the cosine potential in infinite space there is an energy band. Splitting of energies and width of the band obtained with Tamm--Dancoff method and in WKB approximation were compared. In all three cases the agreement is in accord with expectations. For $g=0.16$ there is $7\%$ deviation which decreases when $g$ gets smaller. For nonzero $g$ there are also perturbative corrections which multiply the nonperturbative contributions coming from instantons. We have also shown that the wavefunctions have a Bloch form.

Finally, the anharmonic triple well potential was considered. The exact numerical result exhibited no tunneling from side minima to the central one, as one would expect, but only between left and right minimum. This is because a wavefunction located in the middle minimum has an overlap with only one energy state and thus its evolution is trivial. It was shown numerically that the potential may be perturbed in such way that there is tunneling between all three minima. However, magnitude of the perturbation has to be computed numerically for each $g$ separately which is very time consuming. We performed instanton calculations for the perturbed potential obtaining size of energy splitting and compared with numerical data. $1\%$ agreement was established for $g=0.003$ and becomes better for decreasing $g$.

As we have shown, the cut Fock space allowed us to investigate energies with great precisions in a wide range of coupling constant $g$. It fully confirms results  of WKB approximation. The range of applicability of the latter was quantitatively determined. Still one has to remember that there are some nongeneric examples with nonequivalent minima to which the WKB approximation cannot be directly applied.

The next step in this research would be to investigate next order instanton contributions, i.e. involving instanton interactions. It is known \cite{Bogomolny} that including interaction of instantons has an imaginary ambiguity which is of order $\exp(-2S_0)$ where $S_0$ is action of a single instanton. On the other hand one may calculate perturbation expansion of the ground energy with the Rayleigh--Schr\"odinger theory. The series obtained with this method is asymptotic and Borel sum of this series again has ambiguity of order $\exp(-2S_0)$. It would particularly interesting to check that these imaginary ambiguities cancel as stated by M. \"Unsal in \cite{Unsal}. Preliminary results are already obtained and are to be published \cite{tobepublished}.

\appendix

\chapter{Instanton calculus}\label{ch:instanton_calculus}
\section{Double well potential}\label{sec:double_well_appendix}
We consider first a simple case when the potential $V(x)$ has minima at positions $x=\pm 1$ with equal masses $V''(\pm 1)=1$. Recall that the Hamiltonian is
\begin{align}
H=\frac{1}{2}P^2+a^2V(X/a).
\end{align}
Normalized eigenvectors of the Hamiltonian satisfy orthogonality and completeness relations
\begin{align}
\braket{E_n|E_m}&=\delta_{nm},\\
\sum_n\ket{E_n}\bra{E_n}&=I,
\end{align}
where $\ket{E_n}$ is a bound state of the Hamiltonian $H$ corresponding to energy $E_n$. The potential is positive, so all energies are positive as well. They can be numerated in such way that $E_i<E_{i+1}$.
The key quantity in the instanton calculus is the scalar product
\begin{align}\label{eq:scalar_prod_large_T}
\braket{a|e^{-TH}|-a}
\end{align}
in large $T$ limit. The operator inside the bra-ket is called an evolution operator in Euclidean time.
\begin{align}\label{eq:formula_to_compare_with}\begin{split}
\braket{a|e^{-TH}|-a}&=\sum_{n,m}\braket{a|E_n}\braket{E_n|e^{-TH}|E_m}\braket{E_m|-a}=
\sum_{n}\braket{a|E_n}e^{-TE_n}\braket{E_n|-a}\\
&\approx e^{-TE_0}\braket{a|E_0}\braket{E_0|-a}+e^{-TE_1}\braket{a|E_1}\braket{E_1|-a}
\end{split}\end{align}
where the last approximation is true for large $T$. Once we calculate (\ref{eq:scalar_prod_large_T}) we simply read off energies $E_0$ and $E_1$.

Let $\epsilon=\frac{T}{N+1}$ where $N$ is a natural number. From now on $N$ and $T$ will be (infinitely) large while $a$ large but finite.

\begin{align}\label{eq:scalar_as_a_product}\begin{split}
\braket{a|e^{-TH}|-a}&=\braket{a|(e^{-\epsilon H})^{N+1}|-a}\\
&=\bra{a}e^{-\epsilon H}\int dx_N\ket{x_N}\bra{x_N} e^{-\epsilon H}\ldots e^{-\epsilon H}\int dx_1\ket{x_1}\bra{x_1} e^{-\epsilon H}\ket{-a}\\
&=\int dx_1\ldots dx_N\prod_{i=1}^{N+1}\braket{x_i|e^{-\epsilon H}|x_{i-1}}
\end{split}\end{align}
where $x_0=-a$ and $x_{N+1}=a$. One of amplitudes under the integral can be expressed as follows:
\begin{align}\label{eq:long_conversion}\begin{split}
\braket{x_i|e^{-\epsilon H}|x_{i-1}}&\approx
\int dp_i\braket{x_i|1-\epsilon \left(\frac{p_i^2}{2}+a^2V(x_i/a)\right)|p_i}\braket{p_i|x_{i-1}}\\&\approx
\int dp_i\braket{x_i|e^{-\epsilon\left(\frac{p_i^2}{2}+a^2V(x_i/a)\right)}|p_i}\frac{1}{2\pi}e^{ip_i(x_i-x_{i-i})}\\&=
\frac{1}{2\pi}\int dp_i e^{-\frac{\epsilon}{2}\left(p_i-i\frac{x_i-x_{i-1}}{\epsilon}\right)^2} e^{-\epsilon\left(\frac{1}{2}(\frac{x_i-x_{i-1}}{\epsilon})^2+a^2V(x_i/a)\right)}\\&=
\frac{1}{\sqrt{2\pi\epsilon}}e^{-\epsilon\left(\frac{1}{2}(\frac{x_i-x_{i-1}}{\epsilon})^2+a^2V(x_i/a)\right)}.
\end{split}\end{align}
By substituting (\ref{eq:long_conversion}) to (\ref{eq:scalar_as_a_product}) one gets
\begin{align}
\braket{a|e^{-TH}|-a}&\approx\left(\frac{1}{2\pi\epsilon}\right)^{N/2}\int dx_1\ldots dx_N e^{-\epsilon\sum_{i=1}^{N+1}\left(\frac{1}{2}(\frac{x_i-x_{i-1}}{\epsilon})^2+a^2V(x_i/a)\right)}
\end{align}
Suppose that a function $x_N(t)= x_{i}$ with $t=\frac{iT}{N}-\frac{T}{2}$ is differentiable in the limit $N\to\infty$. It satisfies boundary conditions $x(-T/2)=-a$ and $x(T/2)=a$. Then
\begin{align}\label{eq:euclidean_action_limit}
\epsilon\sum_{i=1}^{N+1}\left(\frac{1}{2}(\frac{x_i-x_{i-1}}{\epsilon})^2+a^2V(x_i/a)\right)\rightarrow S_E[x(\tau)] \equiv \int_{-T/2}^{T/2}d\tau\left(\frac{1}{2}\dot x(\tau)^2+a^2V(x(\tau)/a)\right).
\end{align}
The functional $S_E$ is called Euclidean action. In the $N\to\infty$ limit (\ref{eq:scalar_prod_large_T}) becomes
\begin{align}\label{eq:N_inf_limit}
\braket{a|e^{-TH}|-a}=\mathcal N'\int\mathcal D[x(\tau)]e^{-S_E[x(\tau)]}.
\end{align}
$\mathcal N'$ is an ill-defined normalization constant and we will eliminate it later. Since $x(\tau)$ is usually not even continuous, (\ref{eq:N_inf_limit}) is only a formal expression. On the other hand, only those paths for which left hand side of (\ref{eq:euclidean_action_limit}) is small contribute to the integral and this is true when $x(\tau)$ is close to a differentiable function. It justifies using the expression (\ref{eq:N_inf_limit}).

\subsection{Saddle point approximation}
We will calculate the right hand side of (\ref{eq:N_inf_limit}) in the saddle point approximation. The functional derivative of the Euclidean action is
\begin{align}\label{eq:functional_derivative}
\frac{\delta S_E[x(\tau)]}{\delta x(\tau')}=-\ddot x(\tau')+aV'(x(\tau')/a)
\end{align}
The root of (\ref{eq:functional_derivative}) is called classical solution of the Euclidean action because it satisfies classical equations of motion and is denoted by $\bar x(\tau)$. To be more explicit, we write
\begin{align}\label{eq:critical_point}\begin{split}
-\ddot{\bar x}(\tau)+aV'(\bar x(\tau)/a)&=0,\\
\bar x(-T/2)&=-a,\\
\bar x(T/2)&=a.
\end{split}\end{align}
Expanding $S_E$ around $\bar x(\tau)$ gives
\begin{align}\label{eq:euclid_action_expansion}
S_E[x(\tau)]=S_E[\bar x(\tau)]+\frac{1}{2}\int d\tau' d\tau''\left.\frac{\delta^2S_E[x(\tau)]}{\delta x(\tau')\delta x(\tau'')}\right|_{\delta x(\tau)=0}\delta x(\tau')\delta x(\tau'')+\ldots
\end{align}
where $\delta x(\tau)=x(\tau)-\bar x(\tau)$. Second derivative of the action is
\begin{align}\label{eq:functional_second_derivative}
\left.\frac{\delta^2S_E[x(\tau)]}{\delta x(\tau')\delta x(\tau'')}\right|_{\delta x(\tau)=0}=\delta(\tau'-\tau'')\left(-\frac{d^2}{d\tau'^2}+V''(\bar x(\tau')/a)\right).
\end{align}
Note that the third derivative is
\begin{align}
\left.\frac{\delta^3S_E[x(\tau)]}{\delta x(\tau')\delta x(\tau'')\delta x(\tau''')}\right|_{\delta x(\tau)=0}=a^{-1}\delta(\tau'-\tau'')\delta(\tau'-\tau''')V^{(3)}(\bar x(\tau')/a)=\mathcal O(a^{-1}).
\end{align}
Higher derivatives contain higher powers of $a^{-1}$. In the semiclassical approximation all derivatives beginning from the third are omitted.
Then,
\begin{align}
\braket{a|e^{-TH}|-a}=\mathcal N'e^{-S_E[\bar x(\tau)]}\int\mathcal D[x(\tau)]e^{-\frac{1}{2}\int d\tau\delta x(\tau)\left(-\frac{d^2}{d\tau^2}+V''(\bar x(\tau)/a)\right)\delta x(\tau)}.
\end{align}
The equation
\begin{align}\label{eq:operator}
\left(-\frac{d^2}{d\tau^2}+V''(\bar x(\tau)/a)\right)x(\tau)=\lambda x(\tau)
\end{align}
together with boundary conditions $x(-T/2)=x(T/2)=0$ is a Sturm-Liouville problem and has solutions only for a discrete set $\{\lambda_n\}_{n=0}^\infty$ of the parameter $\lambda$ with $\lambda_i<\lambda_{i+1}$ and $\lim_{n\to\infty}\lambda_n=\infty$. Eigenfunctions $x_n(\tau)$ satisfy completeness and orthogonality relations
\begin{align}\label{eq:orthogonality}
\int_{-T/2}^{T/2}d\tau x_n(\tau) x_m(\tau)&=\delta_{nm},\\
\sum_{n=0}^\infty x_n(\tau)x_n(\tau')&=\delta(\tau-\tau').
\end{align}
Function $\delta x(\tau)$ can be represented in the basis $\{x_n\}$, namely
\begin{align}\label{eq:delta_x_expansion}
\delta x(\tau)=\sum_{n=0}^\infty c_n x_n(\tau).
\end{align}
Finally, inserting (\ref{eq:functional_second_derivative}) and (\ref{eq:delta_x_expansion}) into (\ref{eq:euclid_action_expansion}) yields
\begin{align}\begin{split}
S_E[x(\tau)]&=S_E[\bar x(\tau)]\\&\quad+\frac{1}{2}\sum_{n,m=0}^\infty c_nc_m\int d\tau' d\tau''\delta(\tau'-\tau'')\left(-\frac{d^2}{d\tau'^2}+V''(\bar x(\tau')/a)\right)x_n(\tau')x_m(\tau'')+\mathcal O(a^{-1})\\
&=S_E[\bar x(\tau)]+\frac{1}{2}\sum_{n,m=0}^\infty\lambda_n c_n^2+\mathcal O(a^{-1}).
\end{split}\end{align}
The expression (\ref{eq:N_inf_limit}) can be now approximated by
\begin{align}\label{eq:definition_of_determinant}\begin{split}
\braket{a|e^{-TH}|-a}&=e^{-S_E[\bar x(\tau)]}\mathcal N' \int \mathcal D[x(\tau)]e^{-\frac{1}{2}\sum_{n=0}^\infty \lambda_n c_n^2}\\
&=e^{-S_E[\bar x(\tau)]}\mathcal N \prod_{n=0}^\infty\int \frac{d c_n}{\sqrt{2\pi}}e^{-\frac{1}{2}\lambda_n c_n^2}\\
&=e^{-S_E[\bar x(\tau)]}\mathcal N\prod_{n=0}^\infty\frac{1}{\sqrt {\lambda_n}}\\
&=e^{-S_E[\bar x(\tau)]}\mathcal N\det{}^{-\frac{1}{2}}\left[-\frac{d^2}{d\tau^2}+V''(\bar x(\tau)/a)\right].
\end{split}\end{align}
In the first step the Wiener measure $\mathcal D[x(\tau)]$ was replaced by a product $\prod\frac{dc_n}{\sqrt{2\pi}}$. Additional constant coming from jacobian and coefficient $1/\sqrt{2\pi}$ implies modification of the overall normalization factor $\mathcal N'\to\mathcal N$. The determinant is a notation standing for the product of all eigenvalues of an operator.

\subsection{Action of the classical solution}
We will now calculate action of the classical solution. From (\ref{eq:critical_point}) we have
\begin{align}
\frac{1}{2}\frac{d}{d\tau}(\dot{\bar x}(\tau))^2=\ddot{\bar x}(\tau)\dot{\bar x}(\tau)=\dot{\bar x}(\tau) aV'(\bar x(\tau)/a)=\frac{d}{d\tau}a^2V(\bar x(\tau)/a).
\end{align}
Integrating both sides one obtains
\begin{align}\label{eq:critical_constant}
\frac{1}{2}(\dot{\bar x}(\tau))^2=a^2V(\bar x(\tau)/a)+c,
\end{align}
where $c$ is a constant. Equation (\ref{eq:critical_constant}) transforms to
\begin{align}\label{eq:eq_for_critical_point}
d\tau=\frac{d\bar x}{\sqrt{2a^2V(\bar x/a)+2c}}
\end{align}
Boundary conditions $\bar x(\pm T/2)=\pm a$ determine the constant $c$ which turns out to be positive. Since boundary conditions and the potential are symmetric, there is $\bar x(0)=0$. Hence,
\begin{align}
\frac{T}{2}=\int_0^{T/2}d\tau=\int_0^a \frac{d x}{\sqrt{2a^2V(x/a)+2c}}.
\end{align}
The integral on the right hand side may be divided into two parts:
\begin{align}
\int_0^a dx\left[2a^2V(x/a)+2c\right]^{-1/2}=\int_0^a dx\left[(x-a)^2+2c\right]^{-1/2}+\int_0^a dx r(x)
\end{align}
where $r(x)$ is such that expressions under integrals agree. It is simple to show that $r(x)<b a^{-1}$ for $x\in(0,a)$ where $b$ is some constant depending only on the shape of $V(x)$. The first integral is calculable:
\begin{align}
\frac{T}{2}<\log\left(\frac{\sqrt{2c}}{-a+\sqrt{2c+a^2}}\right)+b.
\end{align}
After simple transformations one arrives to the bound
\begin{align}
0<c<4a^2e^{2b}e^{-T}
\end{align}
for $T$ large enough. Using (\ref{eq:critical_constant}) again,
\begin{align}
S_E[\bar x(\tau)]&=\int_{-T/2}^{T/2}d\tau\left(\frac{1}{2}\dot{\bar x}(\tau)^2+a^2V(\bar x(\tau)/a)\right)\\&=
\int_{-T/2}^{T/2}d\tau\left(\dot{\bar x}(\tau)^2-c\right)\\&=
\int_{-T/2}^{T/2}d\tau\left(\dot{\bar x}(\tau)\sqrt{2a^2V(\bar x(\tau)/a)+2c}\right)-Tc\\&=
\int_{-a}^adx\sqrt{2a^2V(x/a)+2c}-Tc.
\end{align}
For large $T$ the expression $Tc$ is negligible, so Euclidean action of the classical solution may be approximated as follows:
\begin{align}
S_E[\bar x(\tau)] = \int_{-a}^adx\sqrt{2a^2V(x/a)}=a^2\int_{-1}^1dx\sqrt{2V(x)}\equiv S_0.
\end{align}

\subsection{Zero mode}
For large $\tau$ the classical solution $\bar x(\tau)$ is close to $a$ and for $x\approx a$ the potential is approximately $a^2V(x/a)\approx\frac{1}{2}(x-a)^2$. Then,
\begin{align}
\dot{\bar x}=\sqrt{2a^2V(\bar x(\tau)/a)+2c}\approx a-\bar x(\tau).
\end{align}
Solution of this equation yields
\begin{align}\label{eq:asymptotic_behavior_of_zero_mode}
\bar x(\tau)\approx a-Ce^{-\tau}.
\end{align}
That means that for $\tau\gg1$ the solution is very close to the stationary point $a$ of the potential and contribution to the action is small. The same is true for $\tau\ll-1$. Thus we can say that the classical solution jumps from one stationary point to another in time of order $1$. For this reason it is called an instanton. Major contribution to the action is from the region where $|\tau|$ is of the order of $1$ or less. It follows that a trajectory $x_{\tau_1}(\tau)$ satisfying
\begin{align}
x_{\tau_1}(\tau)&=\bar x(\tau-\tau_1)&\text{ for }\frac{T}{2}-|\tau_2|\gg1
\end{align}
and slightly modified near the boundaries (so that is satisfies boundary conditions) gives almost the same value of Euclidean action as $\bar x(\tau)$, i.e. $S_E[x_{\tau_1}(\tau)]\approx S_0$. It means that the lowest eigenvalue $\lambda_0$ of the operator (\ref{eq:operator}) which corresponds to translation of the classical solution tends to zero as $T\to\infty$. This results in a divergence
\begin{align}\label{eq:infinite_integral}
\int\frac{d c_0}{\sqrt{2\pi}}e^{-\frac{1}{2} 0\cdot c_0^2}=\infty.
\end{align}
For infinite $T$ the eigenfunction corresponding to the zero mode $\lambda_0=0$ is $x_0(\tau)=\alpha\frac{d}{d\tau}\bar x(\tau)$ where $\alpha$ is a normalization constant. Indeed,
\begin{align}\label{eq:zero_mode_equation}
\left(-\frac{d^2}{d\tau^2}+V''(\bar x(\tau)/a)\right)\frac{d}{d\tau}\bar x(\tau)&=\frac{d}{d\tau}\Big(-\ddot{\bar x}(\tau)+aV'(\bar x(\tau)/a)\Big)=0
\end{align}
The normalization constant can be obtained from the orthogonality relation (\ref{eq:orthogonality}):
\begin{align}\label{eq:normalization_of_zero_mode}
1=\int_{-T/2}^{T/2}d\tau x_0(\tau)^2=\alpha^2\int_{-T/2}^{T/2}d\tau \dot{\bar x}(\tau)^2=\alpha^2S_0.
\end{align}
Therefore, $\alpha=S_0^{-1/2}$. If $x(\tau)=x_{\tau_1}(\tau)$ is a shifted instanton then
\begin{align}
\delta x(t)=\bar x(\tau-\tau_1)-\bar x(\tau)\approx\tau_1\frac{d}{d\tau}\bar x(\tau).
\end{align}
On the other hand $\delta x(t)=c_0x_0(t)$. Thus,
\begin{align}
c_0=\sqrt{S_0}\tau_1.
\end{align}
Obviously $\tau_1\in(-T/2,T/2)$ and for finite $T$ the integral (\ref{eq:infinite_integral}) is finite since the integration limits are finite:
\begin{align}
\int\frac{d c_0}{\sqrt{2\pi}}=\int_{-T/2}^{T/2}\sqrt{\frac{S_0}{2\pi}}d\tau_1=\sqrt{\frac{S_0}{2\pi}}T.
\end{align}
We pull the zero eigenvalue outside of the determinant defined in (\ref{eq:definition_of_determinant}) and arrive at
\begin{align}
\braket{a|e^{-TH}|-a}&=
e^{-S_0}\sqrt{\frac{S_0}{2\pi}}T\mathcal N \left(\det{}'\left[-\frac{d^2}{d\tau^2}+V''(\bar x(\tau)/a)\right]\right)^{-1/2}.
\end{align}
The symbol $\det{}'$ stands for the product of all eigenvalues apart from the lowest $\lambda_0$.
\subsection{Multi--instanton classical solutions}
There are other approximate classical trajectories that one has to consider. They consist of several instantons glued together so that the resulting function jumps from one minimum of the potential to another $2n+1$ times. The number of glued instantons must be odd so that is starts at $-a$ and ends at $a$. Let us denote such classical solution by $\bar x_{2n+1}(\tau)$. Contribution of each instanton to the action is approximately $S_0$, so $S_E[\bar x_{2n+1}(\tau)]\approx(2n+1)S_0$. Let us denote by $\tau_1,\tau_2,\ldots,\tau_{2n+1}$ times at which $\bar x_{2n+1}(\tau)$ passes zero. We call them positions of instantons.

The gaussian approximation around $\bar x_{2n+1}(\tau)$ yields
\begin{align}
\braket{a|e^{-TH}|-a}_{2n+1}= e^{-(2n+1)S_0}\mathcal N\det{}^{-\frac{1}{2}}\left[-\frac{d^2}{d\tau^2}+V''(\bar x_{2n+1}(\tau)/a)\right].
\end{align}
in analogy to (\ref{eq:definition_of_determinant}). The operator $-\frac{d^2}{d\tau^2}+V''(\bar x_{2n+1}(\tau)/a)$ has now $2n+1$ eigenvalues which are approximately equal $0$. They correspond to freedom of choosing positions of the instantons. We shall treat them in similar way as previously. They are two constraints on $\tau_i$: their order cannot be changed, i.e. $\tau_i<\tau_{i+1}$ and they have to be separated by a distance at least $1$ (which is size of a single instanton).
\begin{align}
\int\frac{d c_0}{\sqrt{2\pi}}\ldots\frac{d c_{2n}}{\sqrt{2\pi}}&=
\left(\sqrt{\frac{S_0}{2\pi}}\right)^{2n+1}\int_{-T/2}^{T/2}d\tau_1\int_{\tau_1+1}^{T/2}d\tau_2\ldots\int_{\tau_{2n}+1}^{T/2}d\tau_{2n+1}\\
&=\left(\sqrt{\frac{S_0}{2\pi}}\right)^{2n+1}\frac{T^{2n+1}}{(2n+1)!}\left(1+\mathcal O\left(T^{-1}\right)\right)
\end{align}
Finally,
\begin{align}
\braket{a|e^{-TH}|-a}_{2n+1}&=
\frac{1}{(2n+1)!} \left(e^{-S_0}\sqrt{\frac{S_0}{2\pi}}T\right)^{2n+1}\mathcal N\left(\det{}'\left[-\frac{d^2}{d\tau^2}+V''(\bar x_{2n+1}(\tau)/a)\right]\right)^{-1/2}.
\end{align}
Here $\det{}'$ stands for the product of all eigenvalues apart from the $2n+1$ lowest ones.

\subsection{Determinant (part 1)}
We will now pass to calculating the determinant. To do this we recall the formula it originated from.
\begin{align}\label{eq:det_as_path_integral}\begin{split}
\mathcal N \left(\det\left[-\frac{d^2}{d\tau^2}+V''(\bar x(\tau)/a)\right]\right)^{-1/2}
&=\mathcal N'\int D[x(\tau)]e^{-\int d\tau\left(\frac{1}{2}(\dot x(\tau))^2+\frac{1}{2}V''(\bar x(\tau)/a)x(\tau)^2\right)}
\end{split}\end{align}
$x(\tau)$ stands for what was earlier denoted by $\delta x(\tau)$ and obeys boundary conditions $x(\pm T/2)=0$. The path integral on the right hand side of (\ref{eq:det_as_path_integral}) is defined as a limit of multiple integral at time slices which can be transformed to scalar products of the form $\braket{x_i|\ldots|x_{i-1}}$ conversely to what was done in (\ref{eq:long_conversion}).
\begin{align}\label{eq:instanton_transformation}\begin{split}
\mathcal N'\int D[x(\tau)]e^{-\int d\tau\left(\frac{1}{2}(\dot x(\tau))^2+\frac{1}{2}V''(\bar x(\tau)/a)x(\tau)^2\right)}
&=\mathcal N'\int dx_1\ldots x_N\prod\braket{x_i|e^{-\int d\tau\left(\frac{1}{2}P^2+\frac{1}{2}V''(\bar x(\tau)/a)X^2\right)}|x_{i-1}}\\
&=\braket{x=0|\hat T\left(e^{-\int d\tau\left(\frac{1}{2}P^2+\frac{1}{2}V''(\bar x(\tau)/a)X^2\right)}\right)|x=0}
\end{split}\end{align}
where $\hat T$ is time ordering operator. The time ordering appears because the last expression was defined a product of operators acting between time slices. The Euclidean evolution operator is defined as
\begin{align}
U\left(\tau_2,\tau_1\right)=\hat T\left(e^{-\int_{\tau_2}^{\tau_1} d\tau\left(\frac{1}{2}P^2+\frac{1}{2}V''(\bar x(\tau)/a)X^2\right)}\right).
\end{align}
Due to the time ordering it satisfies usual properties of an evolution operator
\begin{align}\begin{split}
U(\tau,\tau)&= 1,\\
U(\tau_2,\tau_1)&=U(\tau_2,\tau_3)U(\tau_3,\tau_1).
\end{split}\end{align}
As it was shown before, $V''(\bar x(\tau))\approx 1$ unless $\tau\in(-1,1)$. It indicates that $U(\tau_2,\tau_1)$ may be approximated by the Euclidean evolution operator of the harmonic oscillator
\begin{align}\begin{split}
U_0\left(\tau_2,\tau_1\right)&=\hat T\left(e^{-\int_{\tau_1}^{\tau_2} d\tau\left(\frac{1}{2}P^2+\frac{1}{2}X^2\right)}\right)
=\hat T\left(e^{-(\tau_2-\tau_1)\left(\frac{1}{2}P^2+\frac{1}{2}X^2\right)}\right)
\end{split}\end{align}
Energy states $\{\ket{E_n^0}\}$ of harmonic oscillator are well known and will be used in what follows.
\begin{align}\label{eq:appendix_long_transformation}\begin{split}
U\left(\frac{T}{2},-\frac{T}{2}\right)&=U\left(\frac{T}{2},1\right)U\left(1,-1\right)U\left(-1,-\frac{T}{2}\right)\\
&\approx U_0\left(\frac{T}{2},1\right)U\left(1,-1\right)U_0\left(-1,-\frac{T}{2}\right)\\
&=\sum_{n,m}e^{-(T/2-1)E_n^0}\ket{E_n^0}\bra{E_n^0}U\left(1,-1\right)\ket{E_m^0}\bra{E_m^0}e^{-(T/2-1)E_m^0}\\
&\approx e^{-(T/2-1)E_0^0}\ket{E_0^0}\bra{E_0^0}U\left(1,-1\right)\ket{E_0^0}\bra{E_0^0}e^{-(T/2-1)E_0^0}\\
&=e^{-(T/2-1)E_0^0}\ket{E_0^0}\bra{E_0^0}U_0\left(1,-1\right)\ket{E_0^0}\bra{E_0^0}e^{-(T/2-1)E_0^0}\frac{\bra{E_0^0}U\left(1,-1\right)\ket{E_0^0}}{\bra{E_0^0}U_0\left(1,-1\right)\ket{E_0^0}}\\
&\approx \sum_{n,m} e^{-(T/2-1)E_n^0}\ket{E_n^0}\bra{E_n^0}U_0\left(1,-1\right)\ket{E_m^0}\bra{E_m^0}e^{-(T/2-1)E_m^0}\times\\
&\quad\times\frac{\bra{E_0^0}U\left(1,-1\right)\ket{E_0^0}}{\bra{E_0^0}U_0\left(1,-1\right)\ket{E_0^0}}\\
&=U_0\left(\frac{T}{2},-\frac{T}{2}\right)\frac{\bra{E_0^0}U\left(1,-1\right)\ket{E_0^0}}{\bra{E_0^0}U_0\left(1,-1\right)\ket{E_0^0}}\\
&\equiv U_0\left(\frac{T}{2},-\frac{T}{2}\right)\kappa
\end{split}\end{align}
If we performed the same analysis for the classical trajectory $\bar x_{2n+1}(\tau)$ there would be $2n+1$ intervals of length $2$ at which $V''(\bar x_{2n+1}(\tau))\not\approx1$. This results in $2n+1$ coefficients $\kappa$. Therefore,
\begin{align}\label{eq:def_of_kappa}
\mathcal N \left(\det{}'\left[-\frac{d^2}{d\tau^2}+V''(\bar x_{2n+1}(\tau)/a)\right]\right)^{-1/2}=
\mathcal N \left(\det\left[-\frac{d^2}{d\tau^2}+1\right]\right)^{-1/2}\left(\kappa\sqrt{\lambda_0}\right)^{2n+1}
\end{align}
One may observe that $\mathcal N$ standing in (\ref{eq:definition_of_determinant}) is independent of the structure of the potential or choice of the points $a$ and $-a$ standing on the left hand side of that equation. It is then justified to write
\begin{align}
\mathcal N \left(\det\left[-\frac{d^2}{d\tau^2}+1\right]\right)^{-1/2}=\braket{x=0|e^{-TH_0}|x=0},
\end{align}
where $H_0=\frac{1}{2}P^2+\frac{1}{2}X^2$ is the Hamiltonian of harmonic oscillator. Its ground energy and bound state are
\begin{align}\begin{split}
E_0^0&=\frac{1}{2},\\
\braket{x|E_0^0}&=\pi^{-1/4}e^{-\frac{1}{2}x^2}.
\end{split}\end{align}
Therefore,
\begin{align}\label{eq:simple_determinant}\begin{split}
\mathcal N \left(\det\left[-\frac{d^2}{d\tau^2}+1\right]\right)^{-1/2}&=\braket{x=0|e^{-TH_0}|x=0}=e^{-TE_0^0}\braket{x=0|E_0^0}\braket{E_0^0|x=0}+\ldots\\
&\approx e^{-\frac{T}{2}}\frac{1}{\sqrt\pi}.
\end{split}\end{align}
\subsection{Summing over instantons}
After summing all contributions one obtains
\begin{align}\label{eq:scalar_prod_summed_up}
\braket{a|e^{-TH}|-a}=e^{-\frac{T}{2}}\frac{1}{\sqrt\pi}\sum_{n=0}^N\frac{1}{(2n+1)!}\left(e^{-S_0}\sqrt{\frac{S_0}{2\pi}}\kappa\sqrt{\lambda_0}T\right)^{2n+1}.
\end{align}
There is a question of the range of summation $N$. Since single instantons in the trajectory $\bar x_{2n+1}(\tau)$ have to be well separated, it implies a bound for $N$ which is $T/(2N+1)\gg1$. On the other hand, contributions from multiinstantons should be included until
\begin{align}
\frac{1}{(2N+1)!}\left(e^{-S_0}\sqrt{\frac{S_0}{2\pi}}\kappa\sqrt{\lambda_0}T\right)^{2N+1}\ll1.
\end{align}
By applying the Stirling formula we find
\begin{align}\label{eq:requirement_for_delta}
e^{-S_0}\sqrt{\frac{S_0}{2\pi}}\kappa\sqrt{\lambda_0}T\ll2N+1.
\end{align}
Joining (\ref{eq:requirement_for_delta}) with the condition $2N+1\ll T$ the number of instantons $N$ has to satisfy
\begin{align}
e^{-S_0}\sqrt{\frac{S_0}{2\pi}}\kappa\sqrt{\lambda_0}\ll\frac{2N+1}{T}\ll1.
\end{align}
It is possible to choose such $N$ since $S_0\propto a^2\to\infty$. Terms $\braket{a|e^{-TH}|-a}_{2n+1}$ with $n>N$ do not provide any information about instanton solutions. Yet they are so small that they can be added without introducing a significant error. Thus we set $N=\infty$ and
\begin{align}\label{eq:almost_final}
\braket{a|e^{-TH}|-a}=\mathcal N\left(\det\left[-\frac{d^2}{d\tau^2}+1\right]\right)^{-1/2}\sinh\left(e^{-S_0}\sqrt{\frac{S_0}{2\pi }}\kappa\sqrt{\lambda_0}T\right).
\end{align}
\subsection{Determinant (part 2)}
It is now sufficient to find the term $\kappa\sqrt{\lambda_0}$. From (\ref{eq:def_of_kappa}) it follows that
\begin{align}
\kappa\sqrt{\lambda_0}=\left(\frac{\det\left[-\frac{d^2}{d\tau^2}+1\right]}{\det{}'\left[-\frac{d^2}{d\tau^2}+V''(\bar x(\tau)/a)\right]}\right)^{1/2}.
\end{align}

Let us take a general potential $W(\tau)$. Define $\psi_\lambda$ as a solution of
\begin{align}\label{eq:def_of_psi_lambda}
L_W\psi_{\lambda}(\tau)\equiv\left(-\frac{d^2}{d\tau^2}+W(\tau)\right)\psi_{\lambda}(\tau)&=\lambda\psi_{\lambda}(\tau),&\psi_{\lambda}(-T/2)=0,\ \left.\frac{d}{d\tau}\psi_\lambda(\tau)\right|_{\tau=-T/2}=1.
\end{align}
For a discrete set $\{\lambda_n\}_{n=0}^\infty$ of the parameter $\lambda$ there is $\psi_{\lambda_n}(T/2)=0$. They are eigenvalues of the operator $L_W$. If $W(\tau)=0$ then $\lambda_n=m\left(\frac{n\pi}{T}\right)^2$. $W(\tau)$ is a bounded function on the interval $(-T/2,T/2)$ and can be regarded as a bounded operator added to the hermitian unbounded operator $-\frac{d^2}{d\tau^2}$. Thus, eigenvalues $\lambda_n$ of $L_W$ satisfy $|\lambda_n-\left(\frac{n\pi}{T}\right)^2|\leq M\equiv\max_{(-T/2,T/2)}W(\tau)$. Let us take two arbitrary potentials $W^{(1)}(\tau),\ W^{(2)}(\tau)$ and define the ratio of determinants as follows:
\begin{align}\label{eq:ratio_of_dets}
R(\lambda)\equiv\frac{\det\left[-\frac{d^2}{d\tau^2}+W_1(\tau)-\lambda\right]}{\det\left[-\frac{d^2}{d\tau^2}+W_2(\tau)-\lambda\right]}\equiv
\lim_{N\to\infty}\frac{\prod_{n=0}^{N}(\lambda_n^{(1)}-\lambda)}{\prod_{n=0}^{N}(\lambda_{n}^{(2)}-\lambda)}.
\end{align}
The limit on the right hand side exists. Indeed,
\begin{align}\begin{split}
\frac{\prod_{n=0}^{N}(\lambda_n^{(1)}-\lambda)}{\prod_{n=0}^{N}(\lambda_{n}^{(2)}-\lambda)}
=\exp\left(\sum_{n=0}^N\log\frac{\lambda_n^{(1)}-\lambda}{\lambda_{n}^{(2)}-\lambda}\right)
\approx \exp\left(\sum_{n=0}^N\frac{\lambda_n^{(1)}-\lambda_{n}^{(2)}}{\lambda_{n}^{(2)}-\lambda}\right).
\end{split}\end{align}
$|\lambda_n^{(1)}-\lambda_{n}^{(2)}|\leq M^{(1)}+M^{(2)}$ so for constant $\lambda$ the term in numerator is bounded while the denominator behaves like $n^2$. Thus, the series is convergent for $N\to\infty$. The function $R(\lambda)$ has zeros at $\lambda=\lambda_n^{(1)}$ and poles at $\lambda=\lambda_n^{(2)}$. Observe that for $|\lambda|\to\infty$ each term in the product on the right hand side of (\ref{eq:ratio_of_dets}) tends to $1$. Therefore,
\begin{align}
R(\lambda)\xrightarrow[{|\lambda|\to\infty}]{}1.
\end{align}
Let us define a function
\begin{align}
R'(\lambda)=\frac{\psi_\lambda^{(1)}(T/2)}{\psi_\lambda^{(2)}(T/2)}.
\end{align}
For $|\lambda|$ large enough the potential $W(\tau)$ in (\ref{eq:def_of_psi_lambda}) is negligible compared to $|\lambda|$ so the function $\psi_\lambda(\tau)$ does not depend on the potential significantly. It follows that $R'(\lambda)\to1$ when $|\lambda|\to\infty$.
The function $R'(\lambda)$ has zeros and pole in the same points as function $R(\lambda)$. This is merely because $\psi_\lambda^{(i)}(T/2)=0\iff\lambda=\lambda_n^{(i)}$. Then the function
\begin{align}
g(\lambda)\equiv\frac{R(\lambda)}{R'(\lambda)}
\end{align}
is an entire function. It is also bounded because it converges to $1$ for infinite $|\lambda|$. By Liouville's theorem it is constant, i.e. $R(\lambda)=R'(\lambda)$ for all $\lambda$.
\begin{align}\label{eq:instanton_cal_frac_of_dets}
\frac{\det\left[-\frac{d^2}{d\tau^2}+W_1(\tau)\right]}{\det\left[-\frac{d^2}{d\tau^2}+W_2(\tau)\right]}&=\frac{\psi_0^{(1)}(T/2)}{\psi_0^{(2)}(T/2)},
\end{align}
which implies
\begin{align}
\kappa\sqrt{\lambda_0}=\sqrt{\frac{\lambda_0\psi_0^{0}(T/2)}{\psi_0(T/2)}}.
\end{align}
Let $\psi_0^{0}(\tau)$ be solution of (\ref{eq:def_of_psi_lambda}) with $W(\tau)=1$. It is easy to show that $\psi_0^{0}(\tau)=\sinh(\tau+T/2)$. Thus,
\begin{align}
\psi_0^{0}(T/2)\approx\frac{1}{2}e^{T}.
\end{align}
The function $\psi_0(\tau)$ satisfies
\begin{align}\label{eq:sturm_liouville}
\left(-\frac{d^2}{d\tau^2}+V''(\bar x(\tau)/a)\right)\psi_0(\tau)&=0,\\
\begin{split}\label{eq:sturm_liouville_boundary_conditions}
\psi_0(- T/2)&=0,\\
\dot\psi_0(-T/2)&=1.
\end{split}\end{align}

From (\ref{eq:asymptotic_behavior_of_zero_mode}) it follows that the one--instanton solution has asymptotic behavior
\begin{align}
\dot{\bar x}(\tau)&\approx aA_\pm e^{-|\tau|}&\tau\to\pm\infty.
\end{align}
This is definition of constants $A_\pm$. From (\ref{eq:zero_mode_equation}) it follows that the function $y_1(\tau)=\frac{1}{\sqrt{S_0}}\dot{\bar x}(\tau)$ satisfies the equation (\ref{eq:sturm_liouville}) but not necessarily boundary conditions (\ref{eq:sturm_liouville_boundary_conditions}).
Asymptotic behavior of $y_1(\tau)$ is
\begin{align}
y_1(\tau)&\approx B_\pm e^{-|\tau|}&\tau\to\pm\infty
\end{align}
with $B_\pm=\frac{a}{\sqrt{S_0}}A_\pm$. Constants $B_+$ and $B_-$ can be a priori different. However, we may redefine the solution $y_1(\tau)\to y_1(\tau-\tau_1)$ so that these constants change $B_+\to B_+'=B_+e^{\tau_1}$ and $B_-\to B_-'=B_-e^{-\tau_1}$. We choose $\tau_1$ in such way that $B_+'=B_-'\equiv B$. Wronskian $\mathcal W$ of independent solutions of a linear differential equation is constant. Therefore, one can normalize the other solution $y_2(\tau)$ of equation (\ref{eq:sturm_liouville}) so that is satisfies
\begin{align}
\mathcal W\equiv y_1(\tau)\dot y_2(\tau)-\dot y_1(\tau) y_2(\tau)=2B^2
\end{align}
For large $|\tau|$ it reads
\begin{align}
\dot y_2(\tau)-y_2(\tau)&=2B e^{|\tau|},&\tau\to\pm\infty.
\end{align}
Thus, the asymptotic behavior of $y_2(\tau)$ is
\begin{align}
y_2(\tau)&=\pm B e^{|\tau|}+C_\pm e^{-|\tau|}\approx\pm B e^{|\tau|},&\tau\to\pm\infty.
\end{align}
Finally, $\psi_0(\tau)$ satisfying both (\ref{eq:sturm_liouville}) and (\ref{eq:sturm_liouville_boundary_conditions}) is
\begin{align}
\psi_0(\tau)&=\frac{1}{2 B}\left(e^{T/2} y_1(\tau)+e^{-T/2} y_2(\tau)\right),
\end{align}
implying $\psi_0(T/2)=1$.
The Green function for the differential operator $L_{V''(\bar x(\tau)/a)}$ is
\begin{align}
G(\tau,\tau')=\left\{\begin{array}{ll}
-(\mathcal W)^{-1}y_1(\tau')y_2(\tau)&\tau'<\tau\\
-(\mathcal W)^{-1}y_1(\tau)y_2(\tau')&\tau'\geq\tau
\end{array}\right.
\end{align}
Therefore, the function $\psi_{\lambda_0}$ defined by (\ref{eq:def_of_psi_lambda}) with $W(\tau)=V''(\bar x(\tau))$ fulfills the integral equation
\begin{align}\label{eq:integral_equation}
\psi_{\lambda_0}(\tau)=\psi_0(\tau)+\int_{-T/2}^{T/2}d\tau'G(\tau,\tau')\lambda_0\psi_{\lambda_0}(\tau').
\end{align}
Function $\psi_0(\tau)$ is added so that $\psi_{\lambda_0}$ satisfies initial conditions. As shown previously, $\lambda_0$ is close to $0$. It is then justified to replace $\psi_{\lambda_0}$ with $\psi_0$ on the right hand side of (\ref{eq:integral_equation}) since it introduces higher order error. Secondly, $\psi_{\lambda_0}(T/2)=0$ as mentioned earlier.
\begin{align}\begin{split}
0&\approx\psi_0(T/2)+\int_{-T/2}^{T/2}d\tau'G(T/2,\tau')\lambda_0\psi_{0}(\tau')\\
&=1-\lambda_0\frac{y_2(T/2)}{4B^3}\int_{-T/2}^{T/2}d\tau'y_1(\tau')\left(e^{T/2} y_1(\tau')+e^{-T/2} y_2(\tau')\right)\\
&\approx1-\frac{\lambda_0}{4B^2}\int_{-T/2}^{T/2}d\tau'\left(e^{T} y_1(\tau')^2+y_1(\tau') y_2(\tau')\right)\\
&\approx1-\frac{\lambda_0}{4B^2}\frac{e^{T}}{S_0}\int_{-T/2}^{T/2}d\tau'\dot{\bar x}(\tau')^2\\
&=1-\frac{\lambda_0}{4 B^2}e^{T}
\end{split}\end{align}
The term $\int d\tau' y_1(\tau') y_2(\tau')$ was omitted because it is of order $1$ and is negligible compared to $e^{T}$. The last equality is guaranteed by (\ref{eq:normalization_of_zero_mode}). We obtain
\begin{align}
\lambda_0&=4B^2e^{-T}.
\end{align}
Finally, we get
\begin{align}
\kappa\sqrt{\lambda_0}&=\sqrt{2}B=\sqrt{\frac{2}{S_0}}aA
\end{align}
where $A=\sqrt{A_+A_-}$.
\subsection{Final result}
Inserting this result and (\ref{eq:simple_determinant}) into (\ref{eq:almost_final}) one obtains
\begin{align}
\braket{a|e^{-TH}|-a}=e^{-\frac{T}{2}}\frac{1}{\sqrt\pi}\sinh\left(e^{-S_0}\frac{aA}{\sqrt \pi}T\right).
\end{align}
Comparison of this result with (\ref{eq:formula_to_compare_with}) yields
\begin{align}
E_0&=\frac{1}{2}-e^{-S_0}\frac{aA}{\sqrt\pi},\\
E_1&=\frac{1}{2}+e^{-S_0}\frac{aA}{\sqrt\pi}.
\end{align}
\section{Inequivalent minima}\label{sec:triple_well_appendix}
As demonstrated on the example of triple well anharmonic oscillator, tunneling does not take place always when we naively expect it. The triple well potential is
\begin{align}\begin{split}
V(x)&=\frac{1}{2}x^2+\left(-\frac{85}{24}+\frac{512}{27\pi^2}\right)x^4+\left(\frac{31}{4}-\frac{512}{9\pi^2}\right)x^6
+\left(-\frac{55}{8}+\frac{512}{9\pi^2}\right)x^8
+\left(\frac{13}{6}-\frac{512}{27\pi^2}\right)x^{10}.
\end{split}\end{align}
with the Hamiltonian $H=\frac{1}{2}P^2+a^2V(X/a)$
The potential $V(x)$ has equal derivatives $V''(x)=1$ at all of the three minima $x=0,\ x=-1,\ x=1$. Thus we expect that there are three energies which are close to $E=1/2$. The next supposition is that when one locates a state in one minimum, it will tunnel to other minima. This statement is not true. A state with the lowest energy is localized in the middle minimum. Then there are two states, with even and odd parity, which are localized in left and right minima. It means that there is a tunneling between left and right minimum and the middle minimum is separated. Moreover, energy of the middle minimum is perturbatively different from the two higher energies, which in contrast are split only by a nonperturbative amount. This is clear when one notices that Taylor expansion of the potential $V(x)$ about point $x=0$ is different than the expansion about $x=1$. It turns out that this perturbative inequivalence excludes tunneling.

As presented in chapter \ref{ch:TD_for_three_minima} when one perturbes the potential in the following way:
\begin{align}\begin{split}
V_\delta(x)&=\frac{1 + \delta}{2}x^2+\left(-\frac{85}{24}+\frac{512}{27\pi^2}-\frac{7\delta}{2}\right)x^4+\left(\frac{31}{4}-\frac{512}{9\pi^2}+\frac{15\delta}{2}\right)x^6\\
&\quad+\left(-\frac{55}{8}+\frac{512}{9\pi^2}-\frac{13\delta}{2}\right)x^8
+\left(\frac{13}{6}-\frac{512}{27\pi^2}+2\delta\right)x^{10}
\end{split}\end{align}
then the second derivative in the middle minimum is higher: $V''(0)=1+\delta$ so the energy of the state localized in the middle minimum grows with $\delta$. Energies of states in left and right minima grow slower because the parameter $\delta$ enters expansion of $V(x)$ about $x=\pm a$ with the term $x^3$ rather than with $x^2$. Eventually, energy of the state localized in the middle minimum has to cross energies of the states in left and right minima. It is forbidden by the Wigner non--crossing theorem. Thus, for a particular choice of $\delta$ the tunneling between middle and left and right minima takes place. We introduce notation $\omega^2=(1+\gamma)^2=1+\delta$.

In order to calculate the amplitude $\braket{0|e^{-TH}|a}$ with instanton calculus, we start with formula analogical to (\ref{eq:definition_of_determinant}):
\begin{align}\label{eq:noneq_single_instanton}
\braket{a|e^{-TH}|0}&=e^{-S_E[\bar x(\tau)]}\mathcal N\det{}^{-\frac{1}{2}}\left[-\frac{d^2}{d\tau^2}+V''(\bar x(\tau)/a)\right],
\end{align}
where $\bar x(\tau)$ is a classical trajectory which starts at $0$ and ends at $a$. The potential $V(x)$ satisfies conditions $V''(0)=\omega^2$ and $V''(1)=1$ which gives us information on asymptotic behavior of $\bar x(\tau)$ for infinite $T$. In analogy to formula (\ref{eq:asymptotic_behavior_of_zero_mode}):
\begin{align}\begin{split}
\bar x(\tau)&=a-C e^{-t},\quad\quad t\to\infty\\
\bar x(\tau)&=C' e^{\omega t},\quad\quad t\to-\infty
\end{split}\end{align}
We say that position of an instanton $\bar x(\tau)$ is $\tau=\tau_0$ if  $\bar x(\tau_0)=a/2$. Let $S_0=\lim_{T\to\infty}S_E[\bar x(\tau)]$ be approximate action of a single instanton. Due to the fact that action is symmetric under time reversal and parity transformation $x\to-x$, it is the same for instantons going between neighboring minima: $(a,0),\ (0,a)\ (0,-a),\ (-a,0)$. It yields
\begin{align}
S_0=a^2\int_0^1dx\sqrt{2V(x)}.
\end{align}

The complete amplitude $\braket{a|e^{-TH}|0}$ is a sum over many instanton solutions each of which has a form of RHS of (\ref{eq:noneq_single_instanton}). A trajectory which begins at $x=0$ and ends at $x=a$ jumps between various minima an odd number of times, so it contains and odd number of instantons.
\begin{align}
\braket{a|e^{-TH}|0}&=\sum_n N_{2n+1} e^{-S_E[\bar x_{2n+1}(\tau)]}\mathcal N\det{}^{-\frac{1}{2}}\left[-\frac{d^2}{d\tau^2}+V''(\bar x_{2n+1}(\tau)/a)\right],
\end{align}
where $\bar x_{2n+1}(\tau)$ is a classical trajectory which consists of $2n+1$ glued instantons. Positions of the instantons are $\tau_1<\tau_2<\ldots<\tau_{2n+1}$. For $\tau \in(-\infty,\tau_1)$ the trajectory is exponentially close to the middle minimum $x=0$. Then it jumps to $x=\pm a$ and for $\tau\in(\tau_{2i+1},\tau_{2i+2})$ it is close to one of side minima. From the middle minimum it goes to the left or right minimum, so for $\tau\in(\tau_{2i},\tau_{2i+1})$ we have $\bar x_{2n+1}(\tau)\approx\pm 0$. At last it goes to the right minimum so that $\bar x_{2n+1}(\tau)\approx a$ for $\tau\in(\tau_{2n+1},\infty)$. Because of the freedom of side minimum to which the instanton goes at $\tau_{2i+1}$ there is an additional coefficient $N_{2n+1}=2^n$ which is the number of possible paths. Positions of instantons correspond to zero modes of Euclidean action and they have to be integrated out. Normalization of zero modes gives an additional coefficient $(\sqrt{S_0/2\pi})^{2n+1}$.
\begin{align}\label{eq:appendix_complete_sum}\begin{split}
\braket{a|e^{-TH}|0}&=\sum_n 2^n e^{-(2n+1)S_0}\left(\sqrt\frac{S_0}{2\pi}\right)^{2n+1}\\
&\quad\times\int_{\tau_i<\tau_{i+1}}d\tau_1\ldots d\tau_{2n+1}\mathcal N\left(\det{}'\left[-\frac{d^2}{d\tau^2}+V''(\bar x_{2n+1}(\tau)/a)\right]\right)^{-1/2}.
\end{split}\end{align}
In analogy to formula (\ref{eq:instanton_transformation}) we can write
\begin{align}
N\left(\det\left[-\frac{d^2}{d\tau^2}+V''(\bar x_{2n+1}(\tau)/a)\right]\right)^{-1/2}&=\braket{x=0|U\left(\frac{T}{2},-\frac{T}{2}\right)|x=0},
\end{align}
where $U$ is the evolution operator
\begin{align}
U(\tau_2,\tau_1)=\hat T \exp\left(-\int_{\tau_1}^{\tau_2}d\tau\frac{1}{2}P^2+\frac{1}{2}V''(\bar x_{2n+1}(\tau)/a)X^2\right).
\end{align}
Remember that the zero modes were integrated out, so we will be interested in calculating the expression
\begin{align}\label{eq:appendix_going_to_evolution_op}
\mathcal N\left(\det{}'\left[-\frac{d^2}{d\tau^2}+V''(\bar x_{2n+1}(\tau)/a)\right]\right)^{-1/2}&=\braket{x=0|U\left(\frac{T}{2},-\frac{T}{2}\right)|x=0}\lambda_0^\frac{n+1}{2} {\tilde \lambda_0}^\frac{n}{2}
\end{align}
where $\lambda_0$ is the zero mode corresponding to the instanton connecting the central minimum of $V$ with left or right minimum and $\tilde \lambda_0$ to the instanton ending at the middle minimum.

For $\tau\in(\tau_{2i+1}+1,\tau_{2i+2}-1)$ the trajectory $\bar x_{2n+1}(\tau)$ is in one of the side minima of the potential and $V''(\bar x_{2n+1}(\tau)/a)\approx 1$ and for $\tau\in(\tau_{2i}+1,\tau_{2i+1}-1)$ it is in the central minimum at which $V''(\bar x_{2n+1}(\tau)/a)\approx \omega^2$. Thus, we introduce approximate evolution operators
\begin{align}\begin{split}
U_\omega(\tau_f,\tau_i)=\hat T \exp\left(-(\tau_f-\tau_i)\frac{1}{2}P^2+\frac{\omega ^2}{2}X^2\right),\\
U_1(\tau_f,\tau_i)=\hat T \exp\left(-(\tau_f-\tau_i)\frac{1}{2}P^2+\frac{1}{2}X^2\right)
\end{split}\end{align}
and observe that $U(\tau_{2i+1}+1,\tau_{2i+2}-1)\approx U_1(\tau_{2i+1}+1,\tau_{2i+2}-1)$ and $U(\tau_{2i}+1,\tau_{2i+1}-1)\approx U_\omega(\tau_{2i}+1,\tau_{2i+1}-1)$. Let us denote by $\ket{E_n^1}$ and $\ket{E_n^\omega}$ the eigenbases of harmonic oscillators with frequency $1$ and $\omega$ respectively. We do now analogical transformation to (\ref{eq:appendix_long_transformation}). In this case it is more complicated due to different harmonic frequencies in minima of the potential.
\begin{align}\label{eq:appendix_longest_transformation}\begin{split}
U\left(\frac{T}{2},-\frac{T}{2}\right)&=U\left(\frac{T}{2},\tau_{2n+1}+1\right)\ldots U\left(\tau_{2}+1,\tau_{2}-1\right)U\left(\tau_2-1,\tau_1+1\right)U\left(\tau_1+1,\tau_1-1\right)U\left(\tau_1-1,-\frac{T}{2}\right)\\
&\approx U_1\left(\frac{T}{2},\tau_{2n+1}+1\right)\ldots U\left(\tau_{2}+1,\tau_{2}-1\right)U_1\left(\tau_2-1,\tau_1+1\right)U\left(\tau_1+1,\tau_1-1\right)U_\omega\left(\tau_1-1,-\frac{T}{2}\right)\\
&=U_1\left(\frac{T}{2},\tau_{2n+1}+1\right)\ldots\sum_m\ket{E_m^\omega}\bra{E_m^\omega}U\left(\tau_2+1,\tau_2-1\right)
\sum_m\ket{E_m^1}\bra{E_m^1}U_1\left(\tau_2-1,\tau_1+1\right)\times\\
&\quad\times \sum_m\ket{E_m^1}\bra{E_m^1}U\left(\tau_1+1,\tau_1-1\right)\sum_m\ket{E_m^\omega}\bra{E_m^\omega}U_\omega\left(\tau_1-1,-\frac{T}{2}\right)\\
&\approx U_1\left(\frac{T}{2},\tau_{2n+1}+1\right)\ket{E_0^1}\ldots\ket{E_0^\omega}\bra{E_0^\omega}U\left(\tau_2+1,\tau_2-1\right)\ket{E_0^1}\bra{E_0^1}U_1(\tau_2-1,\tau_1+1)\times\\
&\quad\times \ket{E_0^1}\bra{E_0^1}U(\tau_1+1,\tau_1-1)\ket{E_0^\omega}\bra{E_0^\omega}U_\omega\left(\tau_1-1,-\frac{T}{2}\right)\\
&=U_1\left(\frac{T}{2},\tau_{2n+1}+1\right)\ket{E_0^1}\ldots\ket{E_0^\omega}\bra{E_0^\omega}U_\omega\left(\tau_2+1,\tau_2\right)U_1\left(\tau_2,\tau_2-1\right)\ket{E_0^1}\times\\
&\quad\times \bra{E_0^1}
U_1\left(\tau_2-1,\tau_1+1\right)\ket{E_0^1}\bra{E_0^1}U_1\left(\tau_1+1,\tau_1\right)U_\omega\left(\tau_1,\tau_1-1\right)\ket{E_0^\omega}\times\\
&\quad\times \bra{E_0^\omega}U_\omega\left(\tau_1-1,-\frac{T}{2}\right) \kappa(\tau_{2n+1})\kappa'(\tau_{2n})\ldots \kappa'(\tau_2)\kappa(\tau_1)\\
&=e^{-(\frac{T}{2}-\tau_{2n+1})E_0^1}\ket{E_0^1}\ldots\ket{E_0^\omega}\braket{E_0^\omega|E_0^1}e^{-(\tau_2-\tau_1)E_0^1}\braket{E_0^1|E_0^1}
\braket{E_0^1|E_0^\omega}e^{-(\tau_1+\frac{T}{2})E_0^\omega}\bra{E_0^\omega}\times\\
&\quad \kappa(\tau_{2n+1})\tilde \kappa(\tau_{2n})\ldots \tilde \kappa(\tau_2)\kappa(\tau_1),
\end{split}\end{align}
where
\begin{align}\begin{split}
\kappa(\tau_0)&=\frac{\braket{E_0^1|U(\tau_0+1,\tau_0-1)|E_0^\omega}}{\braket{E_0^1|U_1(\tau_0+1,\tau_0)U_\omega(\tau_0,\tau_0-1)|E_0^\omega}},\\
\tilde \kappa(\tau_0)&=\frac{\braket{E_0^\omega|U(\tau_0+1,\tau_0-1)|E_0^1}}{\braket{E_0^\omega|U_\omega(\tau_0+1,\tau_0)U_1(\tau_0,\tau_0-1)|E_0^1}}.
\end{split}\end{align}
In the last step we used relations
\begin{align}\begin{split}
U_1(\tau_{i+1},\tau_i)\ket{E_0^1}&=e^{-(\tau_{i+1}-\tau_i)E_0^1}\ket{E_0^1},\\
U_\omega(\tau_{i+1},\tau_i)\ket{E_0^\omega}&=e^{-(\tau_{i+1}-\tau_i)E_0^\omega}\ket{E_0^\omega}.
\end{split}\end{align}
The eigenbasis for the harmonic oscillator is known and one can easily calculate scalar products standing in the equation (\ref{eq:appendix_longest_transformation})
\begin{align}\begin{split}
\braket{x=0|E_0^1}&=\pi^{-1/4},\\
\braket{x=0|E_0^\omega}&=\left(\frac{\omega}{\pi}\right)^{1/4},\\
\braket{E_0^1|E_0^\omega}&=\omega^{1/4}\sqrt\frac{2}{1+\omega}.
\end{split}\end{align}
Let us denote by $T_\omega$ the time which the trajectory spends in minimum with $V''(x)=\omega$ and $T_1$ the time which the trajectory spends in minimum with $V''(x)=1$, i.e. left or right minimum, neglecting sizes of instantons:
\begin{align}\begin{split}
T_\omega&=\tau_{2n+1}-\tau_{2n}+\tau_{2n-1}-\tau_{2n-2}+\ldots+\tau_1-(-T/2),\\
T_1&=T/2-\tau_{2n+1}+\tau_{2n}-\tau_{2n-1}+\tau_{2n-2}+\ldots-\tau_1.
\end{split}\end{align}
Then there is $T_1+T_\omega=T$. Then,
\begin{align}\begin{split}
\braket{x=0|U\left(\frac{T}{2},-\frac{T}{2}\right)|x=0}&=\frac{\omega^{1/4}}{\sqrt\pi}e^{-T_\omega E_0^\omega}e^{-T_1 E_0^1}\left(\omega^{1/4}\sqrt\frac{2}{1+\omega}\right)^{2n+1}\prod_{i=0}^n \kappa(\tau_{2i+1})\prod_{i=1}^n \tilde \kappa(\tau_{2i})\\
&=\frac{\omega^{1/4}}{\sqrt\pi}e^{-T/2}e^{-\frac{\gamma}{2}T_\omega}\left(\omega^{1/4}\sqrt\frac{2}{1+\omega}\right)^{2n+1}\prod_{i=0}^n \kappa(\tau_{2i+1})\prod_{i=1}^n\tilde \kappa(\tau_{2i}).
\end{split}\end{align}
Recall that $\omega^2=(1+\gamma)^2$ and $E_0^\omega=\omega/2$. Terms $\kappa$ are corrections coming from the fact that a potential with varying frequency was approximated by a potential which is constant on intervals $(\tau_i,\tau_{i+1})$. They can be written as
\begin{align}
\kappa(\tau_0)=\left(\frac{\det\left[-\frac{d^2}{d\tau^2}+W(\tau)\right]}{\det\left[-\frac{d^2}{d\tau^2}+V''(\bar x_1(\tau)/a)\right]}\right)^{1/2}
\end{align}
where $\bar x_1(\tau)$ is a classical trajectory consisting of one instanton located at $\tau_0$ going from $x=0$ to $x=a$. The function $W(\tau)$ is the frequency which was used for approximation. Coefficient $\kappa$ corresponds to an instanton starting at middle minimum and ending at side minimum while $\tilde \kappa$ corresponds to minimum going from side to central minimum. For an instanton starting at the middle minimum we have
\begin{align}
W(\tau)=\left\{\begin{array}{ll}\omega&\tau<\tau_0,\\1&\tau\geq\tau_0.\end{array}\right.
\end{align}

According to (\ref{eq:instanton_cal_frac_of_dets})
\begin{align}
\kappa(\tau_0)=\sqrt\frac{\psi_0^0(T/2)}{\psi_0(T/2)}
\end{align}
where
\begin{align}\label{eq:noneq_sturm_liouville}
\left(-\frac{d^2}{d\tau^2}+V''(\bar x_1(\tau)/a)\right)\psi_0(\tau)&=0,\\
\begin{split}\label{eq:noneq_sturm_liouville_boundary_conditions}
\psi_0(- T/2)&=0,\\
\dot\psi_0(-T/2)&=1.
\end{split}\end{align}
with $V''(\bar x_1(\tau)/a)$ replaced by $W(\tau)$ for $\psi_0^0(\tau)$. It is easy to show that
\begin{align}
\psi_0^0(\tau)&=\left\{\begin{array}{ll}\frac{1}{\omega}\sinh(\omega(\tau+T/2))&\tau<\tau_0\\
\frac{1}{\omega}\sinh(\omega(\tau_0+T/2))\cosh(\tau-\tau_0)+\cosh(\omega(\tau_0+T/2))\sinh(\tau-\tau_0)&\tau>\tau_0\end{array}\right.
\end{align}
and therefore $\psi_0^0(T/2)\approx\frac{1}{4}(1+\frac{1}{\omega})e^{T+\gamma(\tau_0+T/2)}$.

We will now find value of the function $\psi_0(\tau)$ at point $\tau=T/2$. The one instanton trajectory has asymptotic behavior
\begin{align}\label{eq:definition_of_A}
\dot{\bar x}(\tau)&\approx aA_+e^{-t}&t\to\infty\\
\dot{\bar x}(\tau)&\approx aA_-e^{\omega t}&t\to-\infty
\end{align}
By translating argument of $\bar x(\tau)$ one can make the coefficients $A_+$ and $A_-$ equal: $A_+\to A_+'=A$ and $A_-\to A_-'=A$ where $A=A_+^{\omega/(1+\omega)}A_-^{1/(1+\omega)}$. One of solutions of (\ref{eq:noneq_sturm_liouville}) is $y_1(\tau)=\frac{1}{\sqrt S_0}\dot{\bar x}(\tau)$. Its asymptotic behavior is
\begin{align}
y_1(\tau)&\approx Be^{-t}&t\to\infty,\\
y_1(\tau)&\approx Be^{\omega t}&t\to-\infty
\end{align}
where $B=\frac{aA}{\sqrt S_0}$. We can normalize the other solution $y_2(\tau)$ of (\ref{eq:noneq_sturm_liouville}) so that the Wronskian satisfies relation
\begin{align}
\mathcal W=y_1\dot y_2-\dot y_1 y_2=2B^2.
\end{align}
One can read off from Wronskian the asymptotic behavior of function $y_2(\tau)$:
\begin{align}\begin{split}
y_2(\tau)&\approx Be^{t},\quad t\to\infty,\\
y_2(\tau)&\approx -\frac{1}{\omega}Be^{-\omega t},\quad t\to-\infty.
\end{split}\end{align}
A function $\psi_0(\tau)$ satisfying both, equation (\ref{eq:noneq_sturm_liouville}) and boundary coditions (\ref{eq:noneq_sturm_liouville_boundary_conditions}) can be approximated by a specific linear combination of two solutions $y_1(\tau)$ and $y_2(\tau)$:
\begin{align}
\psi_0(\tau)&=\frac{1}{2\omega B}(e^{\omega T/2}y_1(\tau)+\omega e^{-\omega T/2}y_2(\tau)).
\end{align}
Then,
\begin{align}
\psi_0(T/2)&=\frac{1}{2\omega}(e^{\gamma T/2}+\omega e^{-\gamma T/2}).
\end{align}
From what was shown in Section \ref{sec:double_well_appendix}, we know that $\psi_{\lambda_0}(T/2)=0$. Similarly as before, we have
\begin{align}\begin{split}
0&\approx\psi_0(T/2)+\int_{_T/2}^{T/2}d\tau'G(T/2,\tau') \lambda_0\psi_0(\tau')\\
&=\psi_0(T/2)+\int_{_T/2}^{T/2}d\tau'\frac{-1}{\mathcal W}y_1(\tau')y_2(T/2)\lambda_0\frac{1}{2\omega B}(e^{\omega T/2}y_1(\tau')+\omega e^{-\omega T/2}y_2(\tau'))\\
&\approx\psi_0(T/2)+\int_{-T/2}^{T/2}d\tau'\frac{-1}{2B^2}y_1(\tau')Be^{T/2}\lambda_0\frac{1}{2\omega B}e^{\omega T/2}y_1(\tau')\\
&=\psi_0(T/2)-\frac{1}{2B^2}e^{T+\frac{1}{2}\gamma T}\lambda_0
\end{split}\end{align}
The second approximation holds since $e^{\omega T/2}\int d\tau' y_1(\tau')^2 \gg e^{-\omega T/2}\int d\tau' y_1(\tau')y_2(\tau')$. The last equality holds because of normalization of $y_1(\tau)$. We conclude that
\begin{align}
\ \lambda_0&=4\omega B^2e^{-T-\frac{1}{2}\gamma T}\psi_0(T/2)
\end{align}
We are interested in calculating the quantity
\begin{align}
 \kappa(\tau_0) \sqrt{\lambda_0}=\sqrt\frac{\psi_0^0(T/2)\lambda_0}{\psi_0(T/2)}=Be^{\gamma \tau_0/2}\sqrt{1+\omega}
\end{align}

One can calculate $\tilde\kappa(\tau_0) \sqrt{\tilde\lambda_0}$ precisely in the same manner. The difference is that now
\begin{align}
W(\tau)=\left\{\begin{array}{ll}1&\tau<\tau_0\\\omega&\tau\geq\tau_0\end{array}\right.
\end{align}
so the functions $y_1(\tau)$ and $y_2(\tau)$ are different than previously. The result yields
\begin{align}
\tilde\kappa(\tau_0)\sqrt{\tilde\lambda_0}&=B e^{-\gamma\tau_0/2}\sqrt{2\omega}
\end{align}

Finally, we can calculate right side of (\ref{eq:appendix_going_to_evolution_op})
\begin{align}\begin{split}
\mathcal N\left(\det{}'\left[\ldots\right]\right)^{-1/2}&=\frac{\omega^{1/4}}{\sqrt\pi}e^{-T/2}e^{-\frac{\gamma}{2}T_\omega}\left(\omega^{1/4}\sqrt\frac{2}{1+\omega}\right)^{2n+1}\times\\
&\quad\times\prod_{i=0}^nBe^{\gamma \tau_{2i}/2}\sqrt{1+\omega} \prod_{i=1}^nB e^{-\gamma\tau_{2i+1}/2}\sqrt{2\omega}\\
&=\sqrt\frac{\omega}{\pi}\left(\frac{2}{1+\omega}\right)^{1/4}e^{-\frac{T}{2}(1+\frac{\gamma}{2})}\left(2^{3/4}B\sqrt\omega(1+\omega)^{-1/4}\right)^{2n+1}
\end{split}\end{align}
Applying (\ref{eq:appendix_complete_sum}) we get the final formula
\begin{align}
\braket{a|e^{-TH}|0}&=\sum_n \frac{1}{(2n+1)!}\sqrt\frac{\omega}{2\pi}\left(\frac{2}{1+\omega}\right)^{1/4}e^{-\frac{T}{2}(1+\frac{\gamma}{2})}\left(2^{3/4}\sqrt\frac{\omega S_0}{\pi}e^{-S_0}(1+\omega)^{-1/4}BT\right)^{2n+1}\\
&=\sqrt\frac{\omega}{2\pi}\left(\frac{2}{1+\omega}\right)^{1/4}e^{-\frac{T}{2}(1+\frac{\gamma}{2})}\sinh\left(2^{3/4}\sqrt\frac{\omega S_0}{\pi}e^{-S_0}(1+\omega)^{-1/4}BT\right)
\end{align}
Recall that $B=\frac{aA}{\sqrt S_0}$ where $A=A_+^{\omega/(1+\omega)}A_-^{1/(1+\omega)}$ and $A_\pm$ are defined in (\ref{eq:definition_of_A}). This amplitude reveals two energies:
\begin{align}\label{eq:appendix_side_middle_amplitude}\begin{split}
E_-=\frac{1}{2}(1+\frac{\gamma}{2})-2^{3/4}(1+\omega)^{-1/4}\sqrt\frac{\omega}{\pi}e^{-S_0}aA,\\
E_+=\frac{1}{2}(1+\frac{\gamma}{2})+2^{3/4}(1+\omega)^{-1/4}\sqrt\frac{\omega}{\pi}e^{-S_0}aA.
\end{split}\end{align}

Let us now discuss general expression for the determinant $\mathcal N\big(\det{}'\big[-\frac{d^2}{d\tau^2}+V''(\bar x_{n}(\tau)/a)\big]\big)^{-1/2}$ when the classical trajectory $x_n(\tau)$ connects two minima with $V''(\bar x_n(-\infty))=\omega_1$ and $V''(\bar x_n(+\infty))=\omega_2$. For $\omega_1=\omega_2$ the number of instantons which build the trajectory $\bar x_n(\tau)$ must be even, i.e. $n$ is even. For $\omega_1\neq\omega_2$ number of instantons $n$ is odd. For each instanton going from central minimum to one of side minima there is the lowest eigenvalue corresponding to zero mode $\sqrt{\lambda_0}$ and the term $\kappa(\tau_i)$. For each instanton going the other way there are similar terms $\sqrt{\tilde\lambda_0}$ and $\tilde\kappa(\tau_i)$. For any transition there is additionally $\braket{E_0^1|E_0^\omega}$. The Euclidean evolution of a state in the middle minimum gives $e^{-T_\omega E_0^\omega}$ where $T_\omega$ is the total time which the trajectory spends in the central minimum. Evolution in side minima gives $e^{-T_1 E_0^1}$ where $T_1=T-T_\omega$. Similarly as before there is also term $\braket{x=0|E_0^{\omega_2}}\braket{x=0|E_0^{\omega_1}}=(\omega_1\omega_2)^{1/4}/\sqrt\pi$. If $\omega_1=\omega_2$ then there are $n/2$ instantons going from cental to side minimum and $n/2$ instantons going from side to central minimum. If $\omega_1=1$ and $\omega_2=\omega$ there are $(n+1)/2$ instantons starting at side minimum and ending at central minimum and $(n-1)/2$ the other instantons. Inserting expressions for $\kappa(\tau_i)$ and $\lambda_0$ we get
\begin{align}\label{eq:appendix_tripple_well_determinant}\begin{split}
\mathcal N\left(\det{}'\left[-\frac{d^2}{d\tau^2}+V''(\bar x_n(\tau)/a)\right]\right)^{-\frac{1}{2}}&=\frac{(\omega_1\omega_2)^{1/4}}{\sqrt \pi}
\left(\omega^{1/4}\sqrt\frac{2}{1+\omega}\frac{aA}{\sqrt {S_0}}\right)^ne^{-\frac{T}{2}\frac{\omega_1+\omega_2}{2}}\\
&\quad\times
\left\{\begin{array}{ll}(2\omega(1+\omega))^{n/4}&\omega_1=\omega_2\\
(2\omega(1+\omega))^{(n-1)/4}\sqrt{1+\omega}&\omega_1=\omega,\omega_2=1\\
(2\omega(1+\omega))^{(n-1)/4}\sqrt{2\omega}&\omega_1=1,\omega_2=\omega\\
\end{array}\right.
\end{split}\end{align}

It is now simple to calculate other amplitudes, e.g. $\braket{a|e^{-TH}|a}$. The difference is that now there are even number of instantons. An $2n$ instanton solution has $2^{n-1}$ topologically different trajectories so $N_{2n}=2^{n-1}$. The only exception is $N_0=1$ because there exists a trivial constant trajectory. We may write $N_{2n}=2^{n-1}+\frac{1}{2}\delta_{n,0}$. According to (\ref{eq:appendix_tripple_well_determinant}), with $\omega_1=\omega_2=1$,
\begin{align}\begin{split}
\mathcal N\Bigg(\det{}'\left[-\frac{d^2}{d\tau^2}+V''(\bar x_{2n}(\tau)/a)\right]\Bigg)^{-1/2}
=\frac{1}{\sqrt\pi}e^{-\frac{T}{2}}\left(2^{3/4}\frac{aA}{\sqrt S_0}\sqrt\omega(1+\omega)^{-1/4}\right)^{2n}.
\end{split}\end{align}
In analogy to (\ref{eq:appendix_complete_sum}) there is
\begin{align}
\braket{a|e^{-TH}|a}&=\sum_n N_{2n} e^{-2nS_0}\left(\sqrt\frac{S_0}{2\pi}\right)^{2n}\int_{\tau_i<\tau_{i+1}}d\tau_1\ldots d\tau_{2n}\mathcal N\left(\det{}'\left[-\frac{d^2}{d\tau^2}+V''(\bar x_{2n}(\tau)/a)\right]\right)^{-1/2}\nonumber\\
&=\sum_n\frac{N_{2n}}{(2n)!}\left(e^{-S_0}\sqrt\frac{S_0}{2\pi}T\right)^{2n}\frac{1}{\sqrt\pi}e^{-\frac{T}{2}}\left(2^{3/4}\frac{aA}{\sqrt S_0}\sqrt\omega(1+\omega)^{-1/4}\right)^{2n}\\
&=\frac{1}{2\sqrt\pi}e^{-\frac{T}{2}}\cosh\left(e^{-S_0}\sqrt\frac{\omega}{\pi}2^{3/4}(1+\omega)^{-1/4}aAT\right)+\frac{1}{2\sqrt\pi}e^{-\frac{T}{2}}\nonumber
\end{align}
There are three energies:
\begin{align}\begin{split}
E_0&=\frac{1}{2}-2^{3/4}(1+\omega)^{-1/4}\sqrt\frac{\omega}{\pi}e^{-S_0}aA\\
E_1&=\frac{1}{2}\\
E_2&=\frac{1}{2}+2^{3/4}(1+\omega)^{-1/4}\sqrt\frac{\omega}{\pi}e^{-S_0}aA
\end{split}\end{align}
We can see that this result is different than (\ref{eq:appendix_side_middle_amplitude}). Still, this can be explained. The first thing is that when we calculated $\braket{0|e^{-TH}|a}$ there were two energies. It is because $\braket{E_1|0}=0$ and the expansion of the amplitude $\braket{0|e^{-TH}|a}$ is
\begin{align}
\braket{0|e^{-TH}|a}&=\sum_n e^{-TE_n}\braket{0|E_n}\braket{E_n|0}.
\end{align}
Secondly, energies calculated using different amplitudes differ by a perturbative quantity $\frac{\gamma}{4}$! We know however that the instanton calculus does not take into account perturbative corrections of energies and thus any perturbative terms appearing in semiclassical approximation in Euclidean space can shall be ignored.

\begin{thebibliography}{10}

\bibitem{ITEP}
M.~A. Shifman.
\newblock {ITEP lectures on particle physics and field theory. Vol. 1, 2}.
\newblock {\em World Sci.Lect.Notes Phys.}, 62:1--875, 1999.

\bibitem{BPST}
A.~A. Belavin, Alexander~M. Polyakov, A.~S. Schwartz, and Yu.~S. Tyupkin.
\newblock {Pseudoparticle solutions of the Yang-Mills equations}.
\newblock {\em Phys. Lett.}, B59:85--87, 1975.

\bibitem{Schafer}
T.~Schafer and E.~V. Shuryak.
\newblock {Instantons in QCD}.
\newblock {\em Rev.Mod.Phys.}, 70:323--426, 1998.

\bibitem{Bender}
C.~M. Bender and T.~T. Wu.
\newblock {Anharmonic oscillator}.
\newblock {\em Phys. Rev.}, 184:1231--1260, 1969.

\bibitem{Caswell}
W.~E. Caswell.
\newblock {Accurate energy levels for the anharmonic oscillator and a summable
  series for the double well potential in perturbation theory}.
\newblock {\em Annals Phys.}, 123:153, 1979.

\bibitem{DuncanJones}
A.~Duncan and H.F. Jones.
\newblock {Convergence proof for optimized Delta expansion: The Anharmonic
  oscillator}.
\newblock {\em Phys.Rev.}, D47:2560--2572, 1993.

\bibitem{Meurice}
Y.~Meurice.
\newblock {A Simple method to make asymptotic series of Feynman diagrams
  converge}.
\newblock {\em Phys.Rev.Lett.}, 88:141601, 2002.

\bibitem{ZJ}
J.~Zinn-Justin.
\newblock {Multi - instanton contributions in quantum mechanics}.
\newblock {\em Nucl. Phys.}, B192:125--140, 1981.

\bibitem{ZJ1}
J.~Zinn-Justin.
\newblock {Multi - instanton contributions in quantum mechanics. 2.}
\newblock {\em Nucl.Phys.}, B218:333--348, 1983.

\bibitem{Bogomolny}
E.B. Bogomolny.
\newblock {Calculation if instanton - anti--instanton contributions in quantum
  mechanics}.
\newblock {\em Phys.Lett.}, B91:431--435, 1980.

\bibitem{ZJ2}
U.~D. Jentschura and J.~Zinn-Justin.
\newblock {Higher-order corrections to instantons}.
\newblock {\em J. Phys.}, A34:L253--L258, 2001.

\bibitem{Unsal}
M.~{\"U}nsal.
\newblock {Theta dependence, sign problems and topological interference}.
\newblock 2012.

\bibitem{Wosiek:2002}
J.~Wosiek.
\newblock {Spectra of supersymmetric Yang-Mills quantum mechanics}.
\newblock {\em Nucl.Phys.}, B644:85--112, 2002.

\bibitem{Dancoff}
S.M. Dancoff.
\newblock {Nonadiabatic meson theory of nuclear forces}.
\newblock {\em Phys.Rev.}, 78:382--385, 1950.

\bibitem{Wosiek}
M.~Trzetrzelewski and J.~Wosiek.
\newblock {Quantum systems in a cut Fock space}.
\newblock {\em Acta Phys. Polon.}, B35:1615--1624, 2004.

\bibitem{Trzetrzelewski:2004nz}
M.~Trzetrzelewski.
\newblock {Quantum mechanics in a cut Fock space}.
\newblock {\em Acta Phys. Polon.}, B35:2393--2416, 2004.

\bibitem{Coleman}
S.~Coleman.
\newblock {\em {Aspects of Symmetry: Selected Erice Lectures}}.
\newblock Cambridge University Press, 1988.

\bibitem{Bloch}
F.~Bloch.
\newblock {{\"U}ber die Quantenmechanik der Elektronen in Kristallgittern}.
\newblock {\em Z. Physik}, 52:555--600, 1928.

\bibitem{Wigner}
J.~von Neumann and E.P. Wigner.
\newblock {{\"U}ber merkw{\"u}rdige diskrete eigenwerte}.
\newblock {\em Z.Phys.}, 30:465--467, 1930.

\bibitem{tobepublished}
Z.~Ambrozi\'nski.
\newblock {to be published}.

\end{thebibliography}

\end{document}